\documentclass[12pt,reqno]{amsart}

\usepackage{epsfig}
\usepackage{amsmath, amsthm, amsfonts}
\usepackage{graphicx}

\numberwithin{equation}{section}

\newcommand{\di}{\displaystyle}

\newcommand{\wt}{\widetilde}

\newcommand{\R}{{\mathbb R}}
\newcommand{\C}{{\mathbb C}}

\newcommand{\Z}{{\mathbb Z}}

\renewcommand{\Re}{{\operatorname{Re\,}}}
\renewcommand{\Im}{{\operatorname{Im\,}}}

\newcommand{\Int}{{\operatorname{Int}\,}}

\newcommand{\Tr}{{{\operatorname{Tr}\,}}}

\newcommand{\diag}{{\operatorname{diag}}}

\newcommand{\Ai}{{\operatorname{Ai}}}

\newcommand{\al}{\alpha}
\newcommand{\be}{\beta}
\newcommand{\ga}{\gamma}
\newcommand{\Ga}{\Gamma}

\newcommand{\la}{\lambda}
\newcommand{\ep}{\varepsilon}
\newcommand{\de}{\delta}

\newcommand{\f}{\varphi}
\newcommand{\sg}{\sigma}
\newcommand{\om}{\omega}
\newcommand{\Om}{\Omega}
\newcommand{\z}{\zeta}

\newtheorem{theo}{{\sc \bf Theorem}}[section]
\newtheorem{cor}[theo]{{\sc \bf Corollary}}
\newtheorem{lem}[theo]{{\sc \bf Lemma}}
\newtheorem{prop}[theo]{{\sc \bf Proposition}}

\begin{document}

\title[Lectures on Random Matrix Models]
{Lectures on Random Matrix Models.
The Riemann-Hilbert Approach}

\author{Pavel M. Bleher}
\address{
Department of Mathematical Sciences,
Indiana University-Purdue University Indianapolis,
402 N. Blackford St., Indianapolis, IN 46202, U.S.A.}
\email{bleher@math.iupui.edu}

\thanks{The author is supported in part
by the National Science Foundation (NSF) Grant DMS-0354962.}

\date{\today}

\begin{abstract} This is a review of the Riemann-Hilbert approach
to the large $N$ asymptotics in random matrix models and its
applications. We discuss the following topics: random matrix models and orthogonal polynomials,
the Riemann-Hilbert approach to the large $N$ asymptotics of orthogonal polynomials
and its applications to the problem of universality in random matrix models,
the double scaling limits, the large $N$ asymptotics of the partition function,
 and random matrix models with external source.
\end{abstract}

\maketitle

\section{Introduction}

This article is a review of the Riemann-Hilbert approach to random matrix  
models. It is based on a series of 5 lectures given by  the author at the miniprogram on
``Random Matrices and their Applications'' at the Centre de recherches
math\,ematiques, Universit\'e de Montreal, in June 2005. The review contains  
5 lectures:
\begin{enumerate}
\item []
Lecture 1. Random matrix models and orthogonal polynomials.
\item []
Lecture 2. Large $N$ asymptotics of orthogonal polynomials. The Riemann-Hilbert
approach.
\item []
Lecture 3.
 Double scaling limit in a random matrix model.
\item []
Lecture 4. Large $N$ asymptotics of the partition function
of random matrix models.
\item []
Lecture 5. Random matrix models with external source.
\end{enumerate}
The author would like to thank John Harnad for his invitation to give
the series of lectures at the miniprogram. The lectures are based on the
joint works of the author with several coauthors: Alexander Its, Arno Kuijlaars,
Alexander Aptekarev, and Bertrand Eynard. The author is grateful to his
coauthors for an enjoyable collaboration.

\vskip 5mm

{\Large Lecture 1. Random matrix models and orthogonal polynomials}

\vskip 5mm

The first lecture gives an introduction to random matrix models and 
their relations to orthogonal polynomials. 

\section { Unitary ensembles of random matrices }

\subsection {Unitary ensemble with polynomial interaction}

Let $M=\left( M_{jk}\right)_{j,k=1}^N $ be a random Hermitian
matrix, $M_{kj}=\overline{M_{jk}}$,
 with respect to the probability distribution
\begin{equation}\label{intro1}
d\mu_N(M)=\frac{1}{Z_N}\,e^{-N\Tr V(M)}dM, 
\end{equation}
where 
\begin{equation}\label{intro2}
V(M)=\sum_{i=1}^p t_j M^j,\qquad p=2p_0,\qquad t_p>0,
\end{equation}
is a polynomial,
\begin{equation}\label{intro3}
dM=\prod_{j=1}^N dM_{jj}\prod_{j\not=k}^N d\Re M_{jk}d\Im M_{jk},
\end{equation}
the Lebesgue measure, and
\begin{equation}\label{intro4}
Z_N=\int_{\mathcal H_N}e^{-N\Tr V(M)}dM,
\end{equation}
the partition function. The distribution $\mu_N(dM)$ is invariant
with respect to any unitary conjugation,
\begin{equation}\label{intro4a}
M\to U^{-1}M U,\quad U\in U(N),
\end{equation}
hence the name of the ensemble.

\subsection{ Gaussian unitary ensemble}

For $V(M)=M^2$, the measure $\mu_N$ is the probability distribution
of the Gaussian unitary ensemble (GUE). In this case,
\begin{equation}\label{intro5}
\Tr V(M)=\Tr M^2=\sum_{j,k=1}^N
M_{kj}M_{jk}=
\sum_{j=1}^N
M_{jj}^2+2\sum_{j>k}|M_{jk}|^2,
\end{equation}
hence
\begin{equation}\label{intro6}
\mu_N^{\rm GUE}(dM)=\frac{1}{Z_N^{\rm GUE}}\,\prod_{j=1}^N\left(e^{-NM_{jj}^2}\right)
\prod_{j>k}\left(e^{-2N|M_{jk}|^2}\right)dM,
\end{equation}
so that the matrix elements in GUE are independent Gaussian random variables.
The partition function of GUE is evaluated as
\begin{equation}\label{intro6a}
\begin{aligned}
Z_N^{\rm GUE}&=\int_{\mathcal H_N}\prod_{j=1}^N\left(e^{-NM_{jj}^2}\right)
\prod_{j>k}\left(e^{-2N|M_{jk}|^2}\right)dM=\left(\frac{\pi}{N}\right)^{N/2}
\left(\frac{\pi}{2N}\right)^{N(N-1)/2}\\
&=\left(\frac{\pi}{N}\right)^{N^2/2}\left(\frac{1}{2}\right)^{N(N-1)/2}.
\end{aligned}
\end{equation}
If $V(M)$ is not quadratic then the matrix elements $M_{jk}$ are
dependent.

\subsection{\ Topological large $N$ expansion}

Consider the {\it free energy} of the unitary ensemble of random matrices,
\begin{equation}\label{intro7}
F_N^0= -N^{-2}\ln  Z_N
=-N^{-2}\int_{\mathcal H_N}e^{-N\Tr V(M)}dM,
\end{equation}
and the {\it normalized free energy}, 
\begin{equation}\label{intro8}
F_N= -N^{-2}\ln \frac{Z_N}{Z_N^{\rm GUE}}
=-N^{-2}\ln \frac{\int_{\mathcal H_N}e^{-N\Tr V(M)}dM}
{\int_{\mathcal H_N}e^{-N\Tr M^2}dM}.
\end{equation}
The normalized free energy can be expessed as
\begin{equation}\label{intro9}
F_N=-N^{-2}\ln\left\langle e^{-N\Tr V_1(M)}\right\rangle
\end{equation}
where $V_1(M)=V(M)-M^2$ and
\begin{equation}\label{intro10}
\left\langle f(M) \right\rangle=
\frac{\int_{\mathcal H_N}f(M) e^{-N\Tr M^2}dM}
{\int_{\mathcal H_N}e^{-N\Tr M^2}dM},
\end{equation}
the mathematical expectation of $f(M)$ with respect to GUE.
Suppose that
\begin{equation}\label{intro11}
V(M)=M^2+t_3M^3+\ldots+t_pM^p.
\end{equation}
Then (\ref{intro8}) reduces to
\begin{equation}\label{intro12}
F_N=-N^{-2}\ln\left\langle e^{-N\Tr (t_3M^3+\ldots+t_pM^p)}\right\rangle.
\end{equation} 
$F_N$ can be expanded into the asymptotic series
in negative powers of $N^2$,
\begin{equation}\label{intro13}
F_N\sim F+\sum_{g=1}^\infty \frac{F^{(2g)}}{N^{2g}},
\end{equation} 
which is called the {\it topological large $N$ expansion}.
The Feynman diagrams representing $F^{(2g)}$ are realized 
on a two-dimensional Riemann closed manifold of genus $g$,
and, therefore, $F_N$ serves as a generating function
for enumeration of graphs on Riemannian manifolds, see, e.g.,
the works [BIZ], [BIPZ], [IZ], [DiF], [EM], [EMP]. 
This in turn leads to a fascinating relation between
the matrix integrals and the quantum gravity, see, e.g., the works [DGZ], [Wit],
and others.

\subsection{Ensemble of eigenvalues }

The Weyl integral formula implies, see, e.g., \cite{Meh}, 
that the distribution of eigenvalues of $M$
with respect to the ensemble $\mu_N$ is given as
\begin{equation}\label{intro14}
d\mu_N(\la)=\frac{1}{\tilde Z_N}\,\prod_{j>k}(\la_j-\la_k)^2
\prod_{j=1}^N e^{-NV(\la_j)}d\la,
\end{equation} 
where
\begin{equation}\label{intro15}
\tilde Z_N=\int\prod_{j>k}(\la_j-\la_k)^2
\prod_{j=1}^N e^{-NV(\la_j)}d\la,\qquad
d\la=d\la_1\ldots d\la_N.
\end{equation} 
Respectively, for GUE,
\begin{equation}\label{intro14a}
d\mu_N^{\rm GUE}(\la)=\frac{1}{\tilde Z_N^{\rm GUE}}\,\prod_{j>k}(\la_j-\la_k)^2
\prod_{j=1}^N e^{-N\la_j^2}d\la,
\end{equation} 
where
\begin{equation}\label{intro15a}
\tilde Z_N^{\rm GUE}=\int\prod_{j>k}(\la_j-\la_k)^2
\prod_{j=1}^N e^{-N \la_j^2}d\la.
\end{equation}
The constant $\tilde Z_N^{\rm GUE}$ is a Selberg integral, and its
exact value is
\begin{equation}\label{intro15b}
\tilde Z_N^{\rm GUE}=\frac{(2\pi)^{N/2}}{(2N)^{N^2/2}}\prod_{n=1}^N n!
\end{equation} 
see, e.g., \cite{Meh}. 
The partition functions $Z_N$ and $\tilde Z_N$ are related as follows:
\begin{equation}\label{intro15c}
\frac{\tilde Z_N}{Z_N}=\frac{\tilde Z_N^{\rm GUE}}{Z_N^{\rm GUE}}
=\frac{1}{\pi^{N(N-1)/2}}\prod_{n=1}^N n!
\end{equation} 
One of the main problems is to evaluate the large $N$ asymptotics 
of the partition function $\tilde Z_N$ and of the correlations between  eigenvalues.

The $m$-point correlation function is given as
\begin{equation}\label{intro16}
\begin{aligned}
{}&R_{mN}(x_1,\dots,x_m)
=\frac{N!}{(N-m)!}
\int_{\R^{N-m}}p_N(x_1,\dots,x_N)dx_{m+1}\dots dx_N,
\end{aligned}
\end{equation} 
where
\begin{equation}\label{intro17}
p_N(x_1,\dots,x_N)=
\wt Z_N^{-1}\prod_{j>k}(x_j-x_k)^2
\prod_{j=1}^N e^{-NV(x_j)}.
\end{equation} 
The Dyson determinantal formula for correlation functions, see, e.g., \cite{Meh}, is
\begin{equation}\label{intro18}
R_{mN}(x_1,\dots,x_m)=\det \left( K_N(x_k,x_l)\right)_{k,l=1}^m,
\end{equation} 
where
\begin{equation}\label{intro19}
K_N(x,y)=\sum_{n=0}^{N-1}\psi_n(x)\psi_n(y)
\end{equation} 
and
\begin{equation}\label{intro20}
\psi_n(x)=\frac{1}{h_n^{1/2}}P_n(x)e^{-NV(x)/2},
\end{equation} 
where $P_n(x)=x^n+a_{n-1}x^{n-1}+\dots$ are monic orthogonal polynomials,
\begin{equation}\label{intro21}
\int_{-\infty}^\infty P_n(x)P_m(x)e^{-NV(x)}dx=h_n\de_{nm}.
\end{equation} 
Observe that the functions $\psi_n(x)$, $n=0,1,2,\ldots$, form an orthonormal basis in $L^2(\R^1)$,
and $K_N$ is the kernel of the projection operator on tne $N$ dimensional space
generated by the first $N$ functions $\psi_n$, $n=0,\ldots,N-1$. The kernel $K_N$ can
be expressed in terms of $\psi_{N-1}$, $\psi_N$ only, due to the 
Christoffel-Darboux formula. Consider first recurrent and differential equations for
the functions $\psi_n$.
 
\subsection{Recurrence equations and discrete string equations for orthogonal polynomials}

The orthogonal polynomials satisfy the {\it three term recurrent relation}, see, e.g. [Sze],

\begin{equation}\label{recu_1}
\begin{aligned}
xP_n(x)&=P_{n+1}(x)+\be_n P_n(x)+\ga_n^2 P_{n-1}(x),\\
\ga_n&=\left(\frac{h_n}{h_{n-1}}\right)^{1/2}>0\,,\quad n\ge 1;\qquad
\ga_0=0.
\end{aligned}
\end{equation} 
For the functions $\psi_n$ it reads as
\begin{equation}\label{recu_2}
x\psi_n(x)=\ga_{n+1}\psi_{n+1}(x)+\be_n \psi_n(x)+\ga_n \psi_{n-1}(x).
\end{equation}
This allows the following calculation:
\begin{equation}\label{recu_3}
\begin{aligned}
(x-y)\sum_{n=0}^{N-1}\psi_n(x)\psi_n(y)&=
\sum_{n=0}^{N-1}\left[\left(\ga_{n+1}\psi_{n+1}(x)
+\be_n\psi_n(x)+\ga_n\psi_{n-1}(x)\right)\psi_n(y)\right.\\
{}&\left.-\psi_n(x)
\left(\ga_{n+1}\psi_{n+1}(y)
+\be_n\psi_n(y)+\ga_n\psi_{n-1}(y)\right)
\right]\\
&=\ga_N\left[ \psi_N(x)\psi_{N-1}(y)-\psi_{N-1}(x)\psi_N(y)\right]
\end{aligned}
\end{equation}
(telescopic sum), hence
\begin{equation}\label{recu_4}
\begin{aligned}
K_N(x,y)=\sum_{n=0}^{N-1}\psi_n(x)\psi_n(y)=\ga_N
\frac{\psi_N(x)\psi_{N-1}(y)-\psi_{N-1}(x)\psi_N(y)}{x-y}\,.
\end{aligned}
\end{equation}
which is the {\it Christoffel-Darboux formula}. For the density function we
obtain that
\begin{equation}\label{recu_5}
\begin{aligned}
{}&p_N(x)=\frac{Q_N(x,x)}{N}=\frac{\ga_N}{N}
\left[\psi'_N(x)\psi_{N-1}(x)-\psi'_{N-1}(x)\psi_N(x)\right]\,.
\end{aligned}
\end{equation}

Consider a matrix $Q$ of the operator of multiplication
by $x$, $f(x)\to xf(x)$, in the basis $\{\psi_n(x)\}$.
Then by (\ref{recu_2}), $Q$ is the symmetric tridiagonal Jacobi
matrix,
\begin{equation}\label{recu_6}
Q=\begin{pmatrix}
\be_0 & \ga_1 & 0 & 0 &\dots \\
\ga_1 &\be_1 & \ga_2 & 0 & \dots \\ 
0 & \ga_2 & \be_2 & \ga_3 & \dots \\
0 & 0 & \ga_3 & \be_3 &  \dots \\
\vdots & \vdots & \vdots & \vdots & \ddots
\end{pmatrix}.
\end{equation} 
Let $P=\left( P_{nm}\right)_{n,m=0,1,2,\dots}$ 
be a matrix of the operator $f(z)\to f'(z)$ in the basis
$\psi_n(z)$, $n=0,1,2,\dots$, so that
\begin{equation}\label{recu_7}
P_{nm}=\int_{-\infty}^\infty\psi_n(x)\psi'_m(x)\,dx.
\end{equation} 
Then $P_{mn}=-P_{nm}$ and
\begin{equation}\label{recu_7a}
\begin{aligned}
\psi_n'(z)&=-\frac{NV'(z)}{2}\psi_n(z)+\frac{P'_n(z)}{\sqrt {h_n}}
e^{-NV(z)/2}=-\frac{NV'(z)}{2}\psi_n(z)+\frac{n}{\ga_n}\psi_{n-1}(z)+\dots,
\end{aligned}
\end{equation} 
hence
\begin{equation}\label{recu_8}
\begin{aligned}
{}&\left[ P+\frac{NV'(Q)}{2}\right]_{nn}=0,\qquad
\left[ P+\frac{NV'(Q)}{2}\right]_{n,n+1}=0,\\
{}&\left[ P+\frac{NV'(Q)}{2}\right]_{n,n-1}=\frac{n}{\ga_n}\,.
\end{aligned}
\end{equation} 
Since $P_{nn}=0$, we obtain that 
\begin{equation}\label{recu_9}
[V'(Q)]_{nn}=0.
\end{equation}
In addition,
\begin{equation}\label{recu_10}
\left[ -P+\frac{NV'(Q)}{2}\right]_{n,n-1}=0,\qquad
\left[ P+\frac{NV'(Q)}{2}\right]_{n,n-1}=\frac{n}{\ga_n}\,,
\end{equation}
hence
\begin{equation}\label{recu_11}
\ga_n[V'(Q)]_{n,n-1}=\frac{n}{N}\,.
\end{equation}
Thus, we have the {\it discrete string equations} for the recurrent coefficients,
\begin{equation}\label{recu_12}
\left\{\;
\begin{aligned}
{}&\ga_n[V'(Q)]_{n,n-1}=\frac{n}{N}\,,\\
{}&\\
{}&[V'(Q)]_{nn}=0.
\end{aligned}
\right.
\end{equation}
The string equations can be brought to a variational form. 

\begin{prop}
Define
the infinite Hamiltonian,
\begin{equation}\label{recu_13}
H(\ga,\be)=N\Tr V(Q)-\sum_{n=1}^\infty n \ln \ga_n^2,
\qquad \ga=(\ga_0,\ga_1,\ldots),\quad \be=(\be_0,\be_1,\ldots).
\end{equation}
Then equations (\ref{recu_12}) can be written as
\begin{equation}\label{recu_14}
\frac{\partial H}{\partial \ga_n}=0,
\quad
\frac{\partial H}{\partial \be_n}=0; \qquad n\ge 1,
\end{equation}
which are the Euler-Lagrange equations for the Hamiltonian $H$.
\end{prop}

\begin{proof} We have that
\begin{equation}\label{recu_15}
\frac{\partial H}{\partial \ga_n}=N\Tr\left(V'(Q)\frac{\partial Q}{\partial \ga_n}\right)
-\frac{2n}{\ga_n}=2N[V'(Q)]_{n,n-1}-\frac{2n}{\ga_n},
\end{equation}
and
\begin{equation}\label{recu_16}
\frac{\partial H}{\partial \be_n}=N\Tr\left(V'(Q)\frac{\partial Q}{\partial \be_n}\right)
=N[V'(Q)]_{nn},
\end{equation}
hence equations (\ref{recu_12}) are equivalent to  (\ref{recu_14}).
\end{proof}

\begin{figure}
\scalebox{0.6}{\includegraphics{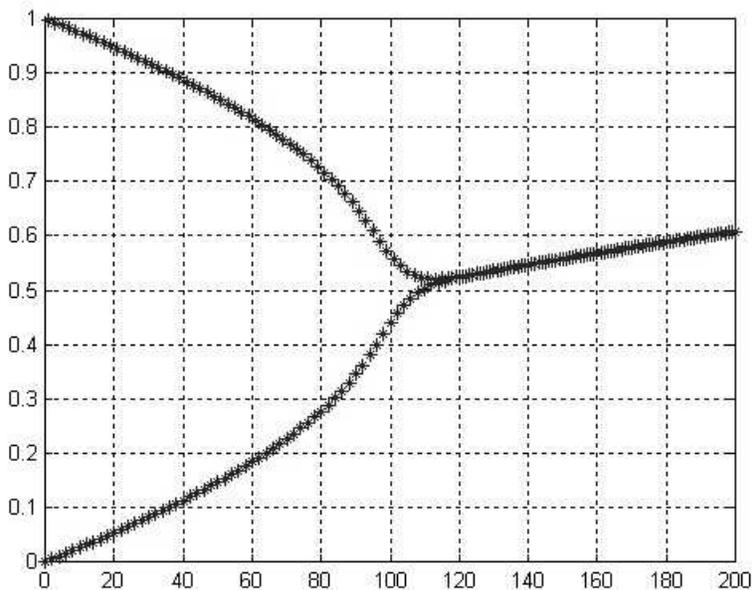}}
\caption{A computer solution, $y=\ga_n^2$, of the string equation for the quartic
model: $g=1$, $t=-1$, $N=400$.}
\label{ga_n_quartic}
\end{figure}

{\bf Example.} The even quartic model,
\begin{equation}\label{recu_17}
V(M)=\frac{t}{2}M^2+\frac{g}{4}M^4.
\end{equation}
In this case, since $V$ is even, $\be_n=0$, and we have one string equation,
\begin{equation}\label{recu_18}
\ga_n^2\left( t+g\ga_{n-1}^2+g\ga_{n}^2+g\ga_{n+1}^2\right)=
\frac{n}{N}\,,
\end{equation}
with the initial conditions: $\ga_0=0$
and
\begin{equation}\label{recu_19}
\ga_1=\frac{\di\int_{-\infty}^\infty z^2 e^{-NV(z)}dz}
{\di\int_{-\infty}^\infty e^{-NV(z)}dz}\,.
\end{equation}
The Hamiltonian is
\begin{equation}\label{recu_20}
H(\ga)= \sum_{n=1}^\infty \left[\frac{N}{2}
\ga_n^2\left( 2t+g\ga_{n-1}^2+g\ga_{n}^2+g\ga_{n+1}^2\right)
-n \ln \ga_n^2\right]\,.
\end{equation}
The minimization of the functional $H$ is a useful procedure for a numerical solution
of the string equations, see [BDJT], [BI2]. 
The problem with the initial value problem for
the string equations, with the initial values $\ga_0=0$ and
(\ref{recu_19}), is  that it is very unstable, while the minimization of $H$
with $\ga_0=0$ and some boundary conditions at $n=N$, say $\ga_N=0$, works very well.
In fact, the boundary condition at $n=N$ creates a narrow boundary layer near $n=N$,
and it does not affect significally the main part of the graph of $\ga_n^2$.
Fig.\ref{ga_n_quartic} presents a  
computer solution, $y=\ga_n^2$, of the string equation for the quartic
model: $g=1$, $t=-1$, $N=400$. For this solution, as shown in [BI1], [BI2],
 there is a critical value,
$\la_c=\frac {1}{4}$, so that for any $\ep>0$, as $N\to\infty$, 
\begin{equation}\label{recu_21a}
\ga_n^2=
R\left(\frac{n}{N}\right)+O(N^{-1}),\quad {\rm if} \quad \frac {n}{N}\ge \la_c+\ep,
\end{equation}
and
\begin{equation}\label{recu_21b}
\ga_n^2=\left\{
\begin{aligned}
&R\left(\frac{n}{N}\right)+O(N^{-1}),\quad n=2k+1,\\
&L\left(\frac{n}{N}\right)+O(N^{-1}),\quad n=2k,
\end{aligned}
\right.\qquad {\rm if} \quad \frac {n}{N}\le \la_c-\ep.
\end{equation}
The functions $R$ for $\la\ge \la_c$ and $R,L$ for $\la\le \la_c$
can be found from string equation (\ref{recu_18}):
\begin{equation}\label{recu_22}
R(\la)=\frac{1+\sqrt{1+12\la}}{6},\qquad \la>\la_c\,,
\end{equation}
and
\begin{equation}\label{recu_23}
R(\la),L(\la)=\frac{1\pm\sqrt{1-4\la}}{2}\,,\qquad \la<\la_c\,.
\end{equation}
We will discuss below how to justify asymptotics (\ref{recu_21a}), (\ref{recu_21b}),
and their extension for a general $V$.

\subsection{Differential equations for the $\psi$-functions.}
Define 
\begin{equation}\label{LP_1}
\vec \Psi_n(z)=
\begin{pmatrix} \psi_n(z) \\ \psi_{n-1}(z)
\end{pmatrix}\,.
\end{equation}
Then
\begin{equation}\label{LP_2}
\vec\Psi'_n(z)=NA_n(z)\vec\Psi_n(z),
\end{equation}
where
\begin{equation}\label{LP_3}
A_n(z)=
\begin{pmatrix}
-\di\frac{V'(z)}{2} -\ga_nu_n(z) & \ga_nv_n(z) \\
-\ga_n v_{n-1}(z) & \di\frac{V'(z)}{2} +\ga_nu_n(z)
\end{pmatrix}
\end{equation}
and
\begin{equation}\label{LP_4}
u_n(z)=[W(Q,z)]_{n,n-1},\qquad
v_n(z)=[W(Q,z)]_{nn},
\end{equation}
where
\begin{equation}\label{LP_5}
W(Q,z)=\frac{V'(Q)-V'(z)}{Q-z}\,.
\end{equation}
Observe that $\Tr A_n(z)=0.$

{\bf Example.} Even quartic model,
$V(M)=\frac{t}{2}M^2+\frac{g}{4}M^4\,.$ 
Matrix $A_n(z)$:
\begin{equation}\label{LP_5a}
A_n(z)=
\begin{pmatrix}
-\frac{1}{2}(tz+gz^3)-g\ga_n^2z
 &  \ga_n(gz^2+\theta_n) \\
-\ga_n(gz^2+\theta_{n-1}) & \frac{1}{2}(tz+gz^3)+g\ga_n^2z
\end{pmatrix}
\end{equation}
where
\begin{equation}\label{LP_5b}
\theta_n=t+g\ga_n^2+g\ga_{n+1}^2.
\end{equation}

\subsection{ Lax pair for the discrete string equations.} 

Three term recurrence relation (\ref{recu_1}) can be written as
\begin{equation}\label{LP_6}
\vec\Psi_{n+1}(z)=U_n(z)\vec\Psi_n(z),
\end{equation}
where
\begin{equation}\label{LP_7}
U_n(z)=
\begin{pmatrix}
\ga^{-1}_{n+1}(z-\be_n) & -\ga^{-1}_{n+1}\ga_n \\
1 & 0 
\end{pmatrix}
\end{equation}
Differential equation (\ref{LP_2}) and recurrence equation (\ref{LP_6}) form
the Lax pair for discrete string equations (\ref{recu_12}).
This means that the compatibility condition of (\ref{LP_2}) and (\ref{LP_6}),
\begin{equation}\label{LP_8}
U_n'=N(A_{n+1}U_n-U_nA_n),
\end{equation}
when written for the matrix elements, implies (\ref{recu_12}).

\section{The Riemann-Hilbert problem for orthogonal polynomials}

\subsection{Adjoint functions.}
Introduce the {\it adjoint functions} to $P_n(z)$ as
\begin{equation}\label{RH_1}
Q_n(z)=\frac{1}{2\pi i}\int_{-\infty}^\infty
\frac{P_n(u)\,w(u)\,du}{u-z}\,,
\quad z\in\C\setminus\R,
\end{equation}
where
\begin{equation}\label{RH_2}
w(z)=e^{-NV(z)}
\end{equation}
is the weight for the orthogonal polynomials $P_n$.
Define
\begin{equation}\label{RH_3}
Q_{n\pm}(x)=\lim_{\substack{z\to x \\ \pm\Im z>0}}Q_n(z),
\quad -\infty< x< \infty.
\end{equation}
Then the well-known formula for the jump of the Cauchy type integral
gives that 
\begin{equation}\label{RH_4}
Q_{n+}(x)-Q_{n-}(x)=w(x)P_n(x).
\end{equation}
The asymptotics of $Q_n(z)$ as $z\to\infty$, $z\in\C$, is given as
\begin{equation}\label{RH_5}
\begin{aligned}
Q_n(z)&=\frac{1}{2\pi i}\int_{-\infty}^\infty
\frac{w(u)P_n(u)\,du}{u-z}
\cong -\frac{1}{2\pi i}\int_{-\infty}^\infty
w(u)P_n(u)\sum_{j=0}^\infty \frac{u^j}{z^{j+1}}\,du\\
&=-\frac{h_n}{2\pi iz^{n+1}}+\sum_{j=n+2}^\infty \frac{\al_j}{z^j}\,,
\end{aligned}
\end{equation}
(due to the orthogonality, the first $n$ terms
cancel out). The sign $\cong$ in (\ref{RH_5}) means an asymptotic expansion, 
so that for any $k\ge n+2$, there
exists a constant $C_k>0$ such that for all $z\in\C$,
\begin{equation}\label{RH_6}
Q_n(z)-
\left(-\frac{h_n}{2\pi iz^{n+1}}+\sum_{j=n+2}^k \frac{\al_j}{z^j}\right)
\le \frac{C_k}{(1+|z|)^{k+1}}\,.
\end{equation}
It can be some doubts in the uniformity of this asymptotics near the real axis, but we assume
that the weight $w(z)$ is analytic in a strip $\{z:\; |\Im z|\le a\}$, $a>0$,
hence the contour of integration in (\ref{RH_5}) can be shifted, and (\ref{RH_6}) holds
uniformly in the complex plane. 

\subsection{The Riemann-Hilbert problem.} Introduce now the matrix-valued function,
\begin{equation}\label{RH_7}
Y_n(z)=\begin{pmatrix}
P_n(z) & Q_n(z) \\
CP_{n-1}(z) & CQ_{n-1}(z) 
\end{pmatrix},
\end{equation}
where the constant,
\begin{equation}\label{RH_8}
C=-\frac{2\pi i}{h_{n-1}}\,,
\end{equation}
is chosen in such a way that
\begin{equation}\label{RH_9}
CQ_{n-1}(z)\cong
\frac{1}{z^n}+\ldots,
\end{equation}
see (\ref{RH_5}).
The function $Y_n$ solves the following Riemann-Hilbert problem (RHP):
\begin{enumerate}
\item $Y_n(z)$ is analytic on $\C^+\equiv \{\Im z\ge 0\}$
and $\C^-\equiv \{ \Im z\le 0\}$ (two-valued on $\R=\C^+\cap\C^-$).
\item For any real $x$,
\begin{equation}\label{RH_10}
Y_{n+}(x)=Y_{n-}(x)j_Y(x),\quad j_Y(x)=
\begin{pmatrix}
1 & w(x) \\
0 & 1 
\end{pmatrix}.
\end{equation}
\item As $z\to\infty,$
\begin{equation}\label{RH_11}
Y_n(z)\cong
\left(\di I+\sum_{k=1}^\infty \frac{Y_k}{z^k}\right)
\begin{pmatrix}
z^n & 0 \\
0 & z^{-n} 
\end{pmatrix}
\end{equation}
where $Y_k,\; k=1,2,\dots$, are some constant $2\times
  2$ matrices.
\end{enumerate}
Observe that (\ref{RH_10}) follows from (\ref{RH_4}), while
(\ref{RH_11}) from (\ref{RH_9}). The RHP (1)--(3) has some nice properties.

First of all, (\ref{RH_7}) is the only
solution of the RHP. Let us sketch a proof of the uniqueness. It follows from 
(\ref{RH_10}), that
\begin{equation}\label{RH_12}
\det Y_{n+}(x)=\det Y_{n-}(x),
\end{equation}
hence $\det Y_n(z)$ has no jump at the real axis, and hence $Y_n(z)$ is
an entire function. At infinity, by (\ref{RH_11}),
\begin{equation}\label{RH_13}
\det Y_n(z)\cong 1+\ldots
\end{equation}
hence 
\begin{equation}\label{RH_13a}
\det Y_n(z)\equiv 1,
\end{equation}
by the Liouville theorem. In particular,
$Y_n(z)$ is invertible for any $z$. Suppose that $\tilde Y_n$
is another solution of the RHP. Then $X_n=\tilde Y_nY_n^{-1}$ satisfies
\begin{equation}\label{RH_14}
X_{n+}(x)=\tilde Y_{n+}(x)Y_{n+}(x)^{-1}=\tilde Y_{n-}(x)j_{Y}(x)j_{Y}(x)^{-1}Y_{n-}^{-1}(x)=X_{n-}(x),
\end{equation}
hence $X_n$ is an entire matrix-valued function. At infinity, by (\ref{RH_11}),
\begin{equation}\label{RH_16}
X_n(z)\cong I+\ldots
\end{equation}
hence $X_n(z)\equiv I$, by the Liouville theorem. This implies that $\tilde Y_n=Y_n$,
the uniqueness.

The recurrent coefficients for the orthogonal polynomials
can be found as
\begin{equation}\label{RH_17}
\ga_n^2=[Y_1]_{21}[Y_1]_{12},
\end{equation}
and
\begin{equation}\label{RH_18}
\be_{n-1}=\frac{[Y_2]_{21}}{[Y_1]_{21}}-[Y_1]_{11}\,.
\end{equation}
Indeed, from (\ref{RH_7}), (\ref{RH_11}),
\begin{equation}\label{RH_19}
Y_n(z)\begin{pmatrix}
z^{-n} & 0 \\
0 & z^n 
\end{pmatrix}
=\begin{pmatrix}
z^{-n}P_n(z) & z^nQ_n(z) \\
Cz^{-n}P_{n-1}(z) & z^nCQ_{n-1}(z) 
\end{pmatrix}\cong 
I+\sum_{k=1}^\infty \frac{Y_k}{z^k}\,,
\end{equation}
hence by (\ref{RH_5}), (\ref{RH_8}), and (\ref{recu_1}),
\begin{equation}\label{RH_20}
[Y_1]_{21}[Y_1]_{12}=\left(-\frac{2\pi i}{h_{n-1}}\right)\left(-\frac{h_n}{2\pi i}\right)
=\frac{h_n}{h_{n-1}}=\ga_n^2,
\end{equation}
which proves (\ref{RH_17}). Also,
\begin{equation}\label{RH_21}
\frac{[Y_2]_{21}}{[Y_1]_{21}}-[Y_1]_{11}=p_{n-1,n-2}-p_{n,n-1}\,,
\end{equation}
where
\begin{equation}\label{RH_22}
P_n(z)=\sum_{j=0}^n p_{nj}z^j.
\end{equation}
From (\ref{recu_1}) we obtain that
\begin{equation}\label{RH_23}
p_{n-1,n-2}-p_{n,n-1}=\be_{n-1}\,,
\end{equation}
hence (\ref{RH_18}) follows. The normalizing constant $h_n$ can be found as
\begin{equation}\label{RH_23a}
h_n=-2\pi i[Y_1]_{12},\quad h_{n-1}=-\frac{2\pi i}{[Y_1]_{21}}\,.
\end{equation}

The reproducing kernel $K_N(x,y)$ of the eigenvalue correlation functions,
see (\ref{intro18}), is expressed in terms of $Y_{N+}(x)$ as follows:
\begin{equation}\label{RH_24}
K_N(x,y)=e^{-\frac{NV(x)}{2}}e^{-\frac{NV(y)}{2}}
\frac{1}{2\pi i(x-y)}
\begin{pmatrix} 0 & 1 \end{pmatrix}
Y_{N+}^{-1}(y)Y_{N+}(x)
\begin{pmatrix} 1 \\ 0 \end{pmatrix}.
\end{equation}
Indeed, by (\ref{recu_4}),
\begin{equation}\label{RH_25}
\begin{aligned}
K_N(x,y)&=\ga_N
\frac{\psi_N(x)\psi_{N-1}(y)-\psi_{N-1}(x)\psi_N(y)}{x-y}\\
&=e^{-\frac{NV(x)}{2}}e^{-\frac{NV(y)}{2}}
\frac{\ga_N}{\sqrt{h_Nh_{N-1}}}\,
\frac{P_N(x)P_{N-1}(y)-P_{N-1}(x)P_N(y)}{x-y}\,.
\end{aligned}
\end{equation}
From (\ref{RH_7}), (\ref{RH_8}) and (\ref{RH_13a}), we obtain that
\begin{equation}\label{RH_26}
\begin{pmatrix} 0 & 1 \end{pmatrix}
Y_{N+}^{-1}(y)Y_{N+}(x)
\begin{pmatrix} 1 \\ 0 \end{pmatrix}
=\frac{2\pi i}{h_{N-1}}\left[P_N(x)P_{N-1}(y)-P_{N-1}(x)P_N(y)\right].
\end{equation}
Also,
\begin{equation}\label{RH_27}
\frac{\ga_N}{\sqrt{h_Nh_{N-1}}}=\frac{1}{h_{N-1}},
\end{equation}
hence equation (\ref{RH_24}) follows.

\section{Distribution of
  eigenvalues and equilibrium measure } 
 
\subsection{Heuristics.}
We begin with some heuristic considerations to explain why we expect
that the limiting distribution of eigenvalues solves a variational problem.
Let us rewrite  (\ref{intro15}) as 
\begin{equation}\label{equi1}
d\mu_N(\la)=\tilde Z_N^{-1}e^{-H_N(\la)}d\la,
\end{equation} 
where
\begin{equation}\label{equi2}
H_N(\la)=-\sum_{j\not=k} \ln |\la_j-\la_k|+N\sum_{j=1}^N
V(\la_j). 
\end{equation}
Given $\la$, introduce the probability measure on $\R^1$,
\begin{equation}\label{equi3}
d\nu_{\la}(x)=N^{-1}\di\sum_{j=1}^N \de(x-\la_j)dx.
\end{equation}
Then (\ref{equi2}) can be rewritten as
\begin{equation}\label{equi4}
H_N(\la)=N^2
\left[-\iint_{x\not=y} \ln |x-y| d\nu_{\la}(x) d\nu_{\la}(y)
+\int V(x) d\nu_{\la}(x)\right].
\end{equation}
Let $\nu$ be an arbitrary probability measure on $\R^1$. Set
\begin{equation}\label{equi5}
I_V(\nu)=-\iint_{x\not=y} \ln |x-y| d\nu(x) d\nu(y)
+\int V(x) d\nu(x).
\end{equation}
Then (\ref{equi1}) reads
\begin{equation}\label{equi6}
d\mu_N(\la)=\tilde Z_N^{-1}e^{-N^2I_V(\nu_{\la})}d\la.   
\end{equation} 
Because of the factor $N^2$ in the exponent,
we expect that for large $N$ the measure $\mu_N$
is concentrated near the minimum
of the functional $ I_V$,
i.e. near the {\it equilibrium measure} $\nu_V$.

\subsection{Equilibrium measure.}
Consider the minimization problem
\begin{equation}\label{equi7}
E_V=\inf_{\nu\in M_1(\R)} I_V(\nu),
\end{equation} 
where
\begin{equation}\label{equi8}
M_1(\R)=\left\{\nu\,:\,\nu\ge 0,\; \int_{\R}d\nu=1\right\},
\end{equation} 
the set of probability measures on the line.

\begin{prop} {\rm (See [DKM].)} The infinum of $I_V(\nu)$
is attained uniquely at
a measure $\nu=\nu_V$, which is called an equilibrium measure. The measure 
\(\nu_V\) is absolutely continuous, and it is supported by a finite union of intervals,
\( J=\cup_{j=1}^q [a_j,b_j] \). On the support, its density has the form
\begin{equation}\label{equi11}
p_V(x)\equiv \frac{d\nu_V}{dx}(x)=\frac{1}{2\pi i} h(x)R^{1/2}_+(x),\qquad
R(x)=\prod_{j=1}^q
(x-a_j)(x-b_j).
\end{equation} 
Here \(R^{1/2}(x) \)
is the branch with cuts on \( J\), which is positive for
 large positive \( x\), and \( R_+^{1/2}(x) \)
is the value of \( R^{1/2}(x) \) on the upper part
of the cut.
The function \( h(x) \) is a polynomial, which
is the polynomial part of the function 
\(\frac{V'(x)}{R^{1/2}(x)}\) at infinity, i.e.
\begin{equation}\label{equi12}
\frac{V'(x)}{R^{1/2}(x)}=h(x)+O(x^{-1}).
\end{equation} 
In particular, \(\deg h=\deg V-1-q\).
\end{prop}

There is a useful formula for the equilibrium density [DKM]:
\begin{equation}\label{equi13}
\frac{d\nu_V(x)}{dx}=\frac{1}{\pi}\sqrt {q(x)},
\end{equation} 
where
\begin{equation}\label{equi14}
q(x)=-\left(\frac{V'(x)}{2}\right)^2+\int\frac{V'(x)-V'(y)}{x-y}d\nu_V(y).
\end{equation}
This, in fact, is an equation on $q$, since the right-hand side contains
an integration with respect to $\nu_V$. Nevertheless, if $V$ is a polynomial of
degree $p=2p_0$, then (\ref{equi14}) determines uniquely more then a half
of the coefficients of the polynomial $q$,
\begin{equation}\label{equi15}
q(x)=-\left(\frac{V'(x)}{2}\right)^2-O(x^{p-2}).
\end{equation}

{\bfseries Example.} If \( V(x)\) is {\it convex}
then \( \nu_V\) is regular, and the support
of \( \nu_V\) consists of a single interval, see e.g. \cite{KuM1}.
For the Gaussian ensemble, $V(x)=x^2$, hence, by (\ref{equi15}),
 $q(x)=a^2-x^2$. Since
\[
\int_{-a}^a \frac{1}{\pi}\sqrt {a^2-x^2}\,dx=1,
\]
we find that $a=\sqrt 2$, hence
\begin{equation}\label{equi15a}
p_V(x)=\frac{1}{\pi}\sqrt {2-x^2}\,,\qquad |x|\le \sqrt 2,
\end{equation}
the Wigner semicircle law.

\subsection{The Euler-Lagrange variational conditions.} A nice and important property
of minimization problem (\ref{equi7}) is that the minimizer is uniquely
determined by the Euler-Lagrange variational conditions: for some real constant 
\( l\),
\begin{align}
{}&2\int_{\R}\log|x-y|d\nu(y)-V(x)=l,\;\text{for}\; x\in J,\label{equi16}\\
{}&2\int_{\R}\log|x-y|d\nu(y)-V(x)\le l,\;\text{for}\; x\in \R\setminus J,
\label{equi16a}
\end{align}
see [DKM].

{\bf Definition.}  (See [DKMVZ2].) {\it The equilibrium measure,
\begin{equation}\label{equi17}
d\nu_V(x)=\frac{1}{2\pi i} h(x)R^{1/2}_+(x)\,dx
\end{equation}
is called {\rm regular}
(otherwise {\rm singular}) if 
\begin{enumerate}
\item \( h(x)\not=0\) on the (closed) set \( J\).
\item Inequality (\ref{equi16a}) is strict,
\begin{equation}\label{equi18}
2\int\log|x-y|d\nu_V(y)-V(x)< l,\;\text{\rm for}\; x\in \R\setminus J.
\end{equation}
\end{enumerate}}

\subsection{Construction of the equilibrium measure: equations on the end-points.}
The strategy to construct the equilibrium measure is the following: first
we find the end-points of the support, and then we use equation (\ref{equi12})
to find $h(x)$ and hence the density. The number $q$ of cuts  is not, in general, known,
and we try different $q$'s. 
Consider the resolvent,
\begin{equation}\label{equi19}
\om(z)=\int_J\frac{d\nu_V (x)}{z-x},\quad
z\in \C\setminus J.
\end{equation}
The Euler-Lagrange variational condition implies that
\begin{equation}\label{equi20}
\om(z)=\frac{V'(z)}{2}-\frac{h(z)R^{1/2}(z)}{2}\,.
\end{equation}
Observe that
as \( z\to\infty\),
\begin{equation}\label{equi21}
\om(z)=\frac{1}{z}+\frac{m_1}{z^2}+\dots,\quad
m_k=\int_J x^k\rho(x)dx.
\end{equation}
The equation
\begin{equation}\label{equi22}
\frac{V'(z)}{2}-\frac{h(z)R^{1/2}(z)}{2}=\frac{1}{z}+O(z^{-2})\,.
\end{equation}
gives \( q+1\) equation on \( a_1,b_1,\dots, a_q,b_q\), if we substitute
formula (\ref{equi12}) for $h$. Remaining \( q-1\) equation are
\begin{equation}\label{equi23}
\int_{b_j}^{a_{j+1}} h(x)R^{1/2}(x)\,dx=0,\quad j=1,\dots, q-1,
\end{equation}
which follow from (\ref{equi20}) and (\ref{equi16}).

{\bfseries Example.} Even quartic model, $V(M)=\frac{t}{2}M^2+\frac{1}{4}M^4$.
For $t\ge t_c=-2$, the support of the equilibrium distribution consists of
one interval $[-a,a]$ where
\begin{equation}\label{equi24}
a=\left(\frac{-2t+2\left(t^2+12\right)^{1/2}}{3}\right)^{1/2}
\end{equation}
and
\begin{equation}\label{equi25}
p_V(x)=\frac{1}{\pi}\left(c+\frac{1}{2}x^2\right)\sqrt{a^2-x^2}
\end{equation}
where
\begin{equation}\label{equi26}
c=\frac{t+\left((t^2/4)+3\right)^{1/2}}{3}\,.
\end{equation}
\begin{figure}
\scalebox{1.0}{\includegraphics{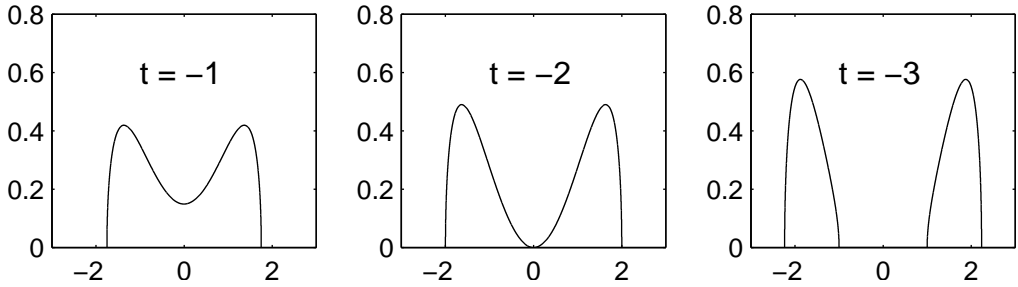}}
\caption{The density function, $p_V(x)$, for the even quartic
potential, $V(M)=\frac{t}{2}M^2+\frac{1}{4}M^4$, for $t=-1,-2,-3$.}
\label{density_quartic}
\end{figure}
In particular, for $t=-2$,
\begin{equation}\label{equi27}
p_V(x)=\frac{1}{2\pi}x^2\sqrt{4-x^2}
\end{equation}
For $t<-2$, the support consists of
two intervals, $[-a,-b]$ and $[b,a]$, where
\begin{equation}\label{equi28}
a=\sqrt{2-t},\quad b=\sqrt{-2-t}\,,
\end{equation}
and
\begin{equation}\label{equi29}
p_V(x)=\frac{1}{ 2\pi} |x|\sqrt{(a^2-x^2)(x^2-b^2)}\,.
\end{equation}
Fig.\ref{density_quartic} shows the density function for the even quartic
potential, for $t=-1,-2,-3$. 

\vskip 5mm

{\Large Lecture 2. Large $N$ asymptotics of orthogonal polynomials. The Riemann-Hilbert
approach}

\vskip 5mm

In this lecture we present the Riemann-Hilbert approach to the large $N$ asymptotics
of orthogonal polynomials. The central point of this approach is a construction of
an asymptotic solution to the RHP, as $N\to\infty$. We call such a solution,
a parametrix.
In the original paper of Bleher and Its [BI1] the RH approach
was developed for an even quartic polynomial $V(M)$ via a semiclassical solution of 
the differential equation for orthogonal polynomials. Then, in a series of
papers, Deift, Kriecherbauer, McLaughlin, Venakides, and Zhou [DKM], [DKMVZ1],
[DKMVZ2] developed the RH approach for a general real analytic $V$, with 
some conditions on the growth at infinity. The DKMVZ-approach is based on the
Deift-Zhou steepest descent method, see \cite{DZ}. In this lecture we present the main steps
of the DKMVZ-approach. For the sake of simplicity,
we will assume that $V$ is regular. In this approach a sequence of transformations of 
the RHP is constructed, which reduces the RHP to a simple RHP which can be
solved by a series of perturbation theory. This sequence of transformations
gives the parametrix of the RHP in different regions on complex plane.
The motivation for the first transformation comes from the Heine formula
for orthogonal polynomials.

\section{Heine's formula for orthogonal polynomials.}

The Heine formula, see, e.g., [Sze], gives the $N$-th orthogonal polynomial as
the matrix integral, 
\begin{equation}\label{heine_1}
P_N(z)=\left\langle \det (z-M) \right\rangle 
\equiv
Z_N^{-1}\int_{\mathcal H_N} \det (z-M) e^{-N\Tr V(M)}dM.
\end{equation}
In the ensemble of eigenvalues,
\begin{equation}\label{heine_2}
P_N(z)=\left\langle \prod_{j=1}^N (z-\la_j) \right\rangle
\equiv
\tilde Z_N^{-1}\int \prod_{j=1}^N (z-\la_j)
\prod_{j>k}(\la_j-\la_k)^2
\prod_{j=1}^N e^{-NV(\la_j)}d\la
\end{equation}
Since \(\nu_{\la}\) is close to the
equilibrium measure \(\nu\) for typical \(\la\),
we expect that
\begin{equation}\label{heine_3}
N^{-1}\log\left\langle \prod_{j=1}^N (z-\la_j) \right\rangle
\approx
\int_J \log(z-x) d\nu_V(x), \notag
\end{equation}
hence by the Heine formula,
\begin{equation}\label{heine_4}
N^{-1}\log P_N(z)\approx
\int_J \log(z-x) d\nu_V(x).
\end{equation}
This gives a {\it heuristic} semiclassical approximation for the 
orthogonal polynomial,
\begin{equation}\label{heine_5}
P_N(z)\approx
\exp\left[N\int_J \log(z-x) d\nu_V(x)\right],
\end{equation}
and it motivates the introduction of the ``$g$-function''.

\subsection{$g$-function}
Define the $g$-function as
\begin{equation}\label{g_1}
g(z)=\int_J \log(z-x) d\nu_V(x),\quad z\in \C\setminus
    (-\infty,b_q],
\end{equation}
where we take the principal branch for logarithm.

{\bfseries Properties of \( g(z)\):}
\begin{enumerate}
\item \( g(z)\) is analytic in \(\C\setminus
    (-\infty,b_q]\).
\item As \( z\to\infty\)
\begin{equation}\label{g_1g}
 g(z)=\log z-\sum_{j=1}^\infty \frac{g_j}{z^j}\,,\qquad 
g_j=\int_J \frac{x^j}{j}\,d\nu_V(x).
\end{equation}
\item By (\ref{equi19}), (\ref{equi20}), 
\begin{equation}\label{g_1a}
 g'(z)=\om(z)=
\frac{V'(z)}{2}-\frac{h(z)R^{1/2}(z)}{2}\,.
\end{equation}
\item By (\ref{equi16}), 
\begin{equation}\label{g_1b}
g_+(x)+g_-(x)=V(x)+l,\quad x\in J.
\end{equation}
\item By (\ref{equi18}), 
\begin{equation}\label{g_1c}
 g_+(x)+g_-(x)< V(x)+l,\quad  x\in \R\setminus J.
\end{equation}
\item Equation (\ref{g_1}) implies that the function
\begin{equation}\label{g_5}
 G(x)\equiv g_+(x)-g_-(x)
\end{equation}
is pure imaginary for all real 
\( x\), and $G(x)$ is constant in each component of 
\( \R\setminus J\),
\begin{equation}\label{g_2}
 G(x)=i\Om_j \;\text{ for}\; b_j< x<a_{j+1},
\quad 1\le j\le q-1,
\end{equation}
where
\begin{equation}\label{g_3}
\Om_j=2\pi
\sum_{k=j+1}^q \int_{a_k}^{b_k} p_V(x)\,dx,\quad 1\le j\le q-1.
\end{equation} 
\item Also, 
\begin{equation}\label{g_4}
 G(x)=i\Om_j-2\pi i\int_{b_j}^x  p_V(s)\,ds\;\text{ for}\; a_j< x<b_j,
\quad 1\le j\le q,
\end{equation}
where we set $\Om_q= 0$.
\end{enumerate}
Observe that from (\ref{g_4}) and (\ref{equi11}) we obtain that $G(x)$
is analytic on $(a_j,b_j)$, and 
\begin{equation}\label{g_6}
 \left. \frac{dG(x+iy)}{dy}\right|_{y=0}=2\pi p_V(x)>0,
\quad x\in (a_j,b_j),\quad 1\le j\le q.
\end{equation}
From (\ref{g_1b}) we have also that
\begin{equation}\label{g_7}
 G(x)=2g_+(x)-V(x)-l=-[2g_-(x)-V(x)-l],\quad x\in J.
\end{equation}

\section{First Transformation of the RH Problem}

Our goal is to construct an asymptotic solution to RHP
(\ref{RH_10}), (\ref{RH_11}) for $Y_N(z)$, as $N\to\infty$. In our construction
we will assume that the equilibrium measure $\nu_V$ is regular.
By (\ref{heine_5}) we expect that
\begin{equation}\label{ft_1}
P_N(z)\approx e^{Ng(z)},
\end{equation}
therefore, we make the following substitution in the RHP:
\begin{equation}\label{ft_2}
Y_N(z)=e^{\frac{Nl}{2}\sg_3}
 T_N(z)e^{N\left [g(z)-\frac{l}{2}\right]\sg_3},\qquad \sg_3=
\begin{pmatrix}
1 & 0 \\
0 & -1
\end{pmatrix}.
\end{equation}
Then \( T_N(z)\) solves the following RH problem:
\begin{enumerate}
\item \( T_N(z)\) is analytic in $\C\setminus\R$.
\item \( T_{N+}(x)=T_{N-}(x)j_T(x)\) for \( x\in\R\), where
\begin{equation}\label{ft_3}
j_T(x)=
\begin{pmatrix}
e^{-N[g_+(x)-g_-(x)]} & e^{N[g_+(x)+g_-(x)-V(x)-l]} \\
0 & e^{N[g_+(x)-g_-(x)]}
\end{pmatrix}.
\end{equation}
\item \( T_N(z)=I+O(z^{-1})\), as \(z\to\infty\).
\end{enumerate}
The above properties of $g(z)$ ensure the following
properties of the jump matrix $j_T$:
\begin{enumerate}
\item $j_T(x)$ is exponentially close to the identity matrix on $(-\infty,a_1)\cup(b_q,\infty)$.
Namely, 
\begin{equation}\label{ft_5}
j_T(x)=
\begin{pmatrix}
1 & O(e^{-Nc(x)})\\
0 & 1
\end{pmatrix},\quad   x\in(-\infty,a_1)\cup(b_q,\infty),
\end{equation}
where $c(x)>0$ is a continuous function such that
\begin{equation}\label{ft_6}
\lim_{x\to\pm\infty}\frac{c(x)}{\ln |x|}=\infty,
\quad \lim_{x\to a_1} c(x)=\lim_{x\to b_q} c(x)=0.
\end{equation}
\item For $1\le j\le q-1$,
\begin{equation}\label{ft_7}
j_T(x)=
\begin{pmatrix}
e^{-iN\Om_j} & O(e^{-Nc(x)})\\
0 & e^{iN\Om_j}
\end{pmatrix},\quad   x\in (b_j,a_{j+1}),
\end{equation}
where $c(x)>0$ is a continuous function such that
\begin{equation}\label{ft_8}
\lim_{x\to b_j} c(x)=\lim_{x\to a_{j+1}} c(x)=0.
\end{equation}
\item
On $J$,
\begin{equation}\label{ft_8a}
j_T(x)=\begin{pmatrix}
e^{-NG(x)} & 1 \\
0 & e^{NG(x)}
\end{pmatrix}\,.
\end{equation}
\end{enumerate}
The latter matrix can be factorized as follows:  
\begin{equation}\label{ft_9}
\begin{aligned}
\begin{pmatrix}
e^{-NG(x)} & 1 \\
0 & e^{NG(x)} 
\end{pmatrix}&=
\begin{pmatrix}
1 & 0 \\
e^{NG(x)} & 1
\end{pmatrix}
\begin{pmatrix}
0 & 1 \\
-1 & 0
\end{pmatrix}
\begin{pmatrix}
1 & 0 \\
e^{-NG(x)} & 1
\end{pmatrix}\\
&\equiv j_-(x)j_M j_+(x),
\end{aligned}
\end{equation}
This leads to the second transformation of the RHP.

\section{ Second transformation of the RHP:
Opening of lenses}

\begin{figure}
\scalebox{0.8}{\includegraphics{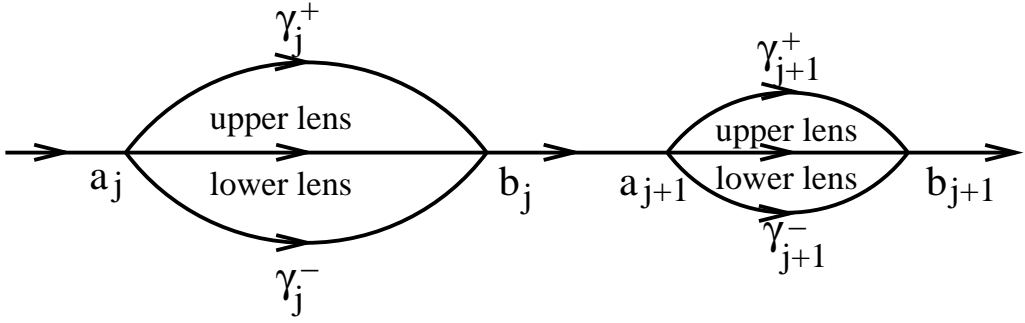}}
\caption{The lenses.}
\label{lenses}
\end{figure}

The function $e^{-NG(x)}$ 
is analytic on each open interval $(a_j,b_j)$. Observe that
$|e^{-NG(x)}|=1$ for real $x\in (a_j,b_j)$, and $e^{-NG(z)}$ is exponentially decaying
for $\Im z>0$. More precisely, by (\ref{g_6}),
there exists $y_0>0$ such that $e^{-NG(z)}$ satisfies the estimate, 
\begin{equation}\label{st_2}
|e^{-NG(z)}|\le e^{-Nc(z)},\quad z\in R_j^+=\{z=x+iy:\;a_j<x<b_j,\; 0<y<y_0\},
\end{equation}  
where $c(z)>0$ is a continuous function in $R_j^+$. Observe that $c(z)\to 0$ as $\Im z\to 0$.
In addition, $|e^{NG(z)}|=|e^{-NG(\bar z)}|$, hence
\begin{equation}\label{st_3}
|e^{NG(z)}|\le e^{-Nc(z)},\quad z\in R_j^-=\{z=x+iy:\;a_j<x<b_j,\; 0<-y<y_0\},
\end{equation} 
where $c(z)=c(\bar z)>0$. 
Consider  a $C^\infty$ curve $\ga_j^+$ from $a_j$ to $b_j$ such that 
\begin{equation}\label{st_4}
\ga_j^+=\{x+iy:\; y=f_j(x)\},
\end{equation}
where $f_j(x)$ is a $C^\infty$ function on $[a_j,b_j]$ such that
\begin{equation}\label{st_5}
 f_j(a_j)=f_j(b_j)=0;
\quad f'_j(a_j)=-f'(b_j)=\sqrt 3;\quad 0<f_j(x)<y_0,\quad a_j<x<b_j.
\end{equation}
Consider the conjugate curve,
\begin{equation}\label{st_6}
\ga_j^-=\overline{\ga_j^+}=\{x-iy:\; y=f_j(x)\},
\end{equation}
see Fig.\ref{lenses}. 
The region bounded by the interval $[a_j,b_j]$ and $\ga_j^+$ ($\ga_j^-$)  is called
the {\it upper (lower) lens}, $\mathcal L_j^{\pm}$, respectively. Define 
for $j=1,\ldots,q$,
\begin{equation}\label{st_7}
S_N(z)=\left\{
\begin{aligned}
{}&T_N(z)j_+^{-1}(z),\quad \text{if $z$ is in the upper lens, $z\in\mathcal L_j^+$},\\
{}&T_N(z)j_-(z),\quad \text{if $z$ is in the lower lens, $z\in\mathcal L_j^-$},\\ 
{}&T_N(z) \quad \text{otherwise,}
\end{aligned}
\right.
\end{equation}  
where 
\begin{equation}\label{st_7a}
j_{\pm}(z)=\begin{pmatrix}
1 & 0 \\
e^{\mp NG(z)} & 1
\end{pmatrix}.
\end{equation}  
Then \( S_N(z)\) solves the following RH problem: 
\begin{enumerate}
\item $S_N(z)$ is analytic in $\C\setminus (\R\cup\Ga)$,
$\Ga=\ga_1^+\cup\ga_1^-\cup\dots\cup \ga_q^+\cup\ga_q^-$.
\item 
\begin{equation}\label{st_8}
 S_{N+}(z)=S_{N-}(z)j_S(z),\quad 
 z\in\R\cup\ga,
\end{equation}  
where the jump matrix \(j_S(z)\) has the following properties:
\begin{enumerate}
\item 
\(j_S(z)=
\begin{pmatrix}
0 & 1 \\
-1 & 0
\end{pmatrix}
\) for \(z\in J\).
\item
\(j_S(z)=j_T(z)=
\begin{pmatrix}
e^{-iN\Om_j} & O(e^{-c(x)N}) \\
0 & e^{iN\Om_j}
\end{pmatrix}\)  for \(z\in (b_j,a_{j+1})\), $j=1,\ldots,q-1$, and
\(j_S(z)=j_T(z)=
\begin{pmatrix}
1 & O(e^{-c(x)N}) \\
0 & 1
\end{pmatrix}\)  for \(z\in (-\infty,a_1)\cup(b_q,\infty).\)
\item
\(j_S(z)=j_{\pm}(z)=
\begin{pmatrix}
1 & 0 \\
O(e^{-c(z)N}) & 1
\end{pmatrix}\)
 for \(z\in \ga_j^{\pm} \), $j=1,\ldots,q$, where $c(z)>0$ is
a continuous function such that $c(z)\to 0$ as $z\to a_j,b_j$.
\end{enumerate}
\item \( S_N(z)=I+O(z^{-1})\), as \(z\to\infty\).
\end{enumerate}
We expect, and this will be justified later, that as $N\to\infty$, $S_N(z)$ 
converges to a solution of the model RHP, in which we drop the $O(e^{-cN})$-terms
in the jump matrix $j_S(z)$. Let us consider the model RHP.

\section{Model RHP}

We are looking for
\( M(z)\) that solves the following model RHP:
\begin{enumerate}
\item $M(z)$ is analytic in $\C\setminus [a_1,b_q]$,
\item 
\begin{equation}\label{MRHP_1}
 M_{+}(z)=M_{-}(z)j_M(z),\quad 
 z\in[a_1,b_q],
\end{equation}  
where the jump matrix \(j_M(z)\) is given by the following formulas:
\begin{enumerate}
\item 
\(j_M(z)=
\begin{pmatrix}
0 & 1 \\
-1 & 0
\end{pmatrix}
\) for \(z\in J\)
\item
\(j_M(z)=
\begin{pmatrix}
e^{-iN\Om_j} & 0 \\
0 & e^{iN\Om_j}
\end{pmatrix}\)  for \(z\in [b_j,a_{j+1}]\), $j=1,\ldots,q-1$.
\end{enumerate}
\item \( M(z)=I+O(z^{-1})\), as \(z\to\infty\).
\end{enumerate}
We will construct a solution to the model RHP by following the work [DKMVZ2].

\subsection{Solution of the model RHP. One-cut case} Assume that \( J\) consist of a single interval 
\([a,b]\). Then the model RH problem reduces to the following:
\begin{enumerate}
\item $M(z)$ is analytic in $\C\setminus [a,b]$,
\item 
\( M_{+}(z)=M_{-}(z)
\begin{pmatrix}
0 & 1 \\
-1 & 0
\end{pmatrix}\) for \(z\in [a,b]\).
\item \( M(z)=I+O(z^{-1})\), as \(z\to\infty\).
\end{enumerate}
This RHP can be reduced  
to a pair of scalar RH problems.  We have 
that
\begin{equation}\label{MRHP_2}
\begin{pmatrix}
0 & 1 \\
-1 & 0
\end{pmatrix}
=\frac {1}{2}
\begin{pmatrix}
1 & 1 \\
i & -i
\end{pmatrix}
\begin{pmatrix}
i & 0 \\
0 & -i
\end{pmatrix}\begin{pmatrix}
1 & -i \\
1 & i
\end{pmatrix}
\end{equation}
Let
\begin{equation}\label{MRHP_3}
\tilde M(z)=
\frac {1}{2}\begin{pmatrix}
1 & -i \\
1 & i
\end{pmatrix}
M(z)
\begin{pmatrix}
1 & 1 \\
i & -i
\end{pmatrix}.
\end{equation}
Then $\tilde M(z)$ solves the following RHP:
\begin{enumerate}
\item $\tilde M(z)$ is analytic in $\C\setminus [a,b]$,
\item 
\(\tilde  M_{+}(z)=\tilde M_{-}(z)
\begin{pmatrix}
i & 0 \\
0 & -i
\end{pmatrix}\) for \(z\in [a,b]\).
\item \(\tilde  M(z)=I+O(z^{-1})\), as \(z\to\infty\).
\end{enumerate}
This is a pair of scalar RH problems, which can be solved by the Cauchy integral:
\begin{equation}\label{MRHP_4}
\begin{aligned}
\tilde M(z)&=
\begin{pmatrix}
e^{\frac{1}{2\pi i}\int_{a}^{b}\frac{\log i}{s-z}ds} & 0 \\
0 & e^{\frac{1}{2\pi i}\int_{a}^{b}\frac{\log (-i)}{s-z}ds}
\end{pmatrix} 
=\begin{pmatrix}
e^{\frac{1}{4}\log\frac{z-b}{z-a}} & 0 \\
0 & e^{-\frac{1}{4}\log\frac{z-b}{z-a}} 
\end{pmatrix} 
=\begin{pmatrix}
\ga^{-1} & 0 \\
0 & \ga
\end{pmatrix},
\end{aligned}
\end{equation}
where
\begin{equation}
\ga(z)=\left(\frac{z-a}{z-b}\right)^{1/4}
\label{MRHP_5}
\end{equation}
with cut on $[a,b]$ and the branch such that $\ga(\infty)=1$.
Thus,
\begin{equation}\label{MRHP_6}
\begin{aligned}
&M(z)=
\frac {1}{2}\begin{pmatrix}
1 & 1 \\
i & -i
\end{pmatrix}
\begin{pmatrix}
\ga^{-1} & 0 \\
0 & \ga
\end{pmatrix}
\begin{pmatrix}
1 & 1 \\
i & -i
\end{pmatrix}^{-1} 
=\begin{pmatrix}
\frac{\ga(z)+\ga^{-1}(z)}{2} & \frac{\ga(z)-\ga^{-1}(z)}{(-2i)} \\
\frac{\ga(z)-\ga^{-1}(z)}{2i} & \frac{\ga(z)+\ga^{-1}(z)}{2}
\end{pmatrix},\\  &\det M(z)=1.
\end{aligned}
\end{equation} 
At infinity we have
\begin{equation}
\ga(z)=1+\frac{b-a}{4z}+O(z^{-2}),
\label{MRHP_7}
\end{equation}
hence
\begin{equation}
M(z)=I+\frac{1}{z}
\begin{pmatrix}
0 & \frac{b-a}{(-4i)} \\
\frac{b-a}{4i}  & 0
\end{pmatrix}
+O(z^{-2}).
\label{MRHP_7a}
\end{equation}

\subsection{Solution of the model RHP. Multicut case}  This will be done in three steps.

{\it Step 1.} Consider the auxiliary RHP,
\begin{enumerate}
\item $Q(z)$ is analytic in $\C\setminus J$, $J=\cup_{j=1}^q[a_j,b_j]$,
\item 
\( Q_{+}(z)=Q_{-}(z)
\begin{pmatrix}
0 & 1 \\
-1 & 0
\end{pmatrix}\) for \(z\in J\).
\item \( Q(z)=I+O(z^{-1})\), as \(z\to\infty\).
\end{enumerate}
Then, similar to the one-cut case, this RHP is reduced to two scalar RHPs,
and the solution is 
\begin{equation}\label{MRHP_8}
Q(z)
=\begin{pmatrix}
\frac{\ga(z)+\ga^{-1}(z)}{2} & \frac{\ga(z)-\ga^{-1}(z)}{(-2i)} \\
\frac{\ga(z)-\ga^{-1}(z)}{2i} & \frac{\ga(z)+\ga^{-1}(z)}{2}
\end{pmatrix},
\end{equation}
where
\begin{equation}\label{MRHP_8a}
\ga(z)=\prod_{j=1}^q \left(\frac{z-a_j}{z-b_j}\right)^{1/4},\qquad \ga(\infty)=1,
\end{equation}
with cuts on $J$.
At infinity we have
\begin{equation}\label{MRHP_8b}
\ga(z)=1+\sum_{j=1}^q \frac{b_j-a_j}{4z}+O(z^{-2}),
\end{equation}
hence
\begin{equation} \label{MRHP_8c}
Q(z)=I+\frac{1}{z}
\begin{pmatrix}
0 & \sum_{j=1}^q \frac{b_j-a_j}{(-4i)} \\
\sum_{j=1}^q \frac{b_j-a_j}{4i}  & 0
\end{pmatrix}
+O(z^{-2}).
\end{equation}
In what follows, we will modify this solution to satisfy part (b) in jump matrix in (\ref{MRHP_1}).
This requires some Riemannian geometry and the theta function.

{\it Step 2.} Let $X$ be the two-sheeted Riemannian surface of the genus
\begin{equation}\label{MRHP_9a}
g=q-1,
\end{equation}
associated to $\sqrt{R(z)}$, where
\begin{equation}\label{MRHP_9}
R(z)=\prod_{j=1}^q (z-a_j)(z-b_j),
\end{equation}
\begin{figure}
\scalebox{0.8}{\includegraphics{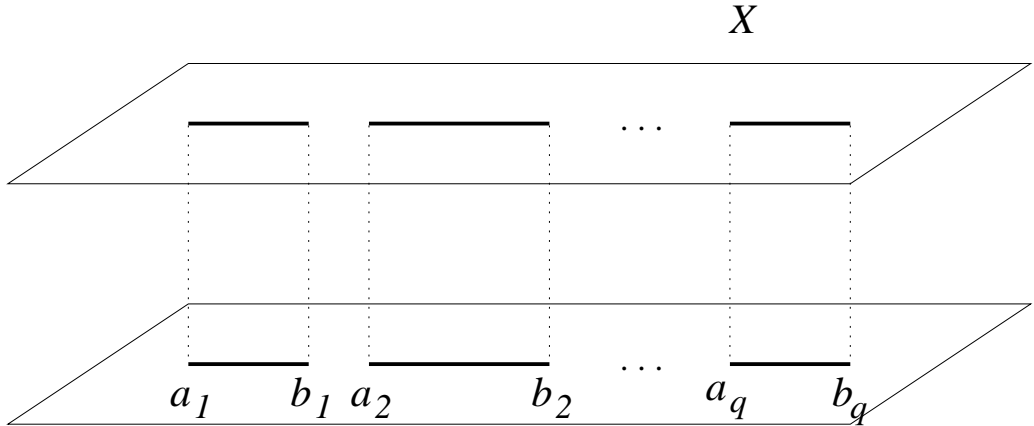}}
\caption{The Riemannian surface associated to $\sqrt{R(z)}$.}
\label{Riemann_surface_DKMVZ}
\end{figure}
with cuts on the intervals $(a_j,b_j)$, $j=1,\ldots,q$, 
see Fig.\ref{Riemann_surface_DKMVZ}.
We fix the first sheet of $X$ by the condition that on this sheet,
\begin{equation}\label{MRHP_11}
\sqrt{R(x)}>0,\qquad x>b_q.
\end{equation}
We would like to introduce $2g$ cycles on $X$, forming a homology basis. To that end,
consider, for $j=1,\ldots, g$, a cycle $A_j$ on $X$, which goes around the interval $(b_{j},a_{j+1})$
in the negative direction, such that the part of $A_j$ in the upper half-plane, 
$A_j^+\equiv A_j\cup\{z:\;\Im z\ge 0\}$,
lies on the first sheet of $X$, and the one in the lower half-plane, $A_j^-\equiv A_j\cup\{z:\;\Im z\le 0\}$,
lies on the second sheet, $j=1,\ldots,g$.
In addition, consider a cycle $B_j$ on $X$, which goes around the interval $(a_1,b_j)$
on the first sheet in the negative direction,
\begin{figure}
\scalebox{0.8}{\includegraphics{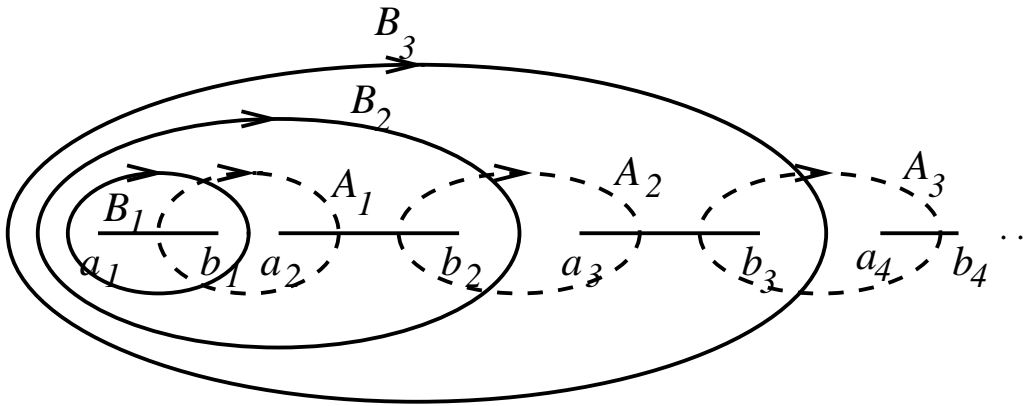}}
\caption{The basis of cycles on $X$.}
\label{cycles}
\end{figure}
see Fig.\ref{cycles}. Then the cycles $(A_1,\ldots,A_{g},B_1,\ldots,B_{g})$ form
a canonical homology basis for $X$.

Consider the linear space $\Om$ of holomorphic one-forms on $X$,
\begin{equation}\label{MRHP_12}
\Om=\left\{\om=\sum_{j=0}^{q-2} \frac{c_jz^jdz}{\sqrt{R(z)}}\right\}.
\end{equation}
The dimension of $\Om$ is equal to $g$. Consider the basis in $\Om$, 
\[
\om=(\om_1,\ldots,\om_{g}),
\]
with normalization
\begin{equation}\label{MRHP_13}
\int_{A_j}\om_k=\de_{jk},\qquad j,k=1,\ldots, g.
\end{equation}
Such a basis exists and it is unique, see \cite{FK}.
Observe that the numbers
\begin{equation}\label{MRHP_14}
m_{jk}=\int_{A_j}\frac{z^kdz}{\sqrt{R(z)}}
=2\int_{b_j}^{a_{j+1}}\frac{x^kdx}{\sqrt{R(x)}}
\,,\qquad 1\le j\le  g,\quad 1\le k\le  g-1,
\end{equation}
are real. This implies that the basis $\om$ is real, i.e., the one-forms,
\begin{equation}\label{MRHP_15}
\om_j=\sum_{k=0}^{g-1} \frac{c_{jk}z^kdz}{\sqrt{R(z)}},
\end{equation}
have real coefficients $c_{jk}$. 

Define the associated Riemann matrix of $B$-periods,
\begin{equation}\label{MRHP_16}
\tau=(\tau_{jk}),\qquad \tau_{jk}=\int_{B_j}\om_k,\qquad j,k=1,\ldots, g.
\end{equation}
Since $\sqrt{R(x)}$ is pure imaginary on $(a_j,b_j)$, the numbers $\tau_{jk}$
are pure imaginary. It is known, see, e.g., \cite{FK},  that the matrix $\tau$
is symmetric and $(-i\tau)$ is positive definite.

The Riemann theta function with the matrix $\tau$ is defined as
\begin{equation}\label{MRHP_17}
\theta(s)=\sum_{m\in\Z^{g}} e^{2\pi i(m,s)+\pi i(m,\tau m)},\quad s\in \C^{g};
\qquad (m,s)=\sum_{j=1}^{g} m_j s_j.
\end{equation}
The quadratic form $i(m,\tau m)$ is negative definite, hence the series
is absolutely convergent and $\theta(s)$ is analytic in $\C^{g}$.
The theta function is an even function,
\begin{equation}\label{MRHP_18a}
\theta(-s)=\theta(s),
\end{equation}
and it has the following periodicity properties:
\begin{equation}\label{MRHP_18}
\theta(s+e_j)=\theta(s);\qquad \theta(s+\tau_j)=e^{-2\pi is_j-\pi i\tau_{jj}}\theta(s),
\end{equation}
where $e_j=(0,\ldots,0,1,0,\ldots,0)$ is the $j$-th basis vector in $\C^{g}$, and $\tau_j=\tau e_j$.
This implies that the function 
\begin{equation}\label{MRHP_19}
f(s)=\frac{\theta(s+d+c)}{\theta(s+d)},
\end{equation}
where $c,d\in \C^g$ are arbitrary constant vectors, has the periodicity properties,
\begin{equation}\label{MRHP_20}
f(s+e_j)=f(s);\qquad f(s+\tau_j)=e^{-2\pi i c_j}f(s).
\end{equation}

Consider now the theta function associated with the Riemann surface $X$. It is defined as follows.
Introduce the vector function,
\begin{equation}\label{MRHP_21}
u(z)=\int_{b_q}^z\omega,\qquad z\in \C\setminus(a_1,b_q),
\end{equation}
where $\om$ is the basis of holomorphic one-forms, determined by equations (\ref{MRHP_13}).
The contour of integration in (\ref{MRHP_21}) lies in $\C\setminus(a_1,b_q)$,
on the first sheet of $X$.
We will consider $u(z)$ as a function with values in $\C^g/\Z^g$,
\begin{equation}\label{MRHP_22}
u:\;\C\setminus(a_1,b_q)\to \C^g/\Z^g\,.
\end{equation}
On $[a_1,b_q]$ the function $u(z)$ is two-valued.
From (\ref{MRHP_16}) we have that
\begin{equation}\label{MRHP_23}
u_+(x)-u_-(x)=\tau_j,\quad x\in [b_j,a_{j+1}];\qquad 1\le j\le q-1.
\end{equation}
Since $\sqrt{R(x)}_-=-\sqrt{R(x)}_+$ on $[a_j,b_j]$, we have that the function
$u_+(x)+u_-(x)$ is constant on $[a_j,b_j]$. It follows from 
(\ref{MRHP_13}) that $\mod \Z^q$,
\begin{equation}\label{MRHP_24}
u_+(b_j)+u_-(b_j)=u_+(a_{j+1})+u_-(a_{j+1}),\qquad 1\le j\le  q-1
\end{equation}
Since $u_+(b_q)=u_-(b_q)=0$, we obtain that
\begin{equation}\label{MRHP_25}
u_+(x)+u_-(x)=0,\quad x\in J=\bigcup_{j=1}^q[a_j,b_j].
\end{equation}
Define
\begin{equation}\label{MRHP_26}
f_1(z)=\frac{\theta(u(z)+d+c)}{\theta(u(z)+d)},
\quad 
f_2(z)=\frac{\theta(-u(z)+d+c)}{\theta(-u(z)+d)},
\qquad z\in \C\setminus(a_1,b_q),
\end{equation}
where $c,d\in \C^g$ are arbitrary constant vectors. Then from 
 (\ref{MRHP_23}) and (\ref{MRHP_20})  we obtain that for $1\le j\le q-1$,
\begin{equation}\label{MRHP_27}
\begin{aligned}
f_1(x+i0)=e^{-2\pi i c_j}f_1(x-i0),\quad
f_2(x+i0)=e^{2\pi i c_j}f_2(x-i0),\quad
 x\in (b_{j},a_{j+1}),
\end{aligned}
\end{equation}
and from (\ref{MRHP_25}) that
\begin{equation}\label{MRHP_28}
\begin{aligned}
f_1(x+i0)=f_2(x-i0), \quad
f_2(x+i0)=f_1(x-i0),\quad
 x\in J.
\end{aligned}
\end{equation}
Let us take
\begin{equation}\label{MRHP_29}
c=\frac{n\Omega}{2\pi},\qquad \Om=(\Om_1,\ldots,\Om_g),
\end{equation}
and define the matrix-valued function,
\begin{equation}\label{MRHP_30}
F(z)=
\begin{pmatrix}
\frac{\theta(u(z)+d_1+c)}{\theta(u(z)+d_1)} & \frac{\theta(-u(z)+d_1+c)}{\theta(-u(z)+d_1)} \\
\frac{\theta(u(z)+d_2+c)}{\theta(u(z)+d_2)} & \frac{\theta(-u(z)+d_2+c)}{\theta(-u(z)+d_2)} 
\end{pmatrix}
\end{equation}
where $d_1,d_2\in \C^g$ are arbitrary constant vectors. Then from 
(\ref{MRHP_28}), we obtain that
\begin{equation}\label{MRHP_31}
\begin{aligned}
&F_+(x)=F_-(x)
\begin{pmatrix}
e^{-iN\Om_j} & 0\\
0 & e^{iN\Om_j} 
\end{pmatrix},
\quad x\in (b_{j},a_{j+1});\qquad j=1,\ldots, q-1,\\
&F_+(x)=F_-(x)
\begin{pmatrix}
0 & 1 \\
1 & 0
\end{pmatrix}, \quad x\in J.
\end{aligned}
\end{equation}

{\it Step 3.} Let us combine formulae (\ref{MRHP_8}) and (\ref{MRHP_30}), and let us set 
\begin{equation}\label{MRHP_32}
M(z)
=F(\infty)^{-1}\begin{pmatrix}
\frac{\ga(z)+\ga^{-1}(z)}{2}\, \frac{\theta(u(z)+d_1+c)}{\theta(u(z)+d_1)}
& \frac{\ga(z)-\ga^{-1}(z)}{(-2i)}\, \frac{\theta(-u(z)+d_1+c)}{\theta(-u(z)+d_1)} \\
\frac{\ga(z)-\ga^{-1}(z)}{2i}\,  \frac{\theta(u(z)+d_2+c)}{\theta(u(z)+d_2)}
& \frac{\ga(z)+\ga^{-1}(z)}{2}\ \frac{\theta(-u(z)+d_2+c)}{\theta(-u(z)+d_2)}
\end{pmatrix},
\end{equation}
where
\begin{equation}\label{MRHP_33}
F(\infty)=
\begin{pmatrix}
\frac{\theta(u(\infty)+d_1+c)}{\theta(u(\infty)+d_1)} & 0 \\
0 & \frac{\theta(-u(\infty)+d_2+c)}{\theta(-u(\infty)+d_2)} 
\end{pmatrix}
\end{equation}
Then $M(z)$ has the following jumps:
\begin{equation}\label{MRHP_34}
\begin{aligned}
&M_+(x)=M_-(x)
\begin{pmatrix}
e^{-iN\Om_j} & 0\\
0 & e^{iN\Om_j} 
\end{pmatrix},
\quad x\in (b_{j},a_{j+1});\qquad j=1,\ldots, q-1,\\
&M_+(x)=M_-(x)
\begin{pmatrix}
0 & 1 \\
-1 & 0
\end{pmatrix}, \quad x\in J,
\end{aligned}
\end{equation}
which fits perfectly to the model RHP, and $M(\infty)=I$. It remains
to find $d_1$, $d_2$ such that $M(z)$ is analytic at the zeros of $\theta(\pm u(z)+d_{1,2})$.
These zeros can be cancelled by the zeros of the functions $\ga(z)\pm \ga^{-1}(z)$.
Let us consider the latter zeros.

The zeros of $\ga(z)\pm \ga^{-1}(z)$ are the ones of $\ga^2(z)\pm 1$, and hence of $\ga^4(z)-1$.
By (\ref{MRHP_8a}), the equation $\ga^4(z)-1=0$ reads
\begin{equation}\label{MRHP_35}
p(z)\equiv\prod_{j=1}^q \frac{z-a_j}{z-b_j}=1\,.
\end{equation}
It is easy to see that
\begin{equation}\label{MRHP_36}
p(b_j+0)=\infty,\quad p(a_{j+1})=0,\quad 1\le j\le q-1,
\end{equation}
hence equation (\ref{MRHP_35}) has a solution $x_j$ on each interval $(b_j,a_{j+1})$,
\begin{equation}\label{MRHP_37}
p(x_j)=1,\quad b_j<x_j<a_{j+1};\quad 1\le j\le q-1.
\end{equation}
Since equation (\ref{MRHP_35}) has $(q-1)$ finite solutions, the numbers $\{x_j,\; 1\le j\le q-1\}$
are all the solutions of (\ref{MRHP_35}). 
The function $\ga(z)$, defined by equation (\ref{MRHP_8a}), with cuts on $J$, is
positive on $\R\setminus J$, hence 
\begin{equation}\label{MRHP_38}
\ga(x_j)=1.
\end{equation}
Thus, we have $(q-1)$ zeros of $\ga(z)- \ga^{-1}(z)$ and no zeros of $\ga(z)+ \ga^{-1}(z)$
on the sheet of $\ga(z)$ under consideration.

Let us consider the zeros of the function $\theta(u(z)-d)$. The vector of Riemann constants
is given by the formula
\begin{equation}\label{MRHP_39}
K=-\sum_{j=1}^{q-1}u(b_j).
\end{equation}
Define
\begin{equation}\label{MRHP_40}
d=-K+\sum_{j=1}^{q-1}u(z_j).
\end{equation}
Then 
\begin{equation}\label{MRHP_41}
\theta(u(x_j)-d)=0,\quad 1\le j\le q-1,
\end{equation}
see [DKMVZ2],
and $\{x_j,\; 1\le j\le q-1\}$ are all the zeros of the function $\theta(u(z)-d)$.
In addition, the function $\theta(u(z)+d)$ has no zeros at all on the
upper sheet of $X$. In fact, all the zeros of $\theta(u(z)+d)$ 
lie on the lower sheet, above the same points $\{x_j,\; 1\le j\le q-1\}$.
Therefore, we set in (\ref{MRHP_32}),
\begin{equation}\label{MRHP_42}
d_1=d,\quad d_2=-d,
\end{equation}
so that
\begin{equation}\label{MRHP_43}
M(z)
=F(\infty)^{-1}\begin{pmatrix}
\frac{\ga(z)+\ga^{-1}(z)}{2}\, \frac{\theta(u(z)+d+c)}{\theta(u(z)+d)}
& \frac{\ga(z)-\ga^{-1}(z)}{(-2i)}\, \frac{\theta(-u(z)+d+c)}{\theta(-u(z)+d)} \\
\frac{\ga(z)-\ga^{-1}(z)}{2i}\,  \frac{\theta(u(z)-d+c)}{\theta(u(z)-d)}
& \frac{\ga(z)+\ga^{-1}(z)}{2}\ \frac{\theta(-u(z)-d+c)}{\theta(-u(z)-d)}
\end{pmatrix},
\end{equation}
where
\begin{equation}\label{MRHP_44}
F(\infty)=
\begin{pmatrix}
\frac{\theta(u(\infty)+d+c)}{\theta(u(\infty)+d)} & 0 \\
0 & \frac{\theta(-u(\infty)-d+c)}{\theta(-u(\infty)-d)} 
\end{pmatrix}.
\end{equation}
This gives the required solution of the model RHP.
As $z\to\infty$,
\begin{equation}\label{MRHP_45}
M(z)
=I+\frac{M_1}{z}+O(z^{-2}),
\end{equation}
where
\begin{equation}\label{MRHP_46}
M_1=\begin{pmatrix}
0
& \frac{\theta(-u+d+c)\theta(u+d)}
{\theta(u+d+c)\theta(-u+d)}\sum_{j=1}^q \frac{(b_j-a_j)}{(-4i)} \\
\frac{\theta(u-d+c)\theta(-u-d)}{\theta(-u-d+c)\theta(u-d)}
\sum_{j=1}^q \frac{(b_j-a_j)}{4i}
& 0
\end{pmatrix},\quad u=u(\infty).
\end{equation}

\section{Construction of  a parametrix at  edge points}

 We consider  small disks $D(a_j,r)$, $D(b_j,r)$, $1\le j\le q$, of radius $r > 0$,
centered at the edge points,
\[
D(a,r)\equiv \{z:\; |z-a|\le r\},
\]
and we look for a local parametrix $U_N(z)$, defined on the union of
these disks, such that
\begin{itemize}
\item $U_N(z)$ is analytic on $D \setminus(\R\cup\Ga)$, where
\begin{equation}\label{edge_0}
D=\bigcup_{j=1}^q (D(a_j,r)\cup D(b_j,r)).
\end{equation}
\item
\begin{equation}\label{edge_1}
U_{N+}(z)= U_{N-}(z)j_S(z),\quad z\in (\R\cup\Ga) \cap D,
\end{equation}
\item as $N\to\infty$,
\begin{equation}\label{edge_2}
U_N(z)=\left(I+O\left(\frac{1}{N}\right) \right) M(z)
\quad \text{\rm uniformly for $z \in \partial D$}.
\end{equation}
\end{itemize}

\begin{figure}[ht]
\scalebox{0.6}{\includegraphics{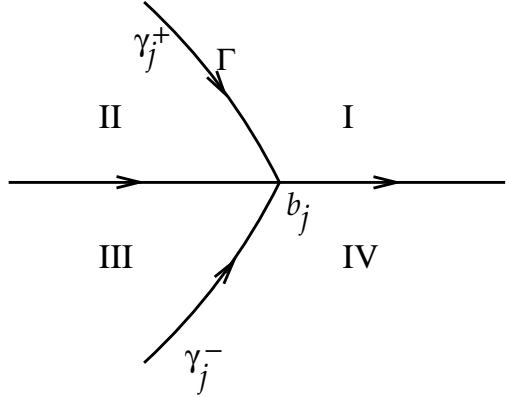}}
\caption{Partition of a neighborhood of the edge point.}
\label{ep_0}
\end{figure}

We consider here the edge point $b_j$, $1\le j\le q$, in detail.
From(\ref{g_1b}) and (\ref{g_4}), we obtain that
\begin{equation} \label{edge_3}
2g_+(x)=V(x)+l+i\Om_j-2\pi i\int_{b_j}^x p_V(s)\,ds,\quad a_j<x<b_j,
\end{equation}
hence
\begin{equation} \label{edge_4}
[2g_+(x)-V(x)]-[2g_+(b_j)-V(b_j)]=-2\pi i\int_{b_j}^x p_V(s)\,ds.
\end{equation}
By using formula (\ref{equi11}), we obtain that
\begin{equation} \label{edge_5}
[2g_+(b_j)-V(b_j)]-[2g_+(x)-V(x)]=\int_{b_j}^x h(s)R^{1/2}_+(s)\,ds,
\quad R(z)=\prod_{j=1}^q(x-a_j)(x-b_j).
\end{equation}
Since both $g(z)$ and $R^{1/2}(z)$ are analytic in the upper half-plane,
we can extend this equation to the upper half-plane,
\begin{equation} \label{edge_5a}
[2g_+(b_j)-V(b_j)]-[2g(z)-V(z)]=\int_{b_j}^z h(s)R^{1/2}(s)\,ds,
\end{equation}
where the contour of integration lies in the upper half-plane.
Observe that
\begin{equation} \label{edge_6}
\int_{b_j}^z h(s)R^{1/2}(s)\,ds=c(z-b_j)^{3/2}+O\left((z-b_j)^{5/2}\right)
\end{equation}
as $z \to b_j$, where $c>0$. Then it follows that
\begin{equation} \label{edge_7}
\beta(z) = \left\{\frac{3}{4}[2g_+(b_j)-V(b_j)]-[2g(z)-V(z)]\right\}^{2/3}
\end{equation}
is analytic at $b_q$, real-valued on the real axis near $b_q$
and $\beta'(b_q) > 0$. So $\beta$ is a conformal map from
$D(b_q, r)$ to a convex neighborhood of the origin, if $r$
is sufficiently small (which we assume to be the case).
We take $\Gamma$ near $b_q$ such that
\[ \beta(\Gamma \cap D(b_q,r)) \subset
\{z \mid \arg(z) = \pm 2\pi/3 \}. \]
Then $\Gamma$ and $\R$ divide the disk $D(z_1,r)$ into
four regions numbered I, II, III, and IV, such
that $0 < \arg \beta(z) < 2\pi/3$, $2\pi/3 < \arg \beta(z) < \pi$,
$-\pi < \arg \beta(z) < -2\pi/3$, and $-2\pi/3 < \arg \beta(z) < 0$ for $z$
in regions I, II, III, and IV, respectively, see Fig.\ref{ep_0}.

Recall that the jumps $j_S$ near $b_q$ are given as
\begin{equation} \label{edge_8}
\begin{aligned}
j_S&=
\begin{pmatrix} 0 & 1 \\
-1 & 0 & \end{pmatrix}
\qquad \text{\rm on } [b_j-r, b_j) \\
j_S&=
\begin{pmatrix} 1 & 0  \\
e^{-NG(z)} & 1  
 \end{pmatrix}
\; \text{\rm on $\ga_q^+$ } \\
j_S&=
\begin{pmatrix} 1 & 0  \\
e^{NG(z)} & 1  
 \end{pmatrix}
\; \text{\rm on $\ga_q^-$ } \\
j_S&=
\begin{pmatrix} e^{-N[g_+(z)-g_-(z)]} & e^{N(g_+(z)+g_-(z)-V(z)-l)}  \\
0 & e^{N[g_+(z)-g_-(z)]} \end{pmatrix}
\qquad \text{\rm on } (b_j, b_j+r].
\end{aligned}
\end{equation}
We look for $U_N(z)$ in the form,
\begin{equation} \label{edge_9}
U_N(z)=Q_N(z)e^{-N\left[g(z)-\frac{V(z)}{2}-\frac{l}{2}\right]\sg_3}.
\end{equation}
Then the jump condition on $U_N(z)$, (\ref{edge_1}), is transformed to the
jump condition on $Q_N(Z)$,
\begin{equation}\label{edge_10}
Q_{N+}(z)= Q_{N-}(z)j_Q(z),
\end{equation}
where
\begin{equation}\label{edge_11}
j_Q(z)= e^{-N\left[g_-(z)-\frac{V(z)}{2}-\frac{l}{2}\right]\sg_3}
j_S(z)e^{N\left[g_+(z)-\frac{V(z)}{2}-\frac{l}{2}\right]\sg_3}.
\end{equation}
From (\ref{edge_8}), (\ref{g_1b}) and (\ref{g_7}) we obtain that
\begin{equation} \label{edge_12}
\begin{aligned}
j_Q&=
\begin{pmatrix} 0 & 1 \\
-1 & 0 
 \end{pmatrix}
\quad \text{\rm on } [b_j-r, b_j), \\
j_Q&=
\begin{pmatrix} 1 & 0  \\
1 & 1  
 \end{pmatrix}
\; \text{\rm on $\ga_q^+$ }, \\
j_Q&=
\begin{pmatrix} 1 & 0  \\
1 & 1  
 \end{pmatrix}
\; \text{\rm on $\ga_q^-$ }, \\
j_Q&=
\begin{pmatrix} 1 & 1  \\
0 & 1 \end{pmatrix}
\qquad \text{\rm on } (b_j, b_j+r].
\end{aligned}
\end{equation}

We construct $Q_N(z)$ with the help of
the Airy function.
The Airy function $\Ai(z)$
solves the equation $y'' = zy$ and for any $\varepsilon >0$, in the
sector $\pi + \varepsilon \leq \arg z \leq \pi - \varepsilon$, it has
the asymptotics as $z \to \infty$,
\begin{equation}\label{edge_13}
\Ai(z)=\frac{1}{2\sqrt\pi z^{1/4}}e^{-\frac{2}{3}z^{3/2}}
\left(1+O(z^{-3/2})\right).
\end{equation}
The functions $\Ai(\om z)$, $\Ai(\om^2 z)$, where
$\om=e^{\frac{2\pi i}{3}}$, also solve the equation $y''=zy$, and we
have the linear relation,
\begin{equation}\label{edge_14}
\Ai(z)+\om\Ai(\om z)+\om^2\Ai(\om^2 z)=0.
\end{equation}
We write
\begin{equation}\label{edge_15}
y_0(z)=\Ai(z), \quad y_1(z)=\om\Ai(\om z),
\quad y_2(z)=\om^2\Ai(\om^2 z),
\end{equation}
and we use these functions to define
\begin{equation}\label{edge_16}
\Phi(z)=
\left\{
\begin{aligned}
{}&\begin{pmatrix}
y_0(z) & -y_2(z)  \\
y_0'(z) & -y_2'(z) 
\end{pmatrix},\quad \mbox{for $0 < \arg z < 2\pi/3$},\\
{}&\begin{pmatrix}
-y_1(z) & - y_2(z) \\
-y_1'(z) & -y_2'(z) \\
\end{pmatrix}, \quad \mbox{for $2\pi/3 < \arg z < \pi$}, \\
{}&\begin{pmatrix}
-y_2(z) & y_1(z) \\
-y_2'(z) & y_1'(z) \\
\end{pmatrix}, \quad \mbox{for $-\pi < \arg z < -2\pi/3$}, \\
{}&\begin{pmatrix}
y_0(z) & y_1(z) \\
y_0'(z) & y_1'(z)  \\
\end{pmatrix},\quad \mbox{for $-2\pi/3 < \arg z < 0$}.
\end{aligned}\right.
\end{equation}
Observe that equation (\ref{edge_14}) reads
\begin{equation}\label{edge_16a}
y_0(z)+y_1(z)+y_2(z)=0,
\end{equation}
and it implies that on the discontinuity rays,
\begin{equation}\label{edge_16b}
\Phi_+(z)=\Phi_-(z)j_Q(z),\quad \arg z=0,\pm\frac{2\pi}{3}\,,\pi.
\end{equation}
Now we set 
\begin{equation}\label{edge_17}
Q_N(z)=E_N(z) \Phi(N^{2/3} \beta(z)),
\end{equation}
so that
\begin{equation}\label{edge_18}
U_N(z)=E_N(z) \Phi(N^{2/3} \beta(z))e^{-N\left[g(z)-\frac{V(z)}{2}-\frac{l}{2}\right]\sg_3},
\end{equation}
where $E_N$ is an analytic prefactor that takes care of the matching
condition (\ref{edge_2}). Since $\Phi(z)$ has the jumps $j_Q$, we 
obtain that $U_N(z)$ has the jumps $j_S$, so that it satisfies jump condition (\ref{edge_1}).
The analytic prefactor $E_N$ is explicitly given by the formula,
\begin{equation} \label{edge_19}
E_N(z) =  M(z)\Theta_N(z)L_N(z)^{-1},
\end{equation}
where $M(z)$ is the solution of the model RHP,
\begin{equation}\label{edge_20}
\Theta_N(z)=e^{\pm \frac{N\Om_j}{2}\,\sg_3},\quad \pm \Im z\ge 0.
\end{equation}
and 
\begin{equation}\label{edge_21}
L_N(z)=\frac{1}{2\sqrt{\pi}}\,\begin{pmatrix} N^{-1/6} \beta^{-1/4}(z) & 0  \\ 0 & N^{1/6}
  \beta^{1/4}(z) 
\end{pmatrix}
\begin{pmatrix} 1 & i  \\ -1 & i 
\end{pmatrix}
\end{equation}
where
for $\be^{1/4}(z)$ we take a branch which is positive for $z\in (b_j,b_j+r]$, with a cut
on $[b_j-r,b_j)$. To prove the analyticity of $E_N(z)$, observe that
\begin{equation}\label{edge_22}
[M(x)\Theta_N(x)]_+=[M(x)\Theta_N(x)]_-j_1(x),\quad b_j-r\le x\le b_j+r,
\end{equation}
where
\begin{equation}\label{edge_23}
j_1(x)=e^{\frac{N\Om_j}{2}\,\sg_3}j_M(x)e^{\frac{N\Om_j}{2}\,\sg_3}.
\end{equation}
From (\ref{MRHP_1}) we obtain that
\begin{equation}\label{edge_23a}
\begin{aligned}
&j_1(x)=\begin{pmatrix} 0 & 1  \\ -1 & 0
\end{pmatrix},\quad b_j-r\le x<b_j,\\
&j_1(x)=I,\quad b_j< x\le b_j+r.
\end{aligned}
\end{equation}
From (\ref{edge_21}),
\begin{equation}\label{edge_24}
L_{N+}(x)=L_{N-}(x)j_2(x),\quad b_j-r\le x\le b_j+r,
\end{equation}
where $j_2(x)=I$ for $b_j<x\le b_j+r$, and
\begin{equation}\label{edge_25}
j_2(x)=\begin{pmatrix} 1 & i  \\ -1 & i 
\end{pmatrix}^{-1}\begin{pmatrix} -i & 0  \\ 0 & i 
\end{pmatrix}\begin{pmatrix} 1 & i  \\ -1 & i 
\end{pmatrix}
=\begin{pmatrix} 0 & 1  \\ -1 & 0
\end{pmatrix},\quad b_j-r\le x<b_j,
\end{equation}
so that $j_2(x)=j_1(x)$, $b_j-r\le x\le b_j+r$. Therefore,
$E_N(z)$ has no jump on $b_j-r\le x\le b_j+r$. Since the entries
of both $M$ and $L$ have at most fourth-root singularities at $b_j$,
the function $E_N(z)$ has a removable singularity at $z=b_j$, hence
it is analytic in $D(b_j,r)$.

Let us prove matching condition (\ref{edge_2}). Consider first $z$ in
domain I on Fig.\ref{ep_0}. From(\ref{edge_13}) we obtain that for 
$0\le\arg z\le \frac{2\pi}{3}$,
\begin{equation}\label{edge_26}
\begin{aligned}
y_0(z)&=\frac{1}{2\sqrt\pi z^{1/4}}\,e^{-\frac{2}{3}z^{3/2}}
\left(1+O(z^{-3/2})\right),\\
-y_2(z)&=\frac{i}{2\sqrt\pi z^{1/4}}\,e^{\frac{2}{3}z^{3/2}}
\left(1+O(z^{-3/2})\right),
\end{aligned}
\end{equation}
hence for $z$ in domain I,
\begin{equation}\label{edge_27}
\Phi(N^{2/3}\be(z))=\frac {1}{2\sqrt{\pi}}\,
N^{-\frac{1}{6}\sg_3}\be(z)^{-\frac{1}{4}\sg_3}
\begin{pmatrix}
1 & i \\
-1 & i
\end{pmatrix}(I+O(N^{-1})) e^{-\frac{2}{3}N\be(z)^{3/2}\sg_3}
\end{equation}
From (\ref{edge_7}),
\begin{equation} \label{edge_28}
\frac{2}{3}\,\beta(z)^{3/2} = \frac{1}{2}\left\{[2g_+(b_j)-V(b_j)]-[2g(z)-V(z)]\right\},
\end{equation}
and from (\ref{edge_3}),
\begin{equation} \label{edge_29}
2g_+(b_j)-V(b_j)=l+i\Om_j, 
\end{equation}
hence
\begin{equation} \label{edge_30}
\frac{2}{3}\,\beta(z)^{3/2} = -g(z)+\frac{V(z)}{2}+\frac{l}{2}+\frac{i\Om_j}{2}\,.
\end{equation}
Therefore, from (\ref{edge_18}) and (\ref{edge_27}) we obtain that
\begin{equation}\label{edge_31}
U_N(z)=E_N(z) \frac {1}{2\sqrt{\pi}}\,
N^{-\frac{1}{6}\sg_3}\be(z)^{-\frac{1}{4}\sg_3}
\begin{pmatrix}
1 & i \\
-1 & i
\end{pmatrix}(I+O(N^{-1})) e^{-\frac{iN\Om_j}{2}}
\end{equation}
Then, from (\ref{edge_19}) and (\ref{edge_21}),
\begin{equation}\label{edge_32}
\begin{aligned}
U_N(z)&=M(z)e^{\frac{iN\Om_j}{2}}L_N(z)^{-1} \frac {1}{2\sqrt{\pi}}\,
N^{-\frac{1}{6}\sg_3}\be(z)^{-\frac{1}{4}\sg_3}
\begin{pmatrix}
1 & i \\
-1 & i
\end{pmatrix}(I+O(N^{-1})) e^{-\frac{iN\Om_j}{2}}\\
&=M(z)e^{\frac{iN\Om_j}{2}}(I+O(N^{-1})) e^{-\frac{iN\Om_j}{2}}=M(z)(I+O(N^{-1})),
\end{aligned}
\end{equation}
which proves (\ref{edge_2}) for $z$ in region I. Similar calculations can be done
for regions II, III, and IV.

\section{ Third and final transformation of the RHP}

\begin{figure}
\scalebox{0.8}{\includegraphics{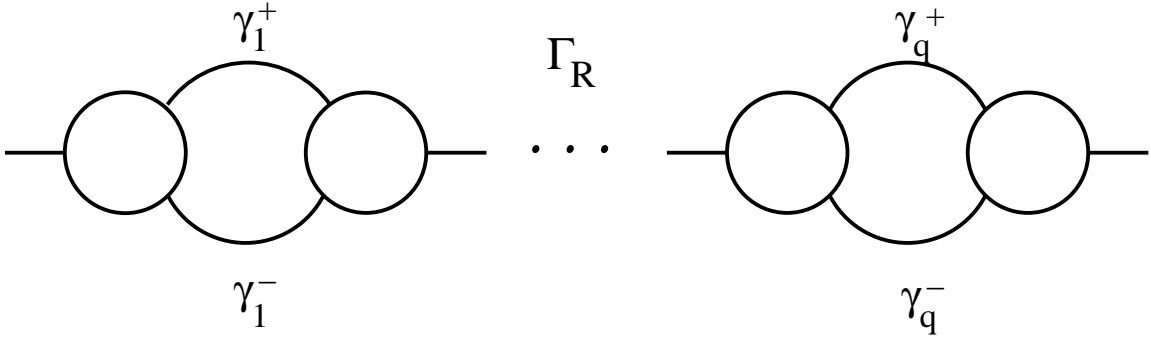}} \caption{The contour
$\Gamma_R$ for $R_N(z)$.} \label{gamma_r_dkmvz}
\end{figure}

In the third and final transformation we put
\begin{equation} \label{r_1}
\begin{aligned}
R_N(z) & = S_N(z) M(z)^{-1}
    \quad \text{\rm for $z$ outside the disks $D( a_j, r)$, $D( b_j, r)$, $1\le j\le q,$} \\
R_N(z) & = S_N(z) U_N(z)^{-1}
    \quad \text{\rm for $z$ inside the disks.}
\end{aligned}
\end{equation}
Then $R_N(z)$ is analytic on $\C \setminus \Ga_R$, where $\Ga_R$ consists of
the circles $\partial D(a_j, r)$, $\partial D( b_j, r)$, $1\le j\le q$, the parts of $\Gamma$
outside of the disks $D( a_j, r)$, $D( b_j, r)$, $1\le j\le q$, and the real intervals $(-\infty, a_1-r)$,
$(b_1+r, a_2-r)$,\ldots, $(b_{q-1},a_q)$, $(b_q+r,\infty)$, see Fig.\ref{gamma_r_dkmvz}.
There are the jump relations,
\begin{equation} \label{r_2}
    R_{N+}(z) = R_{N-}(z) j_R(z),
\end{equation}
where
\begin{equation} \label{r_3}
\begin{aligned}
j_R(z) &= M(z) U_N(z)^{-1}
    \quad \text{\rm on the circles, oriented counterclockwise,} \\
j_R(z) &= M(z) j_S(z) M(z)^{-1}
    \quad \text{\rm on the remaining parts of $\Gamma_R$.}
\end{aligned}
\end{equation}
We have that 
\begin{equation} \label{r_4}
\begin{aligned}
j_R(z) &= I +O(N^{-1})
    \quad \text{\rm uniformly on the circles, } \\
j_R(z) &= I + O(e^{-c(z)N})
    \quad \text{\rm for some $c(z) > 0$, on the remaining parts of $\Gamma_R$.}
\end{aligned}
\end{equation}
In addition, as $x\to\infty$, we have estimate (\ref{ft_6}) on $c(x)$.
As $z\to \infty$, we have
\begin{equation} \label{r_5}
    R_N(z) \cong I + \sum_{j=1}^\infty \frac{R_j}{z^j}\,.
\end{equation}
Thus, $R_N(z)$ solves the following RHP:
\begin{enumerate}
\item
$R_N(z)$ is analytic in $\C\setminus\Ga_R$.  and it is two-valued on $\Ga_R$.
\item
On $\Ga_R$, $R_N(z)$  satisfies jump condition
(\ref{r_2}), where the jump matrix $j_R(z)$ satisfies estimates (\ref{r_4}).
\item 
As $z\to\infty$, $R_N(z)$ has asymptotic expansion (\ref{r_5}).
\end{enumerate}
This RHP can be solved by a perturbation theory series.

\section{Solution of the RHP for $R_N(z)$}

Set
\begin{equation} \label{per_1}
    j_R^0(z)=j_R(z)-I.
\end{equation}
Then by (\ref{r_4}),
\begin{equation} \label{per_2}
\begin{aligned}
j_R^0(z) &= O(N^{-1})
    \quad \text{\rm uniformly on the circles, } \\
j_R^0(z) &=  O(e^{-c(z)N})
    \quad \text{\rm for some $c(z) > 0$, on the remaining parts of $\Gamma_R$,}
\end{aligned}
\end{equation}
where $c(x)$ satisfies (\ref{ft_6}) as $x\to\infty$.
We can apply the following general result.

\begin{prop}  \label{R}
Assume that $v(z)$, $z\in\Ga_R$, solves the
equation
\begin{equation} \label{per_3}
v(z)=I-\frac{1}{2\pi i}\int_{\Ga_R} 
\frac {v(u)j_R^0(u)}{z_- -u}\,du,\quad z\in\Ga_R,
\end{equation}
where $z_-$ means  the value of the integral on the minus side of $\Ga_R$.
Then
\begin{equation} \label{per_4}
R(z)=I-\frac{1}{2\pi i}\int_{\Ga_R} 
\frac {v(u)j_R^0(u)}{z-u}\,du,\quad z\in\C\setminus\Ga_R,
\end{equation}
solves the following RH problem:
\begin{enumerate}
\item[(i)] $ R(z)$ is analytic on $\C \setminus \Ga_R$.
\item[(ii)] {$R_+(z)=R_-(z)j_R(z),$ $z\in\Ga_R$.}
\item[(iii)] {$R(z)= I+O\left(z^{-1}\right),\quad z\to\infty.$}
\end{enumerate}
\end{prop}

{\it Proof.} From (\ref{per_3}), (\ref{per_4}),
\begin{equation} \label{per_5}
R_-(z)=v(z),\quad z\in\Ga_R.
\end{equation}
By the jump property of the Cauchy transform,
\begin{equation} \label{per_6}
R_+(z)-R_-(z)=v(z)j_R^0(z)=R_-(z)j_R^0(z),
\end{equation}
hence $R_+(z)=R_-(z)j_R(z)$. From (\ref{per_4}), $R(z)=I+O(z^{-1})$.
Proposition \ref{R} is proved.

Equation (\ref{per_3}) can be solved by perturbation theory, so that
\begin{equation}
v(z)=I+\sum_{k=1}^\infty v_k(z),
\label {per_7}
\end{equation}
where for $k\ge 1$,
\begin{equation}
v_k(z)=-\frac{1}{2\pi i}\int_{\Ga_R} 
\frac {v_{k-1}(u)j_R^0(u)}{z_- -u}\,du,\quad z\in\Ga_R,
\label {per_8}
\end{equation}
and $v_0(z)=I$. Series (\ref{per_7}) is estimated from above 
by a convergent geometric series, so it is absolutely convergent.
From (\ref{per_2}) we obtain that there exists $C>0$ such that
\begin{equation}
|v_k(z)|\le \frac{C^k}{N^k(1+|z|)}\,.
\label {per_8a}
\end{equation}
Observe that
\begin{equation}
v_1(z)=-\frac{1}{2\pi i}\int_{\Ga_R} 
\frac {j_R^0(u)}{z_- -u}\,du,\quad z\in\Ga_R.
\label {per_9}
\end{equation} 
We apply this solution to find $R_N(z)$.
The function $R_N(z)$ is given then as
\begin{equation}
R_N(z)=I+\sum_{k=1}^\infty R_{Nk}(z),
\label {per_10}
\end{equation}
where
\begin{equation}
R_{Nk}(z)=-\frac{1}{2\pi i}\int_{\Ga_R} 
\frac {v_{k-1}(u)j_R^0(u)}{z -u}\,du.
\label {per_11}
\end{equation}
In particular,
\begin{equation}
R_{N1}(z)=-\frac{1}{2\pi i}\int_{\Ga_R} 
\frac {j_R^0(u)}{z -u}\,du.
\label {per_12}
\end{equation}
From (\ref{per_8a}) we obtain that there exists $C_0>0$ such that
\begin{equation}
|R_{Nk}(z)|\le \frac{C_0 C^k}{N^k(1+|z|)}\,.
\label {per_13}
\end{equation}
Hence from (\ref{per_10}) we obtain that there exists $C_1>0$ such that
for $k\ge 0$,
\begin{equation}
R_N(z)=I+\sum_{j=1}^k R_{Nj}(z)+\ep_{Nk}(z),
\qquad |\ep_{Nk}(z)|\le \frac{C_1 C^k}{N^{k+1}(1+|z|)}\,.
\label {per_14}
\end{equation}
In particular,
\begin{equation} \label{per_15}
    R_N(z) = I + O\left(\frac{1}{N(|z|+1)}\right) \quad \mbox{as } N \to \infty,
\end{equation}
uniformly for $z \in \C \setminus \Ga_R$.

\section{Asymptotics of the recurrent coefficients}

Let us summarize the large $N$ asymptotics of orthogonal polynomials. From 
(\ref{r_1}) and (\ref{per_15}) we obtain that
\begin{equation} \label{main_1}
\begin{aligned}
S_N(z) & =  \left(I + O\left(\frac{1}{N(|z|+1)}\right)\right)M(z),
\quad z\in\C\setminus D,\\
 S_N(z) & = \left(I + O\left(\frac{1}{N(|z|+1)}\right)\right)U_N(z),\quad
z\in D;\\
 D&=\bigcup_{j=1}^q \,[D( a_j, r)\cup D( b_j, r)].
 \end{aligned}
\end{equation}
From (\ref{st_7}) we have that
\begin{equation}\label{main_2}
T_N(z)=\left\{
\begin{aligned}
{}&S_N(z)
\begin{pmatrix}
1 & 0 \\
e^{-NG(z)} & 1
\end{pmatrix},\quad z\in\mathcal L^+=\bigcup_{j=1}^q \mathcal L_j^+,\\
{}&S_N(z)\begin{pmatrix}
1 & 0 \\
-e^{NG(z)} & 1
\end{pmatrix},\quad z\in\mathcal L^-=\bigcup_{j=1}^q \mathcal L_j^-,\\ 
{}&S_N(z), \quad z\in\C\setminus(\mathcal L^+\cup \mathcal L^-).
\end{aligned}
\right.
\end{equation}  
Finally, from (\ref{ft_2}) we obtain that
\begin{equation}\label{main_3}
Y_N(z)=\left\{
\begin{aligned}
{}&e^{\frac{Nl}{2}\sg_3}\left(I + O\left(\frac{1}{N(|z|+1)}\right)\right)M(z)
\begin{pmatrix}
1 & 0 \\
\pm e^{\mp NG(z)} & 1
\end{pmatrix}e^{N\left [g(z)-\frac{l}{2}\right]\sg_3},\quad z\in\mathcal L^{\pm}\setminus D,\\
&e^{\frac{Nl}{2}\sg_3}\left(I + O\left(\frac{1}{N(|z|+1)}\right)\right)U_N(z)
e^{N\left [g(z)-\frac{l}{2}\right]\sg_3},
    \quad z\in  D ,\\
{}&e^{\frac{Nl}{2}\sg_3}\left(I + O\left(\frac{1}{N(|z|+1)}\right)\right)M(z)
e^{N\left [g(z)-\frac{l}{2}\right]\sg_3}, \quad z\in\C\setminus(D\cup\mathcal L^+\cup\mathcal L^-).
\end{aligned}
\right.
\end{equation}  
This gives the large $N$ asymptotics of the orthogonal polynomials and their
adjoint functions on the complex plane. Formulae (\ref{RH_17}) and (\ref{RH_18})
give then the large $N$ asymptotics of the recurrent coefficients. Let us consider
$\ga_N^2$. 

From (\ref{ft_2}) we obtain that for large $z$,
\begin{equation} \label{main_4}
\begin{aligned}
I+\frac{Y_1}{z}+\frac{Y_2}{z^2}+\ldots
&=Y_N(z)z^{-N\sg_3}
=e^{\frac{Nl}{2}\sg_3}T(z)
e^{N\left [g(z)-\frac{l}{2}-\log z\right]\sg_3}\\
&=e^{\frac{Nl}{2}\sg_3}\left(I+\frac{T_1}{z}+\frac{T_2}{z^2}+\ldots\right)
e^{N\left [g(z)-\frac{l}{2}-\log z\right]\sg_3},
 \end{aligned}
\end{equation}
hence
\begin{equation} \label{main_5}
[Y_1]_{12}=e^{Nl}[T_1]_{12},\quad [Y_1]_{21}=e^{-Nl}[T_1]_{21}
\end{equation}
and
\begin{equation} \label{main_6}
\ga_N^2=[Y_1]_{12}[Y_1]_{21}=[T_1]_{12}[T_1]_{21}.
\end{equation}
From (\ref{main_1}), (\ref{main_2}) we obtain further that
\begin{equation} \label{main_6a}
\ga_N^2=[M_1]_{12}[M_1]_{21}+O(N^{-1}),
\end{equation}
and from (\ref{MRHP_46}),
\begin{equation}\label{main_7}
\begin{aligned}
&[M_1]_{12}=\frac{\theta(-u(\infty)+d+c)\theta(u(\infty)+d)}
{\theta(u(\infty)+d+c)\theta(-u(\infty)+d)}\sum_{j=1}^q \frac{(b_j-a_j)}{(-4i)}\,, \\ 
&[M_1]_{21}=\frac{\theta(u(\infty)-d+c)\theta(-u(\infty)-d)}{\theta(-u(\infty)-d+c)\theta(u(\infty)-d)}
\sum_{j=1}^q \frac{(b_j-a_j)}{4i}\,,
 \end{aligned}
\end{equation}
hence
\begin{equation} \label{main_8}
\ga_N^2=\left[\frac{1}{4}\sum_{j=1}^q (b_j-a_j)\right]^2
\frac{\theta^2(u(\infty)+d)\theta(u(\infty)+\frac{N\Om}{2\pi}-d)\theta(-u(\infty)+\frac{N\Om}{2\pi}+d)}
{\theta^2(u(\infty)-d)\theta(-u(\infty)+\frac{N\Om}{2\pi}-d)\theta(u(\infty)+\frac{N\Om}{2\pi}+d)}
+O(N^{-1})\,,
\end{equation}
where $d$ is defined in (\ref{MRHP_40}).
Consider now $\be_{N-1}$. 

From (\ref{main_4}) we obtain that
\begin{equation} \label{main_9}
[Y_1]_{11}=[T_1]_{11}+Ng_1,\quad [Y_2]_{21}=e^{-Nl}([T_2]_{21}+[T_1]_{21}Ng_1),
\end{equation}
hence
\begin{equation} \label{main_10}
\be_{N-1}=\frac{[Y_2]_{21}}{[Y_1]_{21}}-[Y_1]_{11}=\frac{[T_2]_{21}}{[T_1]_{21}}-[T_1]_{11},
\end{equation}
and by (\ref{main_1}), (\ref{main_2}), 
\begin{equation} \label{main_10a}
\be_{N-1}=\frac{[M_2]_{21}}{[M_1]_{21}}-[M_1]_{11}+O(N^{-1}).
\end{equation}
From (\ref{MRHP_43}) we find that
\begin{equation}\label{main_11}
\begin{aligned}
\frac{[M_2]_{21}}{[M_1]_{21}}&=
\frac{\sum_{j=1}^q(b_j^2-a_j^2)}{2\sum_{j=1}^q(b_j-a_j)}
+\frac{\theta(u(\infty)-d)}{\theta(u(\infty)-d+c)}
\left(\left.\nabla_u\frac{\theta(u-d+c)}{\theta(u-d)}\right|_{u=u(\infty)},u'(\infty)\right)\\
&=\frac{\sum_{j=1}^q(b_j^2-a_j^2)}{2\sum_{j=1}^q(b_j-a_j)}
+\left(\frac{\nabla\theta(u(\infty)-d+c)}{\theta(u(\infty)-d+c)}
-\frac{\nabla\theta(u(\infty)-d+c)}{\theta(u(\infty)-d+c)},u'(\infty)\right)
\,,\\
[M_1]_{11}&=\frac{\theta(u(\infty)+d)}{\theta(u(\infty)+d+c)}
\left(\left.\nabla_u\frac{\theta(u+d+c)}{\theta(u+d)}\right|_{u=u(\infty)},u'(\infty)\right)\\
&=\left(\frac{\nabla\theta(u(\infty)+d+c)}{\theta(u(\infty)+d+c)}
-\frac{\nabla\theta(u(\infty)+d+c)}{\theta(u(\infty)+d+c)},u'(\infty)\right).
 \end{aligned}
\end{equation}
Hence,
\begin{equation} \label{main_12}
\begin{aligned}
\be_{N-1}&=\frac{\sum_{j=1}^q(b_j^2-a_j^2)}{2\sum_{j=1}^q(b_j-a_j)}
+\left(
\frac{\nabla\theta(u(\infty)+\frac{N\Om}{2\pi}-d)}{\theta(u(\infty)+\frac{N\Om}{2\pi}-d)}
-\frac{\nabla\theta(u(\infty)+\frac{N\Om}{2\pi}+d)}{\theta(u(\infty)+\frac{N\Om}{2\pi}+d)}\right.\\
 &\left.+\frac{\nabla\theta(u(\infty)+d)}{\theta(u(\infty)+d)}
-\frac{\nabla\theta(u(\infty)-d)}{\theta(u(\infty)-d)},u'(\infty)\right)+O(N^{-1}).
\end{aligned}
\end{equation}
This formula can be also written in the shorter form,
\begin{equation} \label{main_13}
\begin{aligned}
\be_{N-1}&\left.=\frac{\sum_{j=1}^q(b_j^2-a_j^2)}{2\sum_{j=1}^q(b_j-a_j)}
+\frac{d}{dz}\left[\log \frac{\theta(u(z)+\frac{N\Om}{2\pi}-d)\theta(u(z)+d)}
{\theta(u(z)+\frac{N\Om}{2\pi}+d)\theta(u(z)-d)}\right]\right|_{z=\infty}+O(N^{-1}).
\end{aligned}
\end{equation}
In the one-cut case, $q=1$, $a_1=a$, $b_1=b$, formulae (\ref{main_8}), (\ref{main_12}) 
simplify to 
\begin{equation} \label{main_14}
\begin{aligned}
\ga_N=\frac{b-a}{4}+O(N^{-1})\,,\qquad
\be_{N-1}=\frac{a+b}{2}+O(N^{-1})\,.
\end{aligned}
\end{equation}
Formula (\ref{main_8}) is obtained in [DKMVZ2]. Formula (\ref{main_12}) slightly
differs from the formula for $\be_{N-1}$ in [DKMVZ2]: the first term, including $a_j$'s, $b_j$'s,
is missing in [DKMVZ2].

\section {Universality in the random matrix model}

By applying asymptotics (\ref{main_3}) to reproducing kernel (\ref{RH_24}),
we obtain the asymptotics of the eigenvalue correlation functions.
First we consider the eigenvalue correlation functions in the bulk of the spectrum.
Let us fix a point $x_0\in \Int J=\cup_{j=1}^q (a_j,b_j)$. Then the density
$p_V(x_0)>0$. We have the following universal scaling limit of the reproducing
kernel at $x_0$:

\begin{theo} \label{Univ_1}
As $N\to\infty$,
\begin{equation} \label{uni_1}
\lim_{N\to\infty}
    \frac{1}{N p_V(x_0)}
    K_N \left(x_0 + \frac{u}{Np_V(x_0)}, x_0 + \frac{v}{N p_V(x_0)}\right)
    = \frac{\sin  [\pi(u-v)]}{\pi(u-v)}\,.
\end{equation}
\end{theo}

\begin{proof} Assume that for some $1\le j\le q$ and
for some $\ep>0$, we have $\{x_0,x,y\}\in (a_j+\ep,b_j-\ep)$.
By (\ref{RH_24}) and (\ref{ft_2}),
\begin{equation}\label{uni_2}
\begin{aligned}
K_N(x,y)&=
\frac{e^{-\frac{NV(x)}{2}}e^{-\frac{NV(y)}{2}}}{2\pi i(x-y)}
\begin{pmatrix} 0 & 1 \end{pmatrix}
Y_{N+}^{-1}(y)Y_{N+}(x)
\begin{pmatrix} 1 \\ 0 \end{pmatrix}\\
&=
\frac{e^{-\frac{NV(x)}{2}}e^{-\frac{NV(y)}{2}}}{2\pi i(x-y)}
\begin{pmatrix} 0 & e^{N\left[g_+(y)-\frac{l}{2}\right]} \end{pmatrix}
T_{N+}^{-1}(y)T_{N+}(x)
\begin{pmatrix} e^{N\left[g_+(x)-\frac{l}{2}\right]} \\ 0 \end{pmatrix}
\end{aligned}
\end{equation}
Now, from (\ref{st_7}) we obtain that
\begin{equation}\label{uni_3}
\begin{aligned}
K_N(x,y)&=
\frac{e^{-\frac{NV(x)}{2}}e^{-\frac{NV(y)}{2}}}{2\pi i(x-y)}
\begin{pmatrix} 0 & e^{N\left[g_+(y)-\frac{l}{2}\right]} \end{pmatrix}
j_+(y)S_{N+}^{-1}(y)S_{N+}(x)j_+^{-1}(x)
\begin{pmatrix} e^{N\left[g_+(x)-\frac{l}{2}\right]} \\ 0 \end{pmatrix}\\
&=
\frac{e^{-\frac{NV(x)}{2}}e^{-\frac{NV(y)}{2}}}{2\pi i(x-y)}
\begin{pmatrix} e^{N\left[-G(y)+g_+(y)-\frac{l}{2}\right]} & e^{N\left[g_+(y)-\frac{l}{2}\right]}
 \end{pmatrix}
S_{N+}^{-1}(y)\\
&\times  S_{N+}(x)
\begin{pmatrix} e^{N\left[g_+(x)-\frac{l}{2}\right]} \\ -e^{N\left[G(x)+g_+(x)-\frac{l}{2}\right]}
 \end{pmatrix}
\end{aligned}
\end{equation}
By (\ref{g_7}),
\begin{equation} \label{uni_4}
-\frac{V(x)}{2}+g_+(x)-\frac{l}{2}=\frac{G(x)}{2}\,,
\end{equation}
hence
\begin{equation}\label{uni_5}
\begin{aligned}
K_N(x,y)&=
\frac{1}{2\pi i(x-y)}
\begin{pmatrix} e^{\frac{NG(y)}{2}} & e^{-\frac{NG(y)}{2}}
 \end{pmatrix}S_{N+}^{-1}(y) S_{N+}(x)
\begin{pmatrix} e^{-\frac{NG(x)}{2}} \\ -e^{\frac{NG(x)}{2}}
 \end{pmatrix}
\end{aligned}
\end{equation}
By (\ref{r_1}),
\begin{equation} \label{uni_6}
S_{N+}(x)=R_N(x)M_+(x).
\end{equation}
Observe that $M_+(x)$ and $R_N(x)$ are analytic on $(a_j+\ep,b_j-\ep)$
and $R_N(x)$ satisfies estimate (\ref{per_15}). This implies that
as $x-y\to 0$,
\begin{equation} \label{uni_7}
S_{N+}^{-1}(y) S_{N+}(x)=I+O(x-y),
\end{equation}
uniformly in $N$.
Since the function $G(x)$ is pure imaginary for real $x$, we obtain 
from (\ref{uni_5}) and (\ref{uni_7}) that
\begin{equation}\label{uni_8}
K_N(x,y)=
\frac{1}{2\pi i(x-y)}\left[e^{\frac{-N(G(x)-G(y))}{2}}-e^{\frac{N(G(x)-G(y))}{2}}\right]
+O(1).
\end{equation}
By (\ref{g_4}), 
\begin{equation} \label{uni_9}
-\frac{N[G(x)-G(y)]}{2}=\pi iN\int_y^x p_V(s)\,ds=\pi i Np_V(\xi)(x-y),\quad \xi\in[x,y],
\end{equation}
hence
\begin{equation}\label{uni_10}
K_N(x,y)=
\frac{\sin[\pi  Np_V(\xi)(x-y)]}{\pi (x-y)}
+O(1).
\end{equation}
Let
\begin{equation} \label{uni_11}
x=x_0 + \frac{u}{Np_V(x_0)}\,,\quad y=x_0 + \frac{v}{Np_V(x_0)}\,,
\end{equation}
where $u$ and $v$ are bounded. Then
\begin{equation}\label{uni_12}
\frac{1}{Np_V(x_0)}\,K_N(x,y)
=\frac{\sin[\pi  (u-v)]}{\pi (u-v)}
+O(N^{-1}),
\end{equation}
which implies (\ref{uni_1}).
\end{proof}

Consider now the scaling limit at an edge point. Since the density $p_V$
is zero at the edge point, we have to expect a different scaling of the
eigenvalues. 
We have the following universal scaling limit of the reproducing
kernel at the edge point:

\begin{theo} \label{Univ_2}  If  $x_0=b_j$ for some $1\le j\le q$, then for some $c>0$,
as $N\to\infty$,
\begin{equation} \label{uni_13}
\lim_{N\to\infty}
    \frac{1}{(Nc)^{2/3}}
    K_N \left(x_0 + \frac{u}{(Nc)^{2/3}}, x_0 + \frac{v}{(Nc)^{2/3}}\right)
    = \frac{\Ai(u)\Ai'(v)-\Ai'(u)\Ai(v)}{u-v}\,.
\end{equation}
Similarly, if $x_0=a_j$ for some $1\le j\le q$, then for some $c>0$,
as $N\to\infty$,
\begin{equation} \label{uni_13a}
\lim_{N\to\infty}
    \frac{1}{(Nc)^{2/3}}
    K_N \left(x_0 - \frac{u}{(Nc)^{2/3}}, x_0 - \frac{v}{(Nc)^{2/3}}\right)
    = \frac{\Ai(u)\Ai'(v)-\Ai'(u)\Ai(v)}{u-v}\,.
\end{equation}
\end{theo}

The proof is similar to the proof of Theorem \ref{Univ_1}, and we leave it
to the reader.

\vskip 5mm

{\Large Lecture 3.
 Double scaling limit 
in a random matrix model }

\vskip 5mm

\section{Ansatz of the double scaling limit}

This lecture is based on the paper [BI2]. We consider the 
double-well quartic matrix model,
\begin{equation}\label{DSL_1}
\mu_N(M)=Z_N^{-1}e^{-N\Tr V(M)}dM
\end{equation}
(unitary ensemble), with
\begin{equation}\label{DSL_2}
V(M)=\frac {tM^2}{2}+\frac{M^4}{4},\quad t<0.
\end{equation}
The critical point is $t_c=-2$, and the equilibrium measure is one-cut
for $t>-2$ and two-cut for $t<-2$, see Fig.\ref{density_quartic}.

The corresponding monic orthogonal polynomials $P_n(z)=z^n+\dots$ satisfy the orthogonality condition,
\begin{equation}\label{DSL_3a}
\int_{-\infty}^\infty P_n(z)P_m(z)e^{-NV(z)}dz=h_n\de_{nm}.
\end{equation}
 and the recurrence relation,
\begin{equation}\label{DSL_3}
zP_n(z)=P_{n+1}(z)+R_nP_{n-1}(z).
\end{equation}
The string equation has the form,
\begin{equation}\label{DSL_4}
R_n(t+R_{n-1}+R_n+R_{n+1})=\frac{n}{N}\,.
\end{equation}
For any fixed $\ep>0$, the recurrent coefficients $R_n$ have the scaling asymptotics
as $N\to\infty$:
\begin{equation}\label{DSL_5}
R_n=a\left(\frac{n}{N}\right)+(-1)^nb\left(\frac{n}{N}\right)+O(N^{-1}),
\quad \ep\le \frac{n}{N}\le \lambda_c-\ep\,,
\end{equation}
and
\begin{equation}\label{DSL_6}
R_n=a\left(\frac{n}{N}\right)+O(N^{-1}),\quad \ep^{-1}\ge\frac{n}{N}\ge \lambda_c+\ep\,,
\end{equation}
where
\begin{equation}\label{DSL_7}
\lambda_c=\frac{t^2}{4}\,.
\end{equation}
The  scaling functions are: 
\begin{equation}\label{DSL_8}
a(\la)=-\frac{t}{2},\quad
 b(\la)=\frac{\sqrt{t^2-4\la}}{2},
\quad \lambda<\lambda_c,
\end{equation}
and
\begin{equation}\label{DSL_9}
a(\la)=\frac{-t+\sqrt{t^2+3g\la}}{6},\quad \la>\la_c.
\end{equation}
Our goal is to obtain the large $N$ asymptotics of the recurrent coefficients $R_n$,
when $n/N$ is near the critical value $\la_c$. At this point we will assume that $t$ is
an arbitrary (bounded) negative number. In the end we will be interested in the case when
$t$ close to $(-2)$. Let us give first some
heuristic arguments for the critical asymptotics of $R_n$.

We consider $N\to\infty$ with the following scaling behavior of $\frac{n}{N}$:
\begin{equation}\label{DSL_9a}
\frac{n}{N}=\la_c+c_0N^{-2/3}y,\qquad c_0=\left(\frac{t^2}{2}\right)^{1/3},
\end{equation}
where $y\in(-\infty,\infty)$ is a parameter. This limit is called the double
scaling limit. We make the following
Ansatz of the double scaling limit of the recurrent coefficient: 
\begin{equation}\label{DSL_10}
R_n=-\frac{t}{2}+N^{-1/3}(-1)^nc_1u(y)
+N^{-2/3}c_2v(y)+O(N^{-1}),
\end{equation}
where
\begin{equation}\label{DSL_11}
c_1=\left(2|t|\right)^{1/3},\quad
c_2=\frac{1}{2}\left(\frac{1}{2|t|}\right)^{1/3}\,.
\end{equation}
The choice of the constants $c_0,c_1,c_2$ secures that
when we substitute this Ansatz into the left-hand side
of string equation (\ref{DSL_4}), we obtain that
\begin{equation}\label{DSL_12}
R_n(t+R_{n-1}+R_n+R_{n+1})=\frac{n}{N}
+N^{-2/3}c_0\left(v-2u^2-y\right)
+N^{-1}(-1)^n\left(u''-uv\right)+\ldots
\end{equation}
By equating the coefficients at $N^{-2/3}$ and $N^{-1}$ to 0,
we arrive at the equations,
\begin{equation}\label{DSL_13}
v=y+2u^2
\end{equation}
and
\begin{equation}\label{DSL_14}
u''=yu+2u^3,
\end{equation}
the Painlev\'e II equation. The gluing of double scaling asymptotics
(\ref{DSL_10}) with (\ref{DSL_5}) and (\ref{DSL_6}) suggests the boundary conditions:
\begin{equation}\label{DSL_15}
u\sim C\sqrt {-y}, \quad y\to-\infty;\qquad
u\to 0,\quad y\to\infty.
\end{equation}
This selects uniquely the critical, Hastings-McLeod solution to the Painlev\'e II equation.
Thus, in Ansatz (\ref{DSL_10}) $u(y)$ is the Hastings-McLeod solution to  Painlev\'e II
and $v(y)$ is given by (\ref{DSL_13}). The central question is how to prove Ansatz (\ref{DSL_10}).
This will be done with the help of the Riemann-Hilbert approach.

We consider the functions $\psi_n(z)$, $n=0,1,\ldots$, defined in (\ref{intro20}), and their 
adjoint functions,
\begin{equation}\label{DSL_15a}
\f_n(z)=e^{NV(z)/2}\frac{1}{2\pi}\int_{-\infty}^\infty
\frac{e^{-NV(u)/2}\psi_n(u)\,du}{u-z}
\end{equation}
We define the Psi-matrix as
\begin{equation}\label{DSL_15b}
\Psi_n(z)=
\begin{pmatrix}
\psi_n(z) & \f_n(z) \\
\psi_{n-1}(z) & \f_{n-1}(z)
\end{pmatrix}
\end{equation}
The Psi-matrix solves the Lax pair equations:
\begin{align}
&\Psi'_n(z)=NA_n(z)\Psi_n(z),\label{DSL_15c_1}\\
&\Psi_{n+1}(z)=U_n(z)\Psi_n(z).\label{DSL_15c_2}
\end{align}
In the case under consideration, the matrix $A_n$ is given by formula (\ref{LP_5a}),
with $g=1$:
\begin{equation}\label{DSL_16}
A_n(z)=
\begin{pmatrix}
-\left(\frac{tz}{2}+\frac{z^3}{2}+zR_n\right)
 & R_n^{1/2}(z^2+\theta_n) \\
-R_n^{1/2}(z^2+\theta_{n-1}) &  \frac{tz}{2}+\frac{z^3}{2}+zR_n
\end{pmatrix},\qquad \theta_n=t+R_n+R_{n+1}.
\end{equation}
Observe that (\ref{DSL_15c_1}) is a system of two differential equations of the first order.
It can be reduced to the Schr\"odinger equation,
\begin{equation}\label{DSL_WKB_2a}
-\eta''+N^2U\eta=0,\qquad \eta\equiv \frac{\psi_n}{\sqrt{a_{11}}}\,,
\end{equation}
where $a_{ij}$ are the matrix elements of $A_n(z)$, and
\begin{equation}\label{DSL_WKB_2b}
U=-\det A_n+N^{-1}\left[(a_{11})'
-a_{11}\frac{(a_{12})'}{ a_{12}}\right]
-N^{-2}\left[ \frac{(a_{12})''}{ 2a_{12}}-\frac{3((a_{12})')^2}{ 4(a_{12})^2}\right],
\end{equation}
see [BI1], [BI2]. 

The function $\Psi_n(z)$ solves the following RHP:
\begin{enumerate}
\item 
$\Psi_n(z)$ is analytic on $\{\Im z\ge 0\}$
and  on $\{ \Im z\le 0\}$ (two-valued on $\{\Im z=0\}$).
\item
$\Psi_{n+}(x)=\Psi_{n-}(x)
\begin{pmatrix}
1 & - i \\
0 & 1 
\end{pmatrix}$, $x\in\R$.
\item
As $z\to\infty$,
\begin{equation}\label{DSL_16a}
\Psi_n(z)\sim
\left(\di\sum_{k=0}^\infty \frac{\Gamma_k}{z^k}\right)
e^{-\left(\frac{NV(z)}{2}-n\ln z+\la_n\right)\sg_3},
\qquad z\to\infty,
\end{equation}
where $\Ga_k,\; k=0,1,2,\dots$, are some constant $2\times
  2$ matrices, with
\begin{equation}\label{DSL_16b}
\Ga_0=
\begin{pmatrix}
1 & 0 \\
0 & R_n^{-1/2}
\end{pmatrix},\quad
\Ga_1=
\begin{pmatrix}
0 & 1 \\
R_n^{1/2} & 0
\end{pmatrix},
\end{equation}
$\la_n=\frac{\ln h_n}{2}\,,$ and $\sg_3=
\begin{pmatrix}
1 & 0 \\
0 & -1
\end{pmatrix}$ 
is the Pauli matrix,  
\end{enumerate}
Observe that the RHP implies that
$\det \Psi_n(z)$ is an entire function such that $\det \Psi_n(\infty)=R_n^{-1/2}$,
hence
\begin{equation}\label{DSL_16ba}
\det \Psi_n(z)=R_n^{-1/2},\qquad z\in \C.
\end{equation}
We will construct a parametrix, an approximate solution to the RHP.
To that end we use equation (\ref{DSL_15c_1}). We substitute
Ansatz (\ref{DSL_10}) into the matrix elements of $A_n$ and we solve
(\ref{DSL_15c_1}) in the semiclassical approximation,
as $N\to\infty$. First we determine the turning points, the zeros
of $\det A_n(z)$. From (\ref{DSL_16}) we obtain that
\begin{equation}\label{DSL_16c}
\det A_n(z)=a_n(z)\equiv -\frac{tz^4}{2}-\frac{z^6}{4}+\left(\frac{n}{N}-\la_c\right)z^2
+R_n\theta_n\theta_{n+1},
\end{equation}
Ansatz (\ref{DSL_9a}), (\ref{DSL_10}) implies that
\begin{equation}\label{DSL_16d}
\frac{n}{N}-\la_c=c_0N^{-2/3}y     ,\qquad \theta_n=2c_2N^{-2/3}v(y)+O(N^{-1}).
\end{equation}
hence
\begin{equation}\label{DSL_16e}
\det A_n(z)=-\frac{tz^4}{2}-\frac{z^6}{4}+c_0N^{-2/3}yz^2
-2t[c_2v(y)]^2N^{-4/3}+O(N^{-5/3}).
\end{equation}
We see from this formula that
 there are 4 zeros of $\det A_n$, approaching the origin as $N\to\infty$, and 2 zeros,
approaching the points $\pm z_0$, $z_0=\sqrt{-2t}$. Accordingly, we partition
the complex plane into 4 domains: 
\begin{enumerate}
\item a neighborhood of the origin, the critical domain $\Om^{\rm CP}$,
\item 2 neighborhoods of the simple turning points, the turning point
domains $\Om_{1,2}^{\rm TP}$,
\item the rest of the complex plane, the WKB domain $\Om^{\rm WKB}$.
\end{enumerate}
\begin{figure}
\scalebox{0.7}{\includegraphics[width=8in,height=4in]{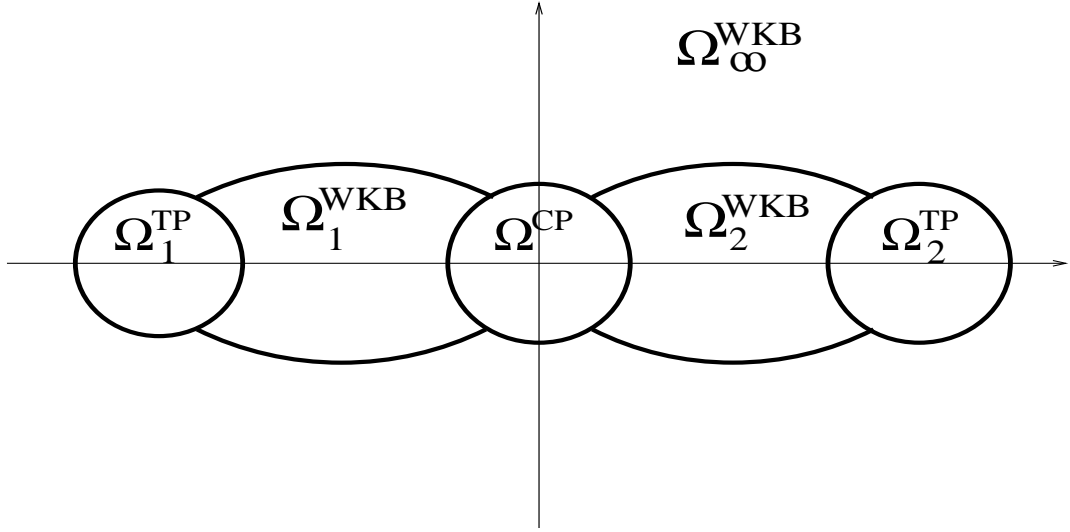}}
\caption{The partition of the complex plane.}
\label{partition}
\end{figure}
We furthermore partition $\Om^{\rm WKB}$ into three domains: $\Om^{\rm WKB}_{1,2}$
and $\Om^{\rm WKB}_{\infty}$, see Fig.\ref{partition}. 

\section {Construction of the  parametrix in $\Om^{\rm WKB}$}

In $\Om^{\rm WKB}_{\infty}$ we define the parametrix by the formula,
\begin{equation}\label{DSL_WKB_1}
\Psi^{\rm WKB}(z)=C_0T(z)e^{-(N\xi(z)+C_1)\sg_3},
\end{equation}
where $C_0\not=0,\,C_1$ are some constants (parameters
of the solution). To introduce $T(z)$ and $\xi(z)$, we need some notations.
We set
\begin{equation}\label{DSL_10_1}
R_n^0=-\frac{t}{2}+N^{-1/3}(-1)^nc_1u(y)
+N^{-2/3}c_2v(y),
\end{equation}
as an approximation to $R_n$, and
\begin{equation}\label{DSL_16_1}
A_n^0(z)=
\begin{pmatrix}
-\left(\frac{tz}{2}+\frac{z^3}{2}+zR_n^0\right)
 & (R_n^0)^{1/2}(z^2+\theta_n^0) \\
-(R_n^0)^{1/2}(z^2+\theta_{n-1}^0) &  \frac{tz}{2}+\frac{z^3}{2}+zR_n^0
\end{pmatrix},\qquad \theta_n^0=t+R_n^0+R_{n+1}^0,
\end{equation}
as an approximation to $A_n(z)$. We set
\begin{equation}\label{DSL_16c_1}
\begin{aligned}
a_n^0(z)&= -\frac{tz^4}{2}-\frac{z^6}{4}+\left(\frac{n}{N}-\la_c\right)z^2
+N^{-4/3}(-t)^{1/3}2^{-5/3}[v(y)^2-4w(y)^2]\\
&-N^{-5/3}(-1)^n(-2t)^{-1/3}w(y)],
\qquad w(y)= u'(y),
\end{aligned}
\end{equation}
as an approximation to $\det A_n(z)$. Finally, we set
\begin{equation}\label{DSL_WKB_2b_1}
U^0=-a_n^0(z)+N^{-1}\left[(a_{11}^0)'
-a_{11}^0\frac{(a_{12}^0)'}{ a_{12}^0}\right],
\end{equation}
as an approximation to the potential $U$ in (\ref{DSL_WKB_2b}).
With these notations, 
\begin{equation}\label{DSL_WKB_2}
\xi(z)=\di\int_{z_N}^z \mu(u)\,du,\qquad \mu(z)=\sqrt{U^0(z)},
\end{equation}
where $z_N$ is the zero of $U^0(z)$ which approaches $z_0$
as $N\to\infty$. Also,
\begin{equation}\label{DSL_WKB_2a_1}
T(z)=
\left(\frac{ a_{12}^0(z)}{ \mu(z)}\right)^{1/2}
\begin{pmatrix}
1 & 0 \\
&\\
-\frac{a^0_{11}(z)}{a^0_{12}(z)} &
\frac{\mu(z)}{a^0_{12}(z)}
\end{pmatrix},\qquad \det T(z)=1.
\end{equation}
From (\ref{DSL_WKB_1}) we obtain the following asymptotics as $z\to\infty$:
\begin{equation}\label{DSL_WKB_infinity}
\Psi^{\rm WKB}(z)=\sqrt 2\,C_0\left(I+z^{-1}(R_n^0)^{1/2}
\begin{pmatrix}
0 & 1 \\
1 & 0
\end{pmatrix}+O(z^{-2})\right)
e^{-\left(\frac{NV(z)}{2}-n\ln z+\la_n^0+C_1\right)\sg_3},\qquad z\to\infty,
\end{equation}
where
\begin{equation}\label{DSL_WKB_infinity_1}
\la_n^0=\lim_{z\to\infty}\left[N\xi(z)-\left(\frac{NV(z)}{2}-n\ln z\right)\right].
\end{equation}
The existence of the latter limit follows from (\ref{DSL_WKB_2}).

In the domains $\Om^{\rm WKB}_{1,2}$ we define
\begin{equation}\label{DSL_WKB_3}
\Psi^{\rm WKB}(z)=\Psi^{\rm WKB}_a(z)S_{\pm},\qquad \pm\Im z\ge 0,
\end{equation}
where $\Psi^{\rm WKB}_a(z)$ is the analytic continuation of $\Psi^{\rm WKB}(z)$
from $\Om^{\rm WKB}_{\infty}$ to $\Om^{\rm WKB}_{1,2}$, from the upper half-plane,
and
\begin{equation}\label{DSL_WKB_4}
S_+=
\begin{pmatrix}
1 & 0 \\
i & 1
\end{pmatrix}
=S_-
\begin{pmatrix}
1 & -i \\
0 & 1
\end{pmatrix}
,\qquad
S_-=
\begin{pmatrix}
1 & i \\
i & 0
\end{pmatrix}.
\end{equation}
Observe that $\Psi^{\rm WKB}(z)$ has jumps:
\begin{equation}\label{DSL_WKB_5}
\Psi^{\rm WKB}_+(z)=(I+O(e^{-cN}))\Psi^{\rm WKB}_-(z),\qquad
z\in \partial\Om^{\rm WKB}_{\infty}\cap\left(\partial\Om^{\rm WKB}_1\cup\partial\Om^{\rm WKB}_2\right),
\end{equation}
and
\begin{equation}\label{DSL_WKB_6}
\Psi^{\rm WKB}_+(z)=\Psi^{\rm WKB}_-(z)
\begin{pmatrix}
1 & -i \\
0 & 1
\end{pmatrix}
,\qquad
z\in \R\cap\left(\Om^{\rm WKB}_1\cup\Om^{\rm WKB}_2\right).
\end{equation}

\section {Construction of the  parametrix near the turning points}

In $\Om_2^{\rm TP}$ we define the parametrix with the help of the Airy matrix-valued functions,
\begin{equation}\label{DSL_Ai_1}
Y_{1,2}(z)=
\begin{pmatrix}
y_0(z) & y_{1,2}(z) \\
y'_0(z) & y'_{1,2}(z) 
\end{pmatrix},
\end{equation}
where
\begin{equation}\label{DSL_Ai_2}
y_0(z)=\Ai(z),\quad
y_1(z)=e^{-\pi i/6}\Ai\left( e^{-2\pi i/3}z\right),\quad
y_2(z)=e^{\pi i/6}\Ai\left( e^{2\pi i/3}z\right).
\end{equation}
Let us remind that $\Ai(z)$ is a solution
to the Airy equation $y''=zy$, which has the following asymptotics as
$z\to\infty$:
\begin{equation}\label{DSL_Ai_3}
\Ai(z)= \frac{1}{ 2\sqrt\pi z^{1/4}}
\exp\left(-\frac{2z^{3/2}}{ 3}+O(|z|^{-3/2})\right),\quad 
-\pi+\ep\le \arg z\le\pi-\ep.
\end{equation}
The functions $y_j(z)$ 
satisfy the relation
\begin{equation}\label{DSL_Ai_4}
y_1(z)-y_2(z)=-iy_0(z).
\end{equation}
We define the parametrix in $\Om_2^{\rm TP}$ by the formula,
\begin{equation}\label{DSL_Ai_5}
\Psi^{\rm TP}(z)=W(z)N^{\sg_3/6}Y_{1,2}(N^{2/3}w(z)),\qquad \pm\Im z\ge 0,
\end{equation}
where 
\begin{equation}\label{DSL_Ai_6}
w(z)=\left[\frac{3}{2}\xi(z)\right]^{2/3},
\end{equation}
with $\xi(z)$ defined in (\ref{DSL_WKB_2}) above. Observe that $w(z)$
is analytic in $\Om_2^{\rm TP}$. The matrix-valued function $W(z)$ is also analytic
in $\Om_2^{\rm TP}$, and it is found from the following condition of the matching $\Psi^{\rm TP}(z)$
to $\Psi^{\rm WKB}(z)$ on $\partial \Om_2^{\rm TP}$:
\begin{equation}\label{DSL_Ai_7}
\Psi^{\rm TP}(z)=(I+O(N^{-1}))\Psi^{\rm WKB}(z),\qquad z\in \partial \Om_2^{\rm TP},
\end{equation}
see [BI1], [BI2]. A similar construction of the parametrix is used in the domain $\Om_1^{\rm TP}$.

\section {Construction of the  parametrix near the critical point}

\subsection{Model solution} The crucial question is, what should be an Ansatz for the parametrix
in the critical domain $\Om^{\rm CP}$? To answer this question, let us construct a normal
form of system of differential equations (\ref{DSL_15c_1}) at the origin.
If we substitute Ansatz (\ref{DSL_10}) into the matrix elements of $A_n(z)$,  
change
\begin{equation}\label{DSL_17}
\Psi(z)=\Phi(CN^{1/3}z),\qquad C=\frac{(2t)^{1/6}}{2}\,,
\end{equation}
and keep the leading terms, as $N\to\infty$, then we obtain the model equation (normal form),
\begin{equation}\label{DSL_18}
\Phi'(s)=A(s)\Phi(s),
\end{equation}
where
\begin{equation}\label{DSL_19}
A(s)=
\begin{pmatrix}
(-1)^n4u(y)s & 4s^2+(-1)^n2w(y)+v(y) \\
-4s^2+(-1)^n2w(y)-v(y)) & -(-1)^n4u(y)s \\
\end{pmatrix},
\end{equation}
and $w(y)=u'(y)$. In fact,
this is one of the equations of the Lax pair for the Hastings-Mcleod solution to Painlev\'e II. 
Equation (\ref{DSL_18}) possesses three special solutions, $\vec\Phi_j$, $j=0,1,2$, 
which are characterized by their asymptotics as $|s|\to\infty$:
\begin{equation}\label{DSL_20}
\begin{aligned}
&\vec \Phi_0(s)
=\begin{pmatrix}
\Phi^1(s)\\ \Phi^2(s)
\end{pmatrix}\sim \begin{pmatrix}\cos\left(\frac{4}{3}\,s^3+ys-\frac{\pi n}{2}\right) \\ 
-\sin\left(\frac{4}{3}\,s^3+ys-\frac{\pi n}{2}\right)
\end{pmatrix},
\quad |\arg s|\le \frac{\pi}{3}-\ep,\\
&\vec \Phi_1(s)=\overline{\vec \Phi_2(\overline s)}\sim \begin{pmatrix}(-i)^{n+1} \\ (-i)^n
\end{pmatrix}
e^{i(\frac{4}{3}\,s^3+ys)},
\quad -\frac{\pi}{ 3}+\ep<\arg s<\frac{4\pi}{ 3}-\ep.
\end{aligned}
\end{equation}
The functions $\Phi^{1,2}(s)$ are real for real $s$ and 
\begin{equation}\label{DSL_20a}
\Phi^1(-s)=(-1)^n\Phi^1(s),\qquad \Phi^2(-s)=-(-1)^n\Phi^2(s).
\end{equation}
We define the $2\times 2$ matrix-valued function on $\C$,
\begin{equation}\label{DSL_21}
\Phi(s)=
 \left(\vec \Phi_0(s),\vec\Phi_{1,2}(s)\right),\qquad \pm\Im s\ge 0,
\end{equation}
which is two-valued on $\R$, and
\begin{equation}\label{DSL_22}
\Phi_+(s)=\Phi_-(s)
\begin{pmatrix}
1 & -i \\
0 & 1
\end{pmatrix},\qquad s\in \R.
\end{equation}
The Ansatz for the parametrix in the critical domain  is
\begin{equation}\label{DSL_23}
\Psi^{\rm CP}_n(z)=\tilde C_0V(z)\Phi\left(N^{1/3}\z(z)\right),
\end{equation}
where $\tilde C_0$ is a constant, a parameter of the solution, 
$\z(z)$ is an analytic scalar function such that $\z'(z)\not=0$ and
$V(z)$ is  an analytic matrix-valued function. We now describe a procedure
of choosing $\z(z)$ and $V(z)$. The essence of the RH approach is that
we don't need to justify this procedure. We need only that
$\z(z)$ and $V(z)$ are analytic in $\Om^{\rm CP}$, and
that on $\partial \Om^{\rm CP}$, Ansatz (\ref{DSL_23}) fits to the WKB Ansatz.
 
\subsection{Construction of $\z(z)$: Step 1}
To find $\z(z)$ and $V(z)$,
let us substitute Ansatz (\ref{DSL_23}) into equation (\ref{LP_2}). 
This gives 
\begin{equation}\label{DSL_24}
V(z)\left[\z'(z)N^{-2/3}A\left(N^{1/3}\z(z)\right)\right]V^{-1}(z)
=A_n(z)-N^{-1}V'(z)V^{-1}(z)
\end{equation}
Let us drop the term of the order of $N^{-1}$ on the right:
\begin{equation}\label{DSL_24a}
V(z)\left[\z'(z)N^{-2/3}A\left(N^{1/3}\z(z)\right)\right]V^{-1}(z)
=A_n(z),
\end{equation}
and take the
determinant of the both sides. This gives an equation on $\z$ only,
\begin{equation}\label{DSL_25}
[\z'(z)]^2 f(\z(z))=a_n(z),
\end{equation}
where
\begin{equation}\label{DSL_26}
f(\z)=N^{-2/3}\det A(N^{1/3}\z)=16\z^4+8N^{-2/3}y\z^2+N^{-4/3}[v^2(y)-4w^2(y)]
\end{equation}
and
\begin{equation}\label{DSL_27}
a_n(z)=\det A_n(z)=-\frac{tz^4}{2}-\frac{z^6}{4}+\left(\frac{n}{N}-\la_c\right)z^2
+R_n\theta_n\theta_{n+1},
\end{equation}
where
\begin{equation}\label{DSL_27a}
\theta_n=t+R_n+R_{n+1}.
\end{equation}
Equation (\ref{DSL_10}) implies that
\begin{equation}\label{DSL_27b}
\theta_n=2c_2N^{-2/3}v(y)+O(N^{-1}).
\end{equation}
At this stage we drop all terms of the order of $N^{-1}$, and,
therefore, we can simplify $f$ and $a_n$ to
\begin{equation}\label{DSL_27c}
f(\z)=16\z^4+8N^{-2/3}y\z^2
\end{equation}
and
\begin{equation}\label{DSL_27d}
a_n(z)=-\frac{tz^4}{2}-\frac{z^6}{4}+\left(\frac{n}{N}-\la_c\right)z^2.
\end{equation}
Equation (\ref{DSL_25}) is separable and we are looking for 
an analytic solution. To construct an analytic solution, let us make
the change of variables,
\begin{equation}\label{DSL_28}
z=CN^{-1/3}s,\qquad \z=N^{-1/3}\sg.
\end{equation}
Then equation (\ref{DSL_25}) becomes 
\begin{equation}\label{DSL_29}
[\sg'(s)]^2f_0(\sg(s))=a_0(s),
\end{equation}
where
\begin{equation}\label{DSL_30}
f_0(\sg)=16\sg^4+8y\sg^2,\qquad
a_0(s)=16s^4+8c_0^{-1}N^{2/3}\left(\frac{n}{N}-\la_c\right)s^2
-N^{-2/3}cs^6.
\end{equation}
When we substitute equation (\ref{DSL_11}) for $y$, we obtain that
\begin{equation}\label{DSL_30a}
a_0(s)=16s^4+8ys^2
-N^{-2/3}cs^6.
\end{equation}
When $y=0$, equation (\ref{DSL_29}) is easy to solve: by taking the square
root of the both sides, we obtain that
\begin{equation}\label{DSL_31}
\sg^2\sg'=s^2\left(1-\frac{1}{16}N^{-2/3}cs^2\right)^{1/2},
\end{equation}
hence
\begin{equation}\label{DSL_32}
\sg(s)=\left[\int_0^s t^2\left(1-\frac{1}{16}N^{-2/3}ct^2\right)^{1/2}dt\right]^{1/3}
\end{equation}
is an analytic solution to (\ref{DSL_29}) in the disk $|s|\le \ep N^{1/3}$, for some $\ep>0$.
This gives an analytic solution $\z(z)=N^{-1/3}\sg(C^{-1}N^{1/3}z)$ to equation (\ref{DSL_27})
in the disk $|z|\le C\ep$.

When $y\not=0$, the situation is more complicated, and in fact, equation (\ref{DSL_29}) 
has no analytic solution in the disk $|s|\le \ep N^{1/3}$.
Consider, for instance, $y>0$. By taking the square
root of the both sides of (\ref{DSL_29}), we obtain that
\begin{equation}\label{DSL_33}
\sg\left(\sg^2+\frac{y}{2}\right)^{1/2}\sg'=s\left(s^2+\frac{y}{2}
-\frac{1}{16}N^{-2/3}cs^4\right)^{1/2},
\end{equation}
The left hand side has simple zeros at $\pm \sg_0=\pm i\sqrt{\frac{y}{2}}$, and
the right hand side has simple zeros at $\pm s_0$, where $s_0=\sg_0+O(N^{-2/3})$. 
The necessary and sufficient condition for the existence of an analytic solution to equation 
(\ref{DSL_29}) in the disk $|s|\le \ep N^{1/3}$ is the equality of the periods,
\begin{equation}\label{DSL_34}
P_1\equiv \int_{-\sg_0}^{\sg_0}\sg\left(\sg^2+\frac{y}{2}\right)^{1/2}d\sg
=P_2\equiv \int_{-s_0}^{s_0} s\left(s^2+\frac{y}{2}
-\frac{1}{16}N^{-2/3}cs^4\right)^{1/2}ds,
\end{equation}
and, in fact, $P_1\not=P_2$. To make the periods equal, we slightly change
equation (\ref{DSL_11}) as follows: 
\begin{equation}\label{DSL_35}
y=c_0^{-1}N^{2/3}\left(\frac{n}{N}-\la_c\right)+\al,
\end{equation}
where $\al$ is a parameter. Then
\begin{equation}\label{DSL_36}
P_2=P_2(\al)\equiv \int_{-s_0(\al)}^{s_0(\al)} s\left(s^2+\frac{y-\al}{2}
-\frac{1}{16}N^{-2/3}cs^4\right)^{1/2}ds.
\end{equation}
It is easy to check that $P_2'(0)\not=0$, and therefore, there exists
an $\al=O(N^{-2/3})$ such that $P_1=P_2$. This gives an analytic 
solution $\sg(s)$, and hence an analytic $\z(z)$.

\subsection{Construction of $V(z)$}
Next, we find a matrix-valued function $V(z)$ from equation (\ref{DSL_24a}). 
Both $V(z)$ and $V^{-1}(z)$ should be analytic at the origin. We have
the following lemma.

\begin{lem} \label{lem_1}
Let $B=\left( b_{ij} \right)$ and $D= \left(d_{ij}\right)$
 be two $2\times 2$ matrices such that
\begin{equation}\label{DSL_37}
\Tr B=\Tr D=0,\qquad \det B=\det D.
\end{equation}
Then the equation $VB=DV$ has the following two explicit solutions:
\begin{equation}\label{DSL_38}
V_1=
\begin{pmatrix}
d_{12} & 0 \\
b_{11}-d_{11} & b_{12}
\end{pmatrix},
\qquad
V_2=\begin{pmatrix}
b_{21} & d_{11}-b_{11} \\
0 & d_{21}
\end{pmatrix}.
\end{equation}
\end{lem}

We would like to apply Lemma \ref{lem_1} to
\begin{equation}\label{DSL_39}
B=\z'(z)N^{-2/3}A\left(N^{1/3}\z(z)\right),\quad D=A_n(z).
\end{equation}
The problem is that we need an
analytic matrix valued function $V(z)$ which is invertible 
in a fixed neighborhood of  the origin, but
neither $V_1$ nor $V_2$ are
invertible there. Nevertheless, we can find a linear combination of
$V_1$ and $V_2$ (plus some negligibly small terms)
which is analytic and invertible. Namely, we take
\begin{equation}\label{DSL_40}
V(z)=\frac{1}{\sqrt{\det W(z)}}W(z)
\end{equation}
where
\begin{equation}\label{DSL_41}
W(z)=\begin{pmatrix}
d_{12}(z)-b_{21}(z)-\al_{11} & 
b_{11}(z)-d_{11}(z)-\al_{12}z \\
b_{11}(z)-d_{11}(z)-\al_{21}z, & 
b_{12}(z)-d_{21}(z)-\al_{22}
\end{pmatrix},
\end{equation}
and the numbers $\al_{ij}=O(N^{-1})$ are chosen in such a way that the
matrix elements of $W$
vanish at the same points $\pm z_0$, $z_0=O(N^{-1/3})$, on the complex plane.
Then $V(z)$ is analytic in a disk $|z|<\ep$, $\ep>0$.

\subsection{Construction of $\z(z)$: Step 2}
The accuracy of $\z(z)$, which is obtained from equation (\ref{DSL_25}),
is not sufficient for the required fit on $|z|=\ep$, of Ansatz (\ref{DSL_23})
to the WKB Ansatz. Therefore, we correct $\z(z)$ slightly by taking into account the term 
$-N^{-1}V'(z)V^{-1}(z)$ in equation (\ref{DSL_24}). We have to solve the equation,
\begin{equation}\label{DSL_42}
[\z'(z)]^2N^{-4/3}\det
A\left(N^{1/3}\z(z)\right)=\det\left[A_n(z)-N^{-1}V'(z)V^{-1}(z)\right].
\end{equation}
By change of variables (\ref{DSL_28}), it reduces to
\begin{equation}\label{DSL_43}
[\sg'(s)]^2f_1(\sg(s))=a_1(s),
\end{equation}
where
\begin{equation}\label{DSL_44}
\begin{aligned}
&f_1(\sg)=16\sg^4+8y\sg^2+[v^2(y)-4w^2(y)];\\
&a_1(s)=16s^4+8(y-\al)s^2
+[v^2(y)-4w^2(y)]+r_N(s),\quad r_N(s)=O(N^{-2/3}).
\end{aligned}
\end{equation}
The function $f_1(\sg)$ has 4 zeros, $\pm \sg_j$ $j=1,2$. The function $a_1(s)$
is a small perturbation of $f_1(s)$, and it has 4 zeros, $\pm s_j$,
such that $|s_j-\sg_j|\to 0$ as $N\to \infty$. Equation (\ref{DSL_43})
has an analytic solution in the disk of radius $\ep N^{2/3}$ if
and only if the periods are equal,
\begin{equation}\label{DSL_45}
P_{1j}\equiv \int_{-\sg_j}^{\sg_0}\sqrt{f_1(\sg)}\,d\sg
=P_{2j}\equiv \int_{-s_j}^{s_j} \sqrt{a_1(s)}\,ds,\quad j=1,2.
\end{equation}
To secure the equality of periods we include $a_1(s)$ into the 2-parameter family
of functions,
\begin{equation}\label{DSL_46}
a_1(s)=16s^4+8(y-\al)s^2
+[v^2(y)-4w^2(y)]+r_N(s)+\be,
\end{equation}
where $-\infty<\al,\be<\infty$ are the parameters. A direct calculation gives that 
\begin{equation}\label{DSL_47}
\det \begin{pmatrix}
\frac{\partial P_{21}}{\partial \al} & \frac{\partial P_{21}}{\partial \be} \\
\frac{\partial P_{22}}{\partial \al} & \frac{\partial P_{22}}{\partial \be}
\end{pmatrix}\not=0,
\end{equation}
 see [BI2], hence, by the implicit function theorem, there exist $\al,\be=O(N^{-2/3})$,
which solve equations of periods (\ref{DSL_45}). This gives an analytic function $\z(z)$,
and hence, by (\ref{DSL_23}), an analytic $\Psi^{\rm CP}(z)$. 

The function $\Psi^{\rm CP}(z)$ matches the WKB-parametrix $\Psi^{\rm WKB}(z)$ on
the boundary of $\Om^{\rm CP}$. Namely, we have the following lemma.

\begin{lem}\label {CP_WKB} {\rm (See [BI2].)} If we take $\tilde C_0=C_0$ and
$C_1=-\frac{1}{4}\,\ln R_n^0$ then
\begin{equation}\label{DSL_48}
\Psi^{\rm CP}(z)=(I+O(N^{-1}))\Psi^{\rm WKB}(z),\qquad z\in \partial \Om^{\rm CP}.
\end{equation}
\end{lem}

We omit the proof of the lemma, because it is rather straightforward, although technical. We refer
the reader to the paper [BI2] for the details.
 
\subsection{Proof of the double scaling asymptotics}
Let us summarize the construction of the parametrix in different domains. We define the
parametrix $\Psi_n^0$ as 
\begin{equation}\label{PAR_1}
\Psi_n^0(z)=\left\{
\begin{aligned}
&\Psi^{\rm WKB}(z),\quad z\in\Om^{\rm WKB}=\Om^{\rm WKB}_{\infty}\cup\Om^{\rm WKB}_1\cup\Om^{\rm WKB}_2,\\
&\Psi^{\rm TP}(z),\quad z\in\Om^{\rm TP}_1\cup\Om^{\rm TP}_2,\\
&\Psi^{\rm CP}(z),\quad z\in\Om^{\rm CP},
\end{aligned}\right.
\end{equation}
where $\Psi^{\rm WKB}(z)$ is given by (\ref{DSL_WKB_1}),
(\ref{DSL_WKB_3}), $\Psi^{\rm TP}(z)$ by (\ref{DSL_Ai_5}), and $\Psi^{\rm CP}$ by 
(\ref{DSL_23}). Consider the quotient,
\begin{equation}\label{PAR_2}
X(z)=\Psi_n(z)[\Psi_n^0(z)]^{-1}.
\end{equation}
\begin{figure}
\scalebox{0.7}{\includegraphics[width=8in,height=2in]{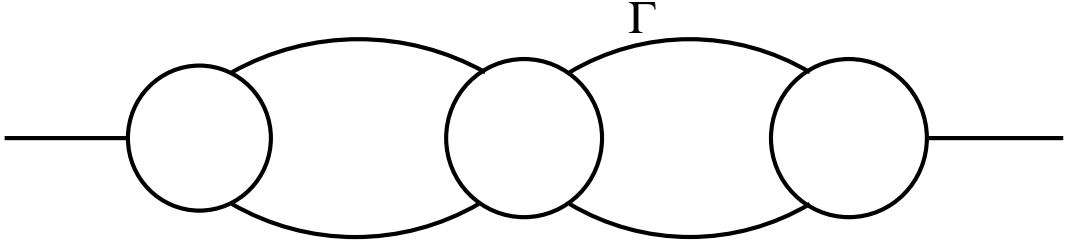}}
\caption{The contour $\Ga$.}
\label{contour_Gamma}
\end{figure}
$X(z)$ has jumps on the contour $\Ga$, depicted on Fig.\ref{contour_Gamma}, such that
\begin{equation}\label{PAR_3}
X_+(z)=X_-(z)\left[I+O(N^{-1}(1+|z|)^{-2})\right],\qquad z\in\Gamma.
\end{equation}
From (\ref{DSL_16a}) and (\ref{DSL_WKB_infinity}) we obtain that 
\begin{equation}\label{PAR_4}
X(z)=X_0+\frac{X_1}{z}+O(z^{-2}),
\qquad z\to\infty,
\end{equation}
where
\begin{equation}\label{PAR_5}
X_0=\frac{1}{\sqrt{2}}C_0^{-1}\Ga_0e^{(C_1+\la_n^0-\la_n)\sg_3}
\end{equation}
and
\begin{equation}\label{PAR_6}
X_1=\frac{1}{\sqrt{2}}C_0^{-1}
\left[\Ga_1e^{(C_1+\la_n^0-\la_n)\sg_3}-\Ga_0e^{(C_1+\la_n^0-\la_n)\sg_3}(R_n^0)^{1/2}\sg_1\right],
\qquad \sg_1=
\begin{pmatrix}
0 & 1 \\
1 & 0
\end{pmatrix}.
\end{equation}
The RHP shares a remarkable property of well-posedness, see, e.g. [BDT], [CG],
[LiS], [Zh]. Namely, equations (\ref{PAR_3}), (\ref{PAR_4}) imply that
\begin{equation}\label{PAR_7}
X_0^{-1}X(z)=I+O(N^{-1}(1+|z|)^{-1}),
\qquad z\in\C.
\end{equation}
This in turn implies that
\begin{equation}\label{PAR_8}
X_0^{-1}X_1=O(N^{-1}),
\end{equation}
or, equivalently,
\begin{equation}\label{PAR_9}
e^{-(C_1+\la_n^0-\la_n)\sg_3}\Ga_0^{-1}\Ga_1e^{(C_1+\la_n^0-\la_n)\sg_3}-(R_n^0)^{1/2}\sg_1=O(N^{-1}).
\end{equation}
By (\ref{DSL_16b}),
\begin{equation}\label{PAR_10}
e^{-(C_1+\la_n^0-\la_n)\sg_3}\Ga_0^{-1}\Ga_1e^{(C_1+\la_n^0-\la_n)\sg_3}
=\begin{pmatrix}
0 & e^{-2(C_1+\la_n^0-\la_n)}\\
R_ne^{2(C_1+\la_n^0-\la_n)} & 0
\end{pmatrix},
\end{equation}
hence (\ref{PAR_9}) reduces to the system of equations,
\begin{equation}\label{PAR_11a}
\left\{
\begin{aligned}
e^{-2(C_1+\la_n^0-\la_n)}&=(R_n^0)^{1/2}+O(N^{-1}),\\
R_ne^{2(C_1+\la_n^0-\la_n)}&=(R_n^0)^{1/2}+O(N^{-1}).
\end{aligned}
\right.
\end{equation}
By multiplying these equations, we obtain that
\begin{equation}\label{PAR_11}
R_n=R_n^0+O(N^{-1}).
\end{equation}
This proves Ansatz (\ref{DSL_10}). Since $C_1=-\frac{1}{4}\,\ln R_n^0$, we obtain 
from (\ref{PAR_11}) that
\begin{equation}\label{PAR_12}
e^{2(\la_n-\la_n^0)}=1+O(N^{-1}),
\end{equation}
or equivalently,
\begin{equation}\label{PAR_13}
h_n=\exp\left( 2N\int_{z_N}^\infty \mu(u)\,du\right)\left(1+O(N^{-1})\right),
\end{equation}
where
\begin{equation}\label{PAR_14}
\int_{z_N}^\infty \mu(u)\,du\equiv \lim_{z\to\infty}\left[\int_{z_N}^z \mu(u)\,du
-\left(\frac{V(z)}{2}-\frac{n\ln z}{N}\right)\right].
\end{equation}
Thus, we have the following theorem.

\begin{theo} {\rm (See [BI2])}. \label{T_1}
The recurrent coefficient $R_n$ under the scaling
(\ref{DSL_9a}) has asymptotics (\ref{DSL_10}). The normalizing coefficient $h_n$
has asymptotics (\ref{PAR_13}).
\end{theo}

Equations (\ref{PAR_2}) and (\ref{PAR_7}) imply that
\begin{equation}\label{PAR_15}
\Psi_n(z)=X_0\left[I+O(N^{-1}(1+|z|)^{-1}\right]\Psi_n^0(z),
\qquad z\in\C.
\end{equation}
The number $C_0$ is a free parameter. Let us take $C_0=1$.
From (\ref{PAR_5}) and (\ref{PAR_11})  we obtain that
\begin{equation}\label{PAR_16}
X_0=\frac{(R_n^0)^{1/4}}{\sqrt 2}\,\left(1+O(N^{-1})\right),
\end{equation}
hence
\begin{equation}\label{PAR_17}
\Psi_n(z)=\frac{(R_n^0)^{1/4}}{\sqrt 2}\,\Psi_n^0(z)\left[I+O(N^{-1})\right],
\qquad z\in\C.
\end{equation}
This gives the large $N$ asymptotics of $\Psi_n(z)$ under scaling (\ref{DSL_9a}),
as well as the asymptotics of the correlation functions. In particular, the asymptotics
near the origin is described as follows.

\begin {theo} {\rm (See [BI2])}. \label{T_2}
Let $\vec\Phi_0(z;y)=
\begin{pmatrix} \Phi^1(z;y) \\ \Phi^2(z;y) \end{pmatrix} $ be the solution for $n=0$
to system (\ref{DSL_18}), with the asymptotics at infinity as in  (\ref{DSL_20}). 
Then the following double scaling limit holds:       
\begin{equation}\label{PAR_18}
\begin{aligned}
\lim_{N\to\infty} \frac{1}{\left(cN^{1/3}\right)^{m-1}}
&K_{Nm}\left(\frac{u_1}{cN^{1/3}},\dots,\frac{u_m}{cN^{1/3}};
t_c+c_0yN^{-2/3}\right)\\
&=\det\left(Q_c(u_i,u_j;y)\right)_{i,j=1,\dots,m},
\end{aligned}
\end{equation}
where $c=\z'(0)>0$, and
\begin{equation}\label{PAR_119}
Q_c(u,v;y)=\frac{\Phi^1(u;y)\Phi^2(v;y)
-\Phi^1(v;y)\Phi^2(u;y)}{\pi(u-v)}\,.
\end{equation}
\end{theo}

Let us mention here some further developments of Theorems \ref{T_1}, \ref{T_2}.
They are extended to a general interaction $V(M)$ in the paper [CK] of Claeys and Kuijlaars.
The double scaling limit of the random matrix ensemble of the form 
$Z_N^{-1}|\det M|^{2\alpha}e^{-N\Tr V(M)}dM$, where $\al>-\frac{1}{2}\,$, is considered
in the papers [CKV] of Claeys, Kuijlaars, and Vahlessen, and  [IKO]
of Its, Kuijlaars, and \"Ostensson. In this case the double scaling limit 
is described in terms of a critical solution to the general Painlev\'e II equation
$q''=sq+2q^3-\al$. The papers, [CK], [CKV], and [IKO] use the RH approach and the Deift-Zhou
steepest descent method, discussed in Lecture 2 above. The double scaling
limit of higher order and Painlev\'e II hierachies is studied in the papers
[PeS], [BE], and others. There are many physical papers
on the double scaling limit related to the Painlev\'e I equation, see e.g., [BrK], [DS], [GM],
[DGZ], [Wit], and others. A rigorous RH approach to the Painlev\'e I double
scaling limit is initiated in the paper [FIK] of Fokas, Its, and Kitaev.
It is continued in the recent paper of Duits and Kuijlaars [DuK], who develop  
the RH approach and the Deift-Zhou
steepest descent method to orthogonal polynomials on contours in complex plane with the
exponential quartic weight, $e^{-N(\frac{z^2}{2}+\frac{t z^4}{4})}$, where $t<0$. Their results
cover both the one-cut case  $-\frac{1}{12}<t<0$ and the 
Painlev\'e I double scaling limit at $t=-\frac{1}{12}\,$.

\vskip 5mm

{\Large  Lecture 4. Large N asymptotics of the partition function
of random matrix models}

\vskip 5mm

\section{Partition function}

The central object of our analysis is the 
partition function of a random matrix model,
\begin{equation}\label{pf_1}
\begin{aligned}
Z_{N} &= \int_{-\infty}^{\infty}\ldots\int_{-\infty}^{\infty}
\prod_{1\leq j< k \leq N}(z_{j}-z_{k})^{2} e^{-N\sum_{j=1}^{N}V(z_{j})}dz_{1}\ldots dz_{N} 
= N!\prod_{n=0}^{N-1}h_{n},
\end{aligned}
\end{equation}
where $V(z)$ is a polynomial,
\begin{equation}\label{pf_2}
V(z) = \sum_{j=1}^{2d}v_jz^{j}, \quad v_{2d} >0,
\end{equation}
and $h_{n}$ are the normalization constants of
the orthogonal polynomials on the line with respect
to the weight $e^{-NV(z)}$,
\begin{equation}\label{pf_3}
\int_{-\infty}^{\infty}P_{n}(z)P_{m}(z)e^{-NV(z)}dz
=h_{n}\delta_{nm},
\end{equation}
where $P_{n}(z) = z^{n} + \ldots$
We will be interested in the asymptotic expansion of the free
energy, 
\begin{equation}\label{pf_4}
F_N=-\frac{1}{N^2}\ln Z_N,
\end{equation}
as $N\to\infty$. 

Our approach will be based on the deformation $\tau_t$
of $V(z)$ to $z^2$,
\begin{equation}\label{pf_5}
\tau_t:\; V(z)\to V(z;t)=(1-t^{-1})z^2+V(t^{-1/2}z),
\end{equation}
$1\le t<\infty,$ so that
\begin{equation}\label{pf_6}
\tau_1 V(z)=V(z),\qquad \tau_{\infty}V(z)=z^2,
\end{equation}
Observe that
\begin{equation}\label{pf_6a}
\tau_t\tau_s=\tau_{ts}.
\end{equation}
We start with the following proposition, which describe the deformation
equations for $h_n$ and the recurrent coefficients of orthogonal polynomials
under the change of the coefficients of $V(z)$.

\begin{prop} \label{Prop1}
We have that 
\begin{equation}\label{pf_7a}
\begin{aligned}
\frac{1}{N}\,\frac{\partial \ln h_{n}}{\partial  v_k}
&= -[Q^k]_{nn},\\
\frac{1}{N}\,\frac{\partial\ga_{n}}{\partial v_k}
&= \frac{\ga_n}{2}\left([Q^k]_{n-1,n-1}-[Q^k]_{nn}\right),\\
\frac{1}{N}\,\frac{\partial\be_{n}}{\partial  v_k}
&= \ga_n[Q^k]_{n,n-1}-\ga_{n+1}[Q^k]_{n+1,n},
\end{aligned}
\end{equation}
where $Q$ is the Jacobi matrix defined in (\ref{recu_6}).
\end{prop}

The proof of Proposition \ref{Prop1} is given in [BI3]. It uses some results of the works
of Eynard and Bertola, Eynard, Harnad [BEH]. For even $V$, it was proved
earlier by Fokas, Its, Kitaev [FIK].

We will be especially interested in the derivatives with respect to $v_2$.
For $k=2$, Proposition \ref{Prop1} gives that
\begin{equation}\label{pf_7}
\begin{aligned}
\frac{1}{N}\,\frac{\partial\ln h_{n}}{\partial  v_2}
&= -\ga_n^2-\be_n^2-\ga_{n+1}^2,\\
\frac{1}{N}\,\frac{\partial\ga_{n}}{\partial  v_2}
&= \frac{\ga_n}{2}\left(\ga_{n-1}^2+\be_{n-1}^2-\ga_{n+1}^2-\be_n^2\right),\\
\frac{1}{N}\,\frac{\partial\be_{n}}{\partial  v_2}
&= \ga_n^2\be_{n-1}+\ga_n^2\be_n
-\ga_{n+1}^2\be_n-\ga_{n+1}^2\be_{n+1}.
\end{aligned}
\end{equation}
Next we describe the $v_2$-deformation of $Z_N$.
 
\begin{prop}\label{Prop2} {\rm (See [BI3].)} We have the following relation: 
\begin{equation}\label{pf_8}
\begin{aligned}
\frac{1}{N^2}\,\frac{\partial^2\ln Z_N}{\partial  v_2^2}
= \ga_N^2(\ga_{N-1}^2+\ga_{N+1}^2
+\be_N^2+2\be_N\be_{N-1}+\be_{N-1}^2).
\end{aligned}
\end{equation}
\end{prop}

Observe that the equation is local in $N$. 
For $V(z)=v_2z^2+v_4z^4$, Proposition \ref{Prop2}
was obtained earlier by Its, Fokas, Kitaev [FIK0].
Proposition \ref{Prop2} can be applied to
deformation (\ref{pf_5}). Let $\ga_n(\tau)$, $\be_n(\tau)$ be the
recurrence coefficients   
for orthogonal polynomials with respect to the weight
$e^{-NV(z;\tau)}$, and let
\begin{equation}\label{pf_9}
Z_{N}^{\rm Gauss} = \int_{-\infty}^{\infty}\ldots\int_{-\infty}^{\infty}
\prod_{1\leq j< k \leq N}(z_{j}-z_{k})^{2} e^{-N\sum_{j=1}^{N}z_{j}^2}dz_{1}\ldots dz_{N} 
\end{equation}
be the partition function for the Gaussian ensemble. It can be evaluated explicitly, and
the corresponding free energy has the form,
\begin{equation}\label{pf_10}
F_N^{\rm Gauss}=-\frac{1}{N^2}\,\ln
\left(\frac{(2\pi)^{N/2}}{(2N)^{N^2/2}} 
\prod_{n=1}^N n!\right)
\end{equation}
By integrating twice formula (\ref{pf_8}), we obtain the following result:

\begin{prop}\label{F_N}
\begin{equation}\label{pf_11}
\begin{aligned}
F_N(t)&=F_N^{\rm Gauss}+\int_t^\infty \frac{t-\tau}{\tau^2}
\left\{\ga_N^2(\tau)\left[\ga_{N-1}^2(\tau)\right.\right.
+\ga_{N+1}^2(\tau)
+\be_N^2(\tau)+2\be_N(\tau)\be_{N-1}(\tau)\\
&\left.\left.
+\be_{N-1}^2(\tau)\right]-\frac{1}{2}\right\}d\tau.
\end{aligned}
\end{equation}
\end{prop}

\section{Analyticity of the free energy for regular $V$}

The basic question of statistical physics is the existence of the free energy
in the thermodynamic limit and the analyticity of the limiting free energy
with respect to various parameters. The values of the parameters at which the
free energy is not analytic are the critical points. When applied to the ``gas'' of eigenvalues,
this question refers to the existence of the limit,
\begin{equation}\label{an_1}
F=\lim_{N\to\infty}F_N,
\end{equation}
and to the analyticity of $F$ with respect to the coefficients $v_j$ of the polynomial $V$.
The existence of limit (\ref{an_1}) is proven under general conditions on $V$, not
only polynomials, see the work of Johansson [Joh] and references
therein. In fact,  $F$ is the energy $E_V$ of minimization problem
(\ref{equi7}), (\ref{equi8}), so that
\begin{equation}\label{an_2}
F=I_V(\nu_V),
\end{equation}
where $\nu_V$ is the equilibrium measure. The following theorem establishes
the analyticity of $F$ for regular $V$. We call $V$ regular, if the corresponding
equilibrium measure $\nu_V$ is regular, as defined in (\ref{equi17}), (\ref{equi18}).
We call $V$, $q$-cut regular, it the measure $\nu_V$ is regular and its support 
consists of $q$ intervals.

\begin{theo} \label{analytic_1}
{\rm (See [BI3].)}
Suppose that $V(z)$ is a $q$-cut
regular polynomial of degree $2d$. Then for any $p\le 2d$, there exists $t_1>0$ such that for any
$t\in[-t_1,t_1]$, 
\begin{enumerate}
\item
the polynomial
 $V(z)+tz^p$ is $q$-cut regular.
\item
The end-points  of the support intervals, $[a_i(t),b_i(t)]$,
    $i=1,\dots,q$, of the equilibrium measure for $[V(z)+tz^p]$ are analytic in $t$. 
\item The free energy $F(t)$ is analytic in $t$.
\end{enumerate}   
\end{theo}

\begin{proof}
Consider for $j=0,\dots,q,$ the quantities
\begin{equation}\label{an_3}
T_j(a,b;t)=\frac{1}{2\pi i}\oint_{\Gamma}
  \frac{V'(z;t)z^j}{\sqrt{R(z)}}\, dz,
\qquad R(z)=\prod_{i=1}^q (z-a_i)(z-b_i),
\end{equation}
where  $\Gamma$ is a contour around $[a_1,b_q]$.
Consider also for $k=1,\dots,q-1,$ the quantities
\begin{equation}\label{an_4}
N_k(a,b;t)=\frac{1}{2\pi i}\oint_{\Gamma_k}h(z;t)
 \sqrt{R(z)}\, dz,
\end{equation}
where $\Gamma_k$ is a contour around $[b_k,a_{k+1}]$.
Then, as shown by Kuijlaars and McLaughlin in \cite{KuM1}, the Jacobian of
the map $\{ [a_i,b_i],\;i=1,\ldots,q\}\to\{T_j,N_k\}$ is nonzero 
at the solution, $\{[a_i(t),b_i(t)],\;i=1,\ldots,q\}$, to the equations
$\{T_j=2\delta_{jq}$, $N_k=0\}$. By the implicit function theorem, this 
implies the analyticity of $[a_i(t),b_i(t)]$ in $t$.
\end{proof}

\section{ Topological expansion}

In the paper [EM],
Ercolani and McLaughlin proves topological expansion (\ref{intro13})
for polynomial $V$ of form (\ref{intro11}), with small values of
the parameters $t_j$. Their proof is based on a construction of the asymptotic 
large $N$ expansion of the parametrix for the RHP. Another  proof
of topological expansion (\ref{intro13}) is given in the paper [BI3].
Here we outline the main steps of the proof of [BI3].
We start with a preliminary result, which follows easily from 
the results of [DKMVZ2].

\begin{prop}  
Suppose $V(x)$ is one-cut regular.
Then there exists $\ep>0$ such 
that for all $n$ in the interval
\[
1-\ep\le \frac{n}{N}\le 1+\ep,
\]
 the recurrence coefficients  admit the uniform asymptotic
representation,   
\begin{equation}\label{asym_1}
\ga_n= \ga\left(\frac{n}{N}\right)+O(N^{-1}),
\quad
\be_n= \be\left(\frac{n}{N}\right)+O(N^{-1}).
\end{equation}
The functions $\ga(s),\,\be(s)$ are expressed as
\begin{equation}\label{asym_2}
\ga(s)=\frac{b(s)-a(s)}{4},\quad \be(s)=\frac{a(s)+b(s)}{2},
\end{equation}
where $[a(s),b(s)]$ is the
support of the equilibrium measure for the polynomial $s^{-1}V(x)$.
\end{prop}

The next theorem gives an asymptotic expansion for the recurrence coefficients.

\begin{theo}  Suppose that $V(x)$ is a one-cut regular
polynomial.
Then there exists $\ep>0$ such that for all $n$ in the interval
\[
1-\ep\le \frac{n}{N}\le 1+\ep,
\]
 the recurrence coefficients  
admit the following
uniform asymptotic expansion as $N\to\infty$:
\begin{equation}\label{asym_3}
\begin{aligned}
\ga_n&\sim \ga\left(\frac{n}{N}\right)+\sum_{k=1}^\infty N^{-2k}
f_{2k}\left(\frac{n}{N}\right)\,,\\ 
\be_n&\sim \be\left(\frac{n+\frac{1}{2}}{N}\right)+\sum_{k=1}^\infty N^{-2k}
g_{2k}\left(\frac{n+\frac{1}{2}}{N}\right)\,, 
\end{aligned}
\end{equation}
where  $f_{2k}(s)$, $g_{2k}(s)$, $k\ge 1$, are analytic functions 
on $[1-\ep,1+\ep]$.
\end{theo}

{\it Sketch of the proof of the theorem.}
The Riemann-Hilbert approach gives an asymptotic expansion 
in powers of $N^{-1}$. We want to show that the odd
coefficients vanish. To prove this,
we use induction in the number of the coefficient
and the invariance of the string equations,
\begin{equation}\label{asym_4}
\ga_n [V'(Q)]_{n,n-1}=\frac{n}{N}\,,\quad [V'(Q)]_{nn}=0,
\end{equation}
with respect to the change of variables
\begin{equation}\label{asym_5}
\{\ga_j\to\ga_{2n-j},\; \be_j\to \be_{2n-j-1},\;j=0,1,
2,\ldots\}.
\end{equation}
This gives the cancellation of the odd coefficients,
which proves the theorem.

The main condition, under which the topological expansion is proved
in [BI3], is the following:

{\bf Hypothesis R.} For all $t\ge 1$ the polynomial $\tau_t V(z)$ 
is one-cut regular.

\begin{theo} {\rm (See [BI3].)}
If a polynomial $V(z)$ satisfies Hypothesis R, then its free energy
admits the asymptotic expansion,
\begin{equation}\label{asym_6}
F_N-F_N^{\rm Gauss}\sim F+N^{-2}F^{(2)}+N^{-4}F^{(4)}+\ldots,
\end{equation}
The leading term of the asymptotic expansion is:
\begin{equation}\label{asym_7}
F=\int_1^\infty \frac{1-\tau}{\tau^2}
\left[2\ga^4(\tau)+4\ga^2(\tau)\be^2(\tau)-\frac{1}{2}\right]
d\tau,
\end{equation}
where 
\begin{equation}\label{asym_8}
\ga(\tau)=\frac{b(\tau)-a(\tau)}{4},\qquad
\be(\tau)=\frac{a(\tau)+b(\tau)}{2},
\end{equation}
and $[a(\tau),b(\tau)]$ is the support of the equilibrium measure
for the polynomial $V(z;\tau)$.
\end{theo}

To prove this theorem, we substitute asymptotic expansions (\ref{asym_3})
into  formula ((\ref{pf_11}) and check that the odd powers of $N^{-1}$
cancel out. See [BI3].

\section{One-sided analyticity at a critical point} 

According
to the definition, see (\ref{equi17})--(\ref{equi18}), the equilibrium measure
is singular in the following three cases:
\begin{enumerate}
\item
$h(c)=0$ where $c\in (a_j,b_j)$, for some $1\le j\le q$,
\item
$h(a_j)=0$ or $h(b_j)=0$, for some $1\le j\le q$,
\item for some $c\not\in J$,
\begin{equation}\label{asym_9}
2\int\log|c-y|d\nu_V(y)-V(c)= l.
\end{equation}
\end{enumerate}
More complicated cases appear as a combination of these three basic ones.
The cases (1) and (3) are generic for a
one-parameter family of polynomials. The case (1) means a split of the support interval
$(a_j,b_j)$ into two intervals. A typical illustration of this case is given in
Fig.2.  Case (3) means a birth of a new support interval at the point $c$.
Case (2) is generic for a two-parameter family of polynomials.

  Introduce the following hypothesis.

{\bf Hypothesis S${}_q$.} $V(z;t)$, $t\in[0,t_0]$, $t_0>0$, is a one-parameter family
of real polynomials  such that
\begin{enumerate}
\item[(i)] $V(z;t)$ is $q$-cut regular for $0< t\le t_0$,
\item[(ii)] $V(z;0)$ is $q$-cut singular and $h(a_i)\not=0$,
$h(b_i)\not=0$, $i=1,\dots,q$, where $\cup_{i=1}^q[a_i,b_i]$
is the support of the equilibrium measure
for $V(z;0)$.
\end{enumerate}

We have the following result, see [BI3].

\begin{theo}\label{analytic_2}
Suppose $V(z;t)$ satisfies Hypothesis S${}_q$. Then the end-points
$a_i(t),b_i(t)$ 
of the equilibrium measure for $V(z;t)$, the density function, and
the free energy
are analytic, as functions of $t$, at $t=0$.
\end{theo}

The proof of this theorem is an extension of the proof of Theorem
\ref{analytic_1}, and it is also based on the work of Kuijlaars
and McLaughlin \cite{KuM1}. Theorem shows that
the free energy can be analytically continued through the critical point, $t=0$,
if conditions (i) and (ii) are fulfilled. If $h(a_j)=0$ or $h(b_j)=0$,
then the free energy is expected to have an algebraic singularity
at $t=0$, but this problem has not been studied yet in details.
As concerns the split of the support interval,
this case was studied for a nonsymmetric quartic polynomial in the
paper of Bleher and Eynard [BE]. To describe the result of [BE], consider the singular
quartic polynomial,
\begin{equation}\label{asym_10}
V'_c(x) =\frac{1}{T_c}\,( x^3 - 4 c_1 x^2 + 2 c_2 x + 8 c_1), 
 \quad T_c=1+4c_1^2\,;\qquad V_c(0)=0\,,
\end{equation}
where we denote
\begin{equation}\label{asym_11}
c_k = \cos{k\pi\epsilon} \,.
\end{equation}
It corresponds to the critical density
\begin{equation}\label{asym_12}
\rho_c(x)=\frac{1}{2\pi T_c}(x-2c_1)^2\sqrt{4-x^2}.
\end{equation}
Observe that $0<\epsilon<1$ is a parameter of the problem
 which determines the location of the critical point, 
\begin{equation}\label{asym_13}
-2<2c_1= 2\cos\pi\epsilon<2\,.
\end{equation}
We include $V_c$ into the one-parameter family of quartic polynomials,
$\{V(x;T),\;T>0\}$, where
\begin{equation}\label{asym_14}
V'(x;T) =\frac{1}{T}\,( x^3 - 4 c_1 x^2 + 2 c_2 x + 8 c_1);\qquad V(0;T)=0\,.
\end{equation}
Let $F(T)$ be the free energy corresponding to $V(x;T)$.

\begin{theo}\label{analytic_3}
The function $F(T)$ can be analytically continued through $T=T_c$
both from $T\ge T_c$ and from $T\le T_c$. At $T=T_c$, $F(T)$
is continuous, as well as its first two derivatives, but the
third derivative jumps.
\end{theo}

This corresponds to the third order phase transition. Earlier
the third order phase transition was observed in a circular
ensemble of random matrices by Gross and Witten [GW].

\section{Double Scaling Limit of the Free Energy}

Consider an even quartic critical potential,
\begin{equation}\label{dsfe_1}
V(z) = \frac{1}{4}\,z^4-z^2,
\end{equation}
and its deformation,
\begin{equation}\label{dsfe_2}
\tau_{t}V(z) \equiv V(z;t) 
=\frac{1}{4t^2}\,z^4+\left(1-\frac{2}{t}\right)z^2 \,.
\end{equation}
Introduce the scaling,
\begin{equation}\label{dsfe_3}
t=1+N^{-2/3}2^{-2/3}x\,.
\end{equation}
The Tracy-Widom distribution function defined
by the formula
\begin{equation}\label{dsfe_4}
F_{\rm TW}(x) = \exp \left [ \int_{x}^{\infty}(x-y)u^{2}(y)dy\right ],
\end{equation}
where $u(y)$ is the Hastings-McLeod solution to the Painlev\'e II,
see (\ref{DSL_14}), (\ref{DSL_15}).

\begin{theo}\label{analytic_3a}{\rm (See [BI3].)}
For every $\ep > 0$,
\begin{equation}\label{dsfe_5}
F_{N}(t) - F_{N}^{\rm Gauss} = F^{\rm reg}_{N}(t) 
- N^{-2}\log F_{\rm TW}\Big((t-1)2^{2/3}N^{2/3}\Big)
+ O(N^{-7/3 + \ep}), 
\end{equation}
as $N \to \infty$ and $|(t-1)N^{2/3}| < C$, where 
\begin{equation}\label{dsfe_6}
F^{\rm reg}_{N}(t) \equiv F(t) + N^{-2} F^{(2)}(t)
\end{equation}
is the sum of the first two terms of the topological expansion.
\end{theo}

\vskip 5mm

{\Large  Lecture 5. Random matrix model with external source}

\vskip 5mm

\section{Random matrix model with external source
and multiple orthogonal polynomials}

We consider the Hermitian random matrix ensemble with an external source,
\begin{equation}\label{es_1}
d\mu_n(M)=\frac{1}{Z_n}\,e^{-n\Tr(V(M)-AM)}dM,
\end{equation}
where
\begin{equation}\label{es_2}
Z_n=\int \,e^{-n\Tr(V(M)-AM)}dM,
\end{equation}
and $A$ is a fixed Hermitian matrix. Without loss of generality
we may assume that $A$ is diagonal,
\begin{equation}\label{es_3}
A=\diag(a_1,\dots,a_n),\qquad a_1\le \ldots\le a_n.
\end{equation}
Define the monic polynomial
\begin{equation}\label{es_3a}
 P_n(z) =\int \det(z-M)\,d\mu_n(M). 
\end{equation}

\begin{prop} \label{prop21}
\begin{enumerate}
\item[\rm (a)] There is a constant $\tilde{Z}_n$ such that
\begin{equation}\label{es_4}
P_n(z) = \frac{1}{\tilde{Z}_n}
    \int \prod_{j=1}^n (z-\lambda_j)
    \prod_{j=1}^n e^{-(V(\lambda_j) - a_j \lambda_j)} \Delta(\lambda) d\lambda,
\end{equation}
where
\begin{equation}\label{es_5}
\Delta(\lambda) = \prod_{i>j} (\lambda_i-\lambda_j) 
\end{equation}
and $d \lambda = d\lambda_1 d\lambda_2 \cdots d\lambda_n$.
\item[\rm (b)] Let
\begin{equation}\label{es_6}
 m_{jk} = \int_{-\infty}^{\infty} x^k e^{-(V(x) - a_j x)} d x. 
\end{equation}
Then we have the determinantal formula
\begin{equation}\label{es_7}
    P_n(z) = \frac{1}{\tilde{Z}_n}
    \begin{vmatrix}
    m_{10} & m_{11} & \cdots & m_{1n} \\
    \vdots & \vdots & \ddots & \vdots \\
    m_{n0} & m_{n1} & \cdots & m_{nn} \\
    1 & z &  \cdots &  z^n
    \end{vmatrix}.
\end{equation}
\item[\rm (c)] For $j=1, \ldots, n$,
\begin{equation}\label{es_8}
    \int_{-\infty}^{\infty} P_n(x) e^{-(V(x)-a_jx)} dx = 0,
\end{equation}
and these equations uniquely determine the monic polynomial $P_n$.
\end{enumerate}
\end{prop}

Proposition \ref{prop21} can be extended to the case of
multiple $a_j$'s as follows.

\begin{prop} \label{prop22}
Suppose $A$ has distinct eigenvalues $a_i$, $i=1,\ldots,p$, with
respective multiplicities $n_i$, so that $n_1 + \cdots + n_p = n$.
Let $n^{(i)} = n_1 + \cdots + n_i$ and $n^{(0)} = 0$. Define
\begin{equation}\label{es_9}
w_j(x) = x^{d_j-1} e^{-(V(x)-a_ix)}, \qquad j=1,\ldots, n, 
\end{equation}
where $i = i_j$ is such that $n^{(i-1)} < j \leq n^{(i)}$ and
$d_j = j-n^{(i-1)}$. Then the following hold.
\begin{enumerate}
\item[\rm (a)] There is a constant $\tilde{Z}_n > 0$ such that
\begin{equation}\label{es_10}
P_n(z) = \frac{1}{\tilde{Z}_n}
    \int \prod_{j=1}^n (z-\lambda_j)
    \prod_{j=1}^n w_j(\lambda_j)  \Delta(\lambda) d\lambda.
\end{equation}
\item[\rm (b)] Let
\begin{equation}\label{es_11}
m_{jk} = \int_{-\infty}^{\infty} x^k w_j(x) dx. 
\end{equation}
Then we have the determinantal formula
\begin{equation}\label{es_12}
    P_n(z) = \frac{1}{\tilde{Z}_n}
    \begin{vmatrix}
    m_{10} & m_{11} & \cdots & m_{1n} \\
    \vdots & \vdots & \ddots & \vdots \\
    m_{n0} & m_{n1} & \cdots & m_{nn} \\
    1 & z &  \cdots &  z^n
    \end{vmatrix}.
\end{equation}
\item[\rm (c)] For $i=1, \ldots, p$,
\begin{equation}\label{es_13}
    \int_{-\infty}^{\infty} P_n(x) x^j e^{-(V(x)-a_ix)} dx = 0,
    \qquad j=0, \ldots, n_i-1,
\end{equation}
and these equations uniquely determine the monic polynomial $P_n$.
\end{enumerate}
\end{prop}

The relations (\ref{es_13}) can be viewed as
multiple orthogonality conditions for the polynomial $P_n$.
There are $p$ weights $e^{-(V(x)-a_jx)}$, $j=1,\ldots, p$,
and for each weight there are a number of orthogonality
conditions, so that the total number of them is $n$.
This point of view is especially useful in the case when $A$
has only a small number of distinct eigenvalues.

\subsection{Determinantal Formula for Eigenvalue Correlation Functions}

P. Zinn-Justin proved in [ZJ1], [ZJ2] a determinantal formula for the
eigenvalue correlation functions of the random matrix model with 
external source. We relate the determinantal
kernel to the multiple orthogonal polynomials.

Let $\Sigma_n$ be the collection of functions
\begin{equation} \label{es_14}
    \Sigma_n := \{ x^{j} e^{a_i x} \mid i=1, \ldots, p,
 \, j=0, \ldots, n_i-1 \}.
\end{equation}
We start with a lemma.
\begin{lem} \label{lemma31}
There exists a unique function $Q_{n-1}$ in the linear span of $\Sigma_n$ such
that
\begin{equation} \label{charQn1}
    \int_{-\infty}^{\infty} x^j Q_{n-1}(x) e^{-V(x)} dx = 0,
\end{equation}
$ j =0, \ldots, n-2,$
and
\begin{equation} \label{charQn2}
    \int_{-\infty}^{\infty} x^{n-1} Q_{n-1}(x) e^{-V(x)}dx = 1.
\end{equation}
\end{lem}

Consider $P_0,\dots,P_n$, a sequence of multiple orthogonal
polynomials such that $\deg P_k=k$, with an increasing sequence
of the multiplicity vectors, so that $k_i\le l_i$,
$i=1,\dots,p$, when $k\le l$.  
Consider the biorthogonal system of functions, $\{ Q_k(x),
\; k=0,\dots,n-1\}$, 
\begin{equation} \label{biorthogonal}
    \int_{-\infty}^{\infty} P_j(x) Q_k(x) e^{-V(x)} dx = \delta_{jk},
\end{equation}
for $j,k=0, \ldots, n-1$.
Define the kernel
\begin{equation} \label{defKn}
    K_n(x,y) = e^{- \frac{1}{2} (V(x)+ V(y))}
    \sum_{k=0}^{n-1} P_k(x) Q_k(y).
\end{equation}

\begin{theo}
The $m$-point correlation function of eigenvalues has the 
determinantal form
\begin{equation} \label{Rm}
    R_m(\lambda_1, \ldots, \lambda_m) =
    \det (K_n(\lambda_j,\lambda_k))_{1\leq j,k\leq m}
\end{equation}
\end{theo}

\subsection{Christoffel-Darboux formula}

We will assume that there are only two distinct eigenvalues,
$a_1=a$ and $a_2=-a$, with multiplicities $n_1$ and $n_2$, respectively.
We redenote $P_n$ by $P_{n_1,n_2}$.
Set
\begin{equation} \label{h-numbers}
    h_{n_1,n_2}^{(j)} =
    \int_{-\infty}^{\infty} P_{n_1,n_2}(x) x^{n_j} w_j(x) dx, 
\end{equation}
$ j=1,2$, which are non-zero numbers.

\begin{theo} \label{theorem41}
With the notation introduced above, 
\begin{equation} \label{CDformula} 
\begin{aligned}
   (x-y) e^{\frac{1}{2} (V(x) + V(y))} K_n(x,y) &= 
      P_{n_1,n_2}(x) Q_{n_1,n_2}(y)  - \frac{h_{n_1,n_2}^{(1)}}{h_{n_1-1,n_2}^{(1)}}
 P_{n_1-1,n_2}(x) Q_{n_1+1,n_2}(y) \\
  &- \frac{h_{n_1,n_2}^{(2)}}{h_{n_1,n_2-1}^{(2)}}
 P_{n_1,n_2-1}(x) Q_{n_1,n_2+1}(y)
\end{aligned}
\end{equation}
\end{theo}

\subsection{ Riemann-Hilbert problem}

The Rieman-Hilbert problem is to find
$Y : \mathbb C \setminus \mathbb R \to \mathbb C^{3\times 3}$
such that
\begin{itemize}
\item $Y$ is analytic on $\mathbb C \setminus \mathbb R$,
\item for $x \in \mathbb R$, we have
\begin{equation} \label{jumpY}
    Y_+(x)  = Y_-(x) \begin{pmatrix} 1 & w_1(x) & w_2(x) \\
    0 & 1 & 0 \\ 0 & 0 & 1 \end{pmatrix}
\end{equation}
where $Y_+(x)$ ($Y_-(x)$) denotes the limit of $Y(z)$ as $z \to x$ from
the upper (lower) half-plane,
\item as $z \to \infty$, we have
\begin{equation} \label{asympY}
    Y(z) = \left(I + O\left(\frac{1}{z}\right)\right)
    \begin{pmatrix} z^{n} & 0 & 0 \\
    0 & z^{-n_1} & 0 \\ 0 & 0 & z^{-n_2} \end{pmatrix}
\end{equation}
where $I$ denotes the $3\times 3$ identity matrix.
\end{itemize}

\begin{prop}
There exists a unique solution to the RH problem,
\begin{equation} \label{rmef_1}
Y = 
\begin{pmatrix}
    P_{n_1,n_2} & C(P_{n_1,n_2} w_1) & C(P_{n_1,n_2} w_2) \\
    c_1 P_{n_1-1,n_2} & c_1 C(P_{n_1-1,n_2}w_1) & c_1 C(P_{n_1-1,n_2} w_2) \\
    c_2 P_{n_1,n_2-1} & c_2 C(P_{n_1,n_2-1}w_1) & c_2 C(P_{n_1,n_2-1} w_2)
    \end{pmatrix}
\end{equation}
with constants
\begin{equation} \label{rmef_2}
c_1 = -2\pi i \left( h_{n_1-1,n_2}^{(1)} \right)^{-1}, \qquad
c_2 = -2\pi i \left(h_{n_1,n_2-1}^{(2)}\right)^{-1},
\end{equation}
 and where
$Cf$ denotes the Cauchy transform of $f$, i.e.,
\begin{equation} \label{rmef_3}
Cf(z) = \frac{1}{2\pi i} \int_{\mathbb R} \frac{f(s)}{s-z} ds. 
\end{equation}
\end{prop}

The Christoffel-Darboux formula, (\ref{CDformula}),
can be expressed in terms of the solution of RH Problem:

\begin{equation} \label{rmef_4}
   K_n(x,y) = e^{-\frac{1}{2} (V(x) + V(y))}
     \frac{e^{a_1y} [Y^{-1}(y) Y(x)]_{21}
        + e^{a_2y} [Y^{-1}(y) Y(x)]_{31}
        }{2\pi i(x-y)}\,.
\end{equation}

\subsection{Recurrence and differential equations} 

The recurrence and differential equations are
nicely formulated in terms of the function 
\begin{equation}\label{m13}
\Psi(z)=
\begin{pmatrix}
1 & 0 & 0 \\
0 & c_1^{-1} & 0 \\
0 & 0 & c_2^{-1}
\end{pmatrix} Y(z)
\begin{pmatrix}
w(z) & 0 & 0 \\
0 & e^{-Naz} & 0 \\
0 & 0 & e^{Naz}
\end{pmatrix},
\end{equation}
where
\begin{equation}\label{m13a}
w(z)=e^{-NV(z)}.
\end{equation}
The function $\Psi$ solves the following RH problem:
\begin{itemize}
\item $\Psi$ is analytic on $\C\setminus\R$,
\item for $x\in\R$, 
\begin{equation}\label{m14}
\Psi_+(x)=\Psi_-(x)
\begin{pmatrix}
1 & 1 & 1 \\
0 & 1 & 0 \\
0 & 0 & 1
\end{pmatrix},
\end{equation}
\item as $z\to\infty$, 
\begin{equation}\label{m15}
\begin{aligned}
{}&\Psi(z)\sim\left(I+\frac{\Psi^{(1)}_{n_1,n_2}}{z}+\ldots\right)
\\
&\times
\begin{pmatrix}
z^n w& 0 & 0 \\
0 & c_1^{-1}z^{-n_1}e^{-Naz} & 0 \\
0 & 0 & c_2^{-1}z^{-n_2}e^{Naz}
\end{pmatrix}.
\end{aligned}
\end{equation}
\end{itemize}
The recurrence equation for $\Psi$ has the form:
\begin{equation}\label{rec10}
\Psi_{n_1+1,n_2}(z)=U_{n_1,n_2}(z)\Psi_{n_1,n_2}(z),
\end{equation}
where
\begin{equation}\label{rec11}
U_{n_1,n_2}(z)=
\begin{pmatrix}
z-b_{n_1,n_2} & -c_{n_1,n_2} & -d_{n_1,n_2}  \\
1 & 0 & 0 \\
1 & 0 & e_{n_1,n_2} 
\end{pmatrix}
\end{equation}
and
\begin{equation}\label{rec12}
c_{n_1,n_2}
=\frac{h^{(1)}_{n_1,n_2}}{h^{(1)}_{n_1-1,n_2}}\not=0,
\quad
d_{n_1,n_2}
=\frac{h^{(2)}_{n_1,n_2}}{h^{(2)}_{n_1,n_2-1}}\not=0,
\quad
e_{n_1,n_2}=\frac{h^{(2)}_{n_1+1,n_2-1}}{h^{(2)}_{n_1,n_2-1}}\not=0.
\end{equation}
Respectively, the recurrence equations for the multiple orthogonal polynomials are
\begin{equation}\label{rec11a}
\begin{aligned}
{}&P_{n_1+1,n_2}(z)=(z-b_{n_1,n_2})P_{n_1,n_2}(z)
-c_{n_1,n_2}P_{n_1-1,n_2}(z)
-d_{n_1,n_2}P_{n_1,n_2-1}(z),
\end{aligned}
\end{equation}
and
\begin{equation}\label{rec12a}
P_{n_1+1,n_2-1}(z)=P_{n_1,n_2}(z)
+e_{n_1,n_2}P_{n_1,n_2-1}(z).
\end{equation}
The differential equation for $\Psi$ is 
\begin{equation}\label{de5}
\Psi_{n_1,n_2}'(z)=N A_{n_1,n_2}(z)\Psi_{n_1,n_2}(z),
\end{equation}
where
\begin{equation}\label{de4}
\begin{aligned}
A_{n_1,n_2}(z)&=-\left[\left(I+\frac{\Psi^{(1)}_{n_1,n_2}}{z}
+\ldots\right)
\begin{pmatrix}
V'(z) & 0 & 0 \\
0 & 0 & 0 \\
0 & 0 & 0 
\end{pmatrix}
\left(I+\frac{\Psi^{(1)}_{n_1,n_2}}{z}+\ldots\right)^{-1}\right]_{\rm pol}\\
&+\begin{pmatrix}
0 & 0 & 0 \\
0 & -a & 0 \\
0 & 0 & a 
\end{pmatrix},
\end{aligned}
\end{equation}
where $[f(z)]_{\rm pol}$ means the polynomial part of $f(z)$ at 
infinity.

For the Gaussian model,
\(
V(x)=\frac{x^2}{2}\,,
\)
the recurrence equation reduces to
\begin{equation}\label{Lax1}
\begin{aligned}
\Psi_{n_1+1,n_2}&=\begin{pmatrix}
z-a & -\frac{n_1}{n} 
& -\frac{n_2}{n}  \\
1 & 0 & 0 \\ 
1 & 0 & -2a
\end{pmatrix}\Psi_{n_1,n_2},
\end{aligned}
\end{equation}
where $n=n_1+n_2,$
and the
differential equation reads
\begin{equation}\label{Lax2}
\Psi_{n_1,n_2}'=n \begin{pmatrix}
-z
 & \frac{n_1}{n} & \frac{n_2}{n}  \\
{}-1 & -a & 0 \\
{}-1 & 0 & a 
\end{pmatrix}\Psi_{n_1,n_2}.
\end{equation}
In what follows, we will restrict ourselves to 
the case when $n$ is even and
\begin{equation}\label{n1}
n_1=n_2=\frac{n}{2}\,,
\end{equation}
so that
\begin{equation}\label{es_1a}
A=\diag(\underbrace{-a,\ldots,-a}_{\frac{n}{2}},\underbrace{a,\ldots,a}_{\frac{n}{2}}).
\end{equation}

\section{Gaussian matrix model with external source and 
non-intersecting Brownian bridges}

\begin{figure}[ht]
\scalebox{0.5}{\includegraphics{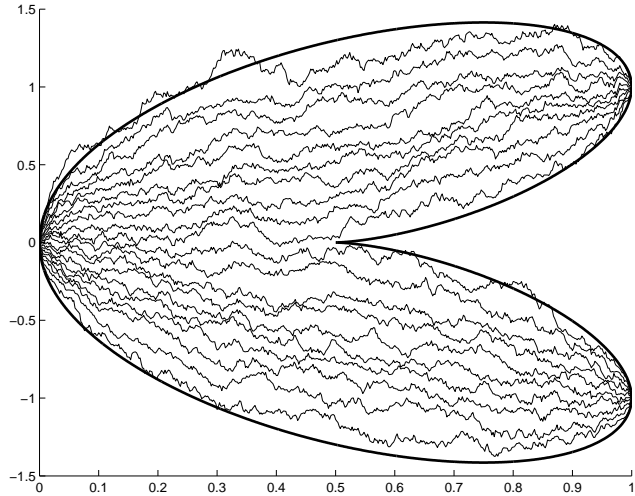}}
\caption{Non-intersecting Brownian paths that start at one point and end
at two points. At any intermediate time the positions of the paths are
distributed as the eigenvalues of a Gaussian random matrix ensemble with external
source. As their number increases the paths fill out a region whose boundary has a cusp.}
\label{figpaths}
\end{figure}

Consider $n$ independent Brownian motions
(Brownian bridges) $x_j(t)$, $j=1,\dots,n$,
on the line, starting at the origin
 at time $t=0$, half ending at $x=1$ and half at $x=-1$
 at time $t=1$, and conditioned not to
intersect for  $t \in (0,1)$.
Then at any time $t \in (0,1)$ the positions of $n$ non-intersecting
Brownian bridges  are
distributed as the scaled eigenvalues,
\[
\la_j=\frac{x_j}{\sqrt{t(1-t)}}\,,
\]
 of a Gaussian random matrix with
the external source
\[
a(t)=\sqrt{\frac{t}{1-t}}\,.
\]
Fig.\ref{figpaths} gives an illustration of the non-intersecting Brownian bridges.
See also the paper \cite{TW4} of Tracy and Widom on
nonintersecting Brownian excursions.

In the Gaussian model the value $a=1$ is critical, and we will discuss 
its large $n$ asymptotics in the three cases:
\begin{enumerate}
\item $a>1$, two cuts,
\item $a<1$, one cut,
\item $a=1$, double scaling limit.
\end{enumerate} 
In the picture of the non-intersecting Brownian bridges this transforms to a critical
time $t=\frac{1}{2}$, and there are two cuts for $t>\frac{1}{2}$, one cut for 
$t<\frac{1}{2}$, and the double scaling limit appears in a scaled
neighborhood of $t=\frac{1}{2}$. 

\section {Gaussian model with external source. Main results}

\begin{figure}
\scalebox{0.35}{\includegraphics{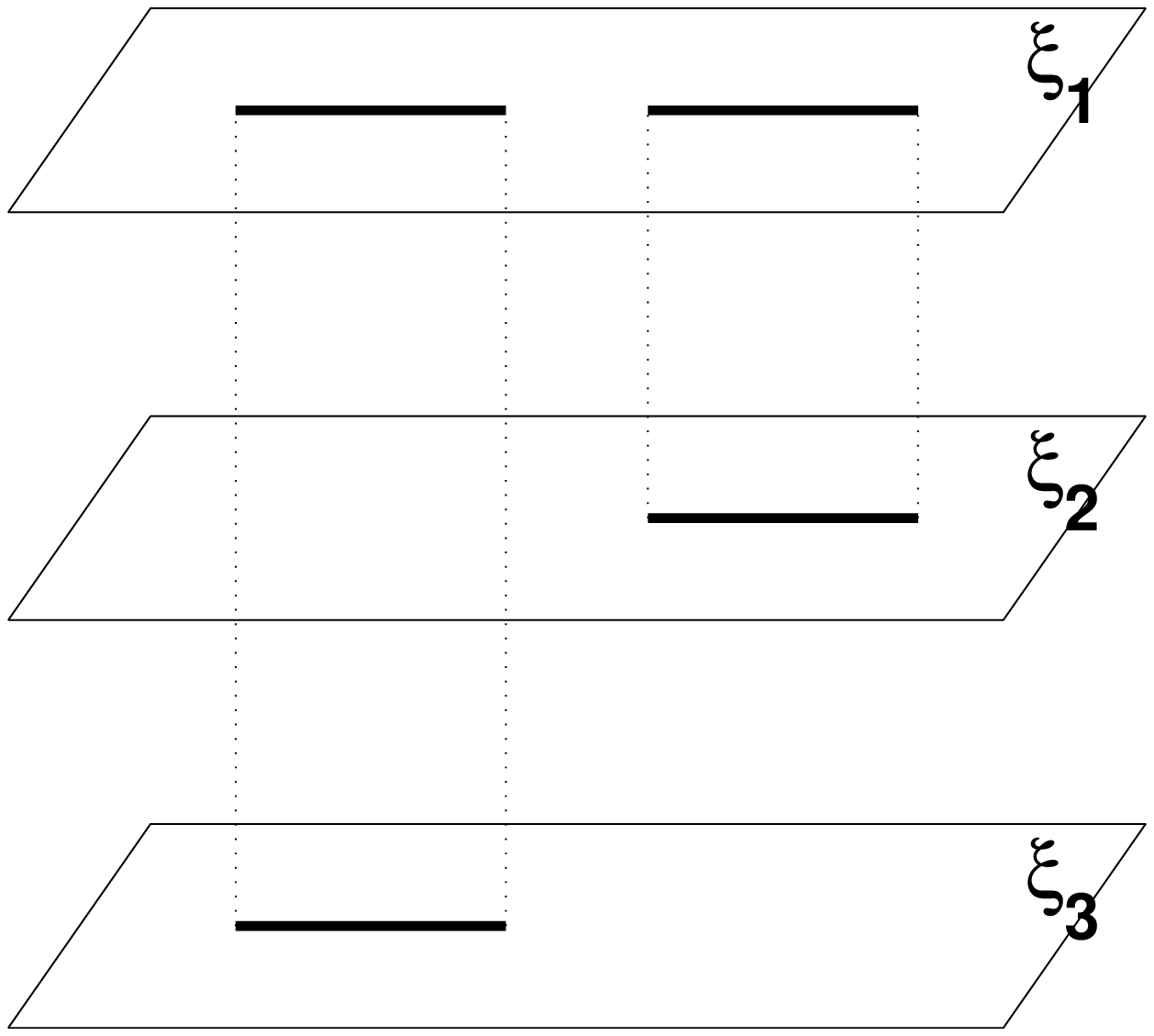}}
\scalebox{0.35}{\includegraphics{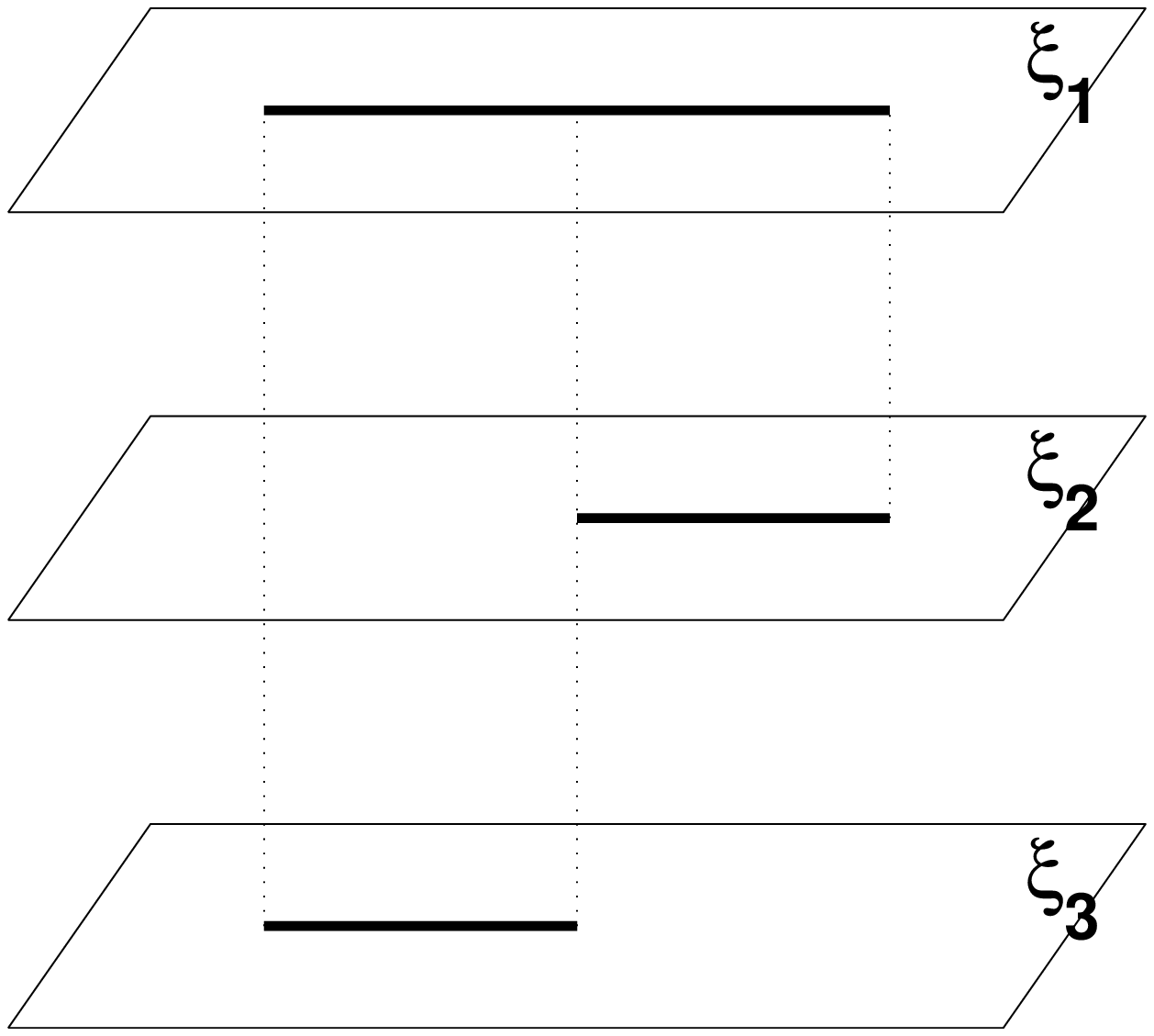}}
\scalebox{0.35}{\includegraphics{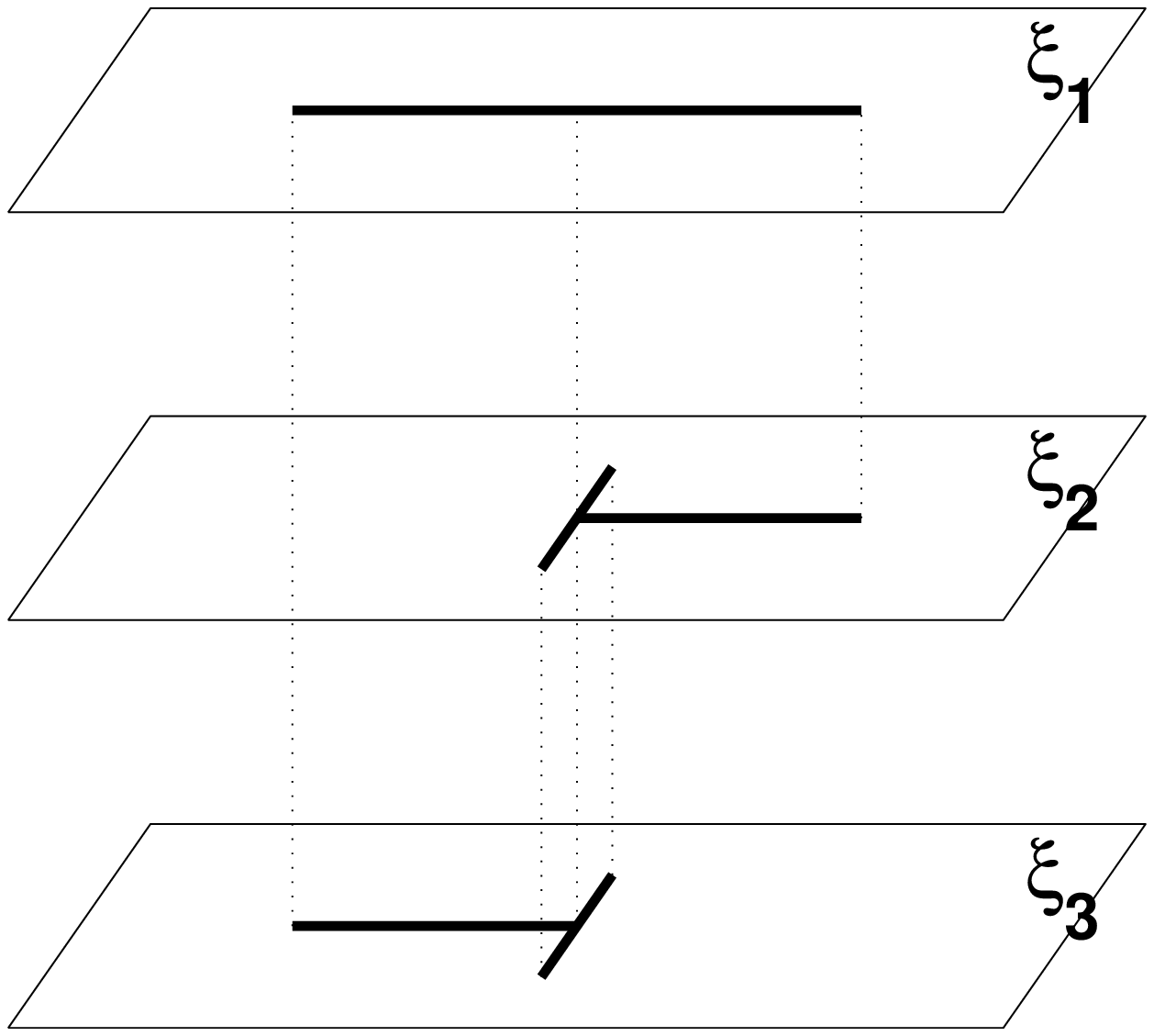}}
\caption{The structure of the Riemann surface for the equation (\ref{PasturEquation})
for the values $a> 1$ (left), $a=1$ (middle) and $a<1$ (right).
In all cases the eigenvalues of $M$ accumulate on the interval(s)
of the first sheet with a density given by (\ref{eigdensity}).}
\label{Riemann_surface}
\end{figure}

First we describe the limiting mean
density of eigenvalues.
The limiting mean density  follows
from earlier work of Pastur \cite{Pas}.
It is based on an analysis of the equation (Pastur equation)
\begin{equation} \label{PasturEquation}
    \xi^3 - z \xi^2 + (1-a^2) \xi + a^2 z = 0,
\end{equation}
which yields an algebraic function $\xi(z)$ defined on a
three-sheeted Riemann surface. The restrictions of $\xi(z)$
to the three sheets are denoted by $\xi_j(z)$, $j=1,2,3$. There are four
real branch points if $a > 1$ which determine two real intervals.
The two intervals come together for $a=1$, and for $0 < a < 1$, there are
two real branch points, and two purely imaginary branch points.
Fig.\ref{Riemann_surface} depicts the structure of the
Riemann surface $\xi(z)$ for $a>1$, $a=1$, and $a<1$.

In all cases we have that the limiting mean eigenvalue density
$\rho(x) = \rho(x;a)$ 
is given by
\begin{equation} \label{eigdensity}
    \rho(x;a) = \frac{1}{\pi} \Im \xi_{1+}(x), \qquad x \in \mathbb R,
\end{equation}
where $\xi_{1+}(x)$ denotes the limiting value of $\xi_1(z)$ as $z \to x$
with $\Im z > 0$.
For $a = 1$ the limiting mean eigenvalue density vanishes
at $x=0$ and $\rho(x;a) \sim |x|^{1/3}$ as $x \to 0$.

We note that this behavior at the closing (or opening) of a gap
is markedly different from the behavior that occurs in the usual
unitary random matrix ensembles $Z_n^{-1} e^{-n \Tr V(M)} dM$ where a closing of
the gap in the spectrum typically leads to a limiting mean eigenvalue
density $\rho$ that satisfies $\rho(x) \sim (x-x^*)^2$ as $x \to x^*$
if the gap closes at $x= x^*$. In that case the local eigenvalue correlations
can be described in terms of $\psi$-functions associated with the
Painlev\'e II equation, see above and \cite{BI2,CK}.
The phase transition for the model under consideration is different,
and it cannot be realized in a unitary random matrix ensemble.

\begin{theo} \label{maintheo1}
The limiting mean density of eigenvalues
\begin{equation} \label{rho1}
 \rho(x) = \lim_{n \to \infty} \frac{1}{n} K_n(x,x)
\end{equation}
exists for every $a > 0$. It satisfies
\begin{equation} \label{pastur1}
    \rho(x)=\frac{1}{\pi} \left|\, \Im\xi(x)\right|,
\end{equation}
where $\xi=\xi(x)$ is a solution of the cubic equation,
\begin{equation} \label{pastur2}
    \xi^3-x\xi^2-(a^2-1)\xi+xa^2=0.
\end{equation}
The support of $\rho$ consists of those $x \in \mathbb R$
for which {\rm (\ref{pastur2})} has a non-real solution.
\begin{enumerate}
\item[\rm (a)] For $0 < a < 1$,
the support of $\rho$ consists of one interval
$[-z_1, z_1]$,
and $\rho$ is real analytic and positive on $(-z_1, z_1)$, and it
vanishes like a square root at the edge points $\pm z_1$, i.e.,
there exists a constant $\rho_1 > 0$ such that
\begin{equation} \label{density1}
    \rho(x) = \frac{\rho_1}{\pi} |x \mp z_1|^{1/2} (1+o(1))
    \qquad \mbox{as } x \to \pm z_1, \, x \in (-z_1, z_1).
\end{equation}
\item[\rm (b)] For $a=1$,
the support of $\rho$ consists of one interval
$[-z_1, z_1]$,
and $\rho$ is real analytic and positive on $(-z_1, 0) \cup (0, z_1)$, it
vanishes like a square root at the edge points $\pm z_1$, and it
vanishes like a third root at $0$, i.e., there exists a constant $c > 0$
such that
\begin{equation} \label{density2}
    \rho(x) = c |x|^{1/3} \left(1+ o(1) \right), \qquad
        \mbox{as } x \to 0.
\end{equation}
\item[\rm (c)] For $a > 1$,
the support of $\rho$ consists of two disjoint intervals
$[-z_1, -z_2] \cup [z_2, z_1]$ with $0 < z_2 < z_1$,
$\rho$ is real analytic and positive on $(-z_1, -z_2) \cup (z_2, z_1)$,
and it vanishes like a square root at the edge points $\pm z_1$, $\pm z_2$.
\end{enumerate}
\end{theo}

To describe the universality of local eigenvalue correlations
in the large $n$ limit, we use  a rescaled version of the kernel $K_n$:
\begin{equation}\label{a1_4}
\hat{K}_n(x,y) = e^{n(h(x)-h(y))} K_n(x,y)
\end{equation}
for some function $h$. The rescaling  is allowed because it
does not affect the correlation functions $R_m$, which are
expressed as determinants of the kernel.
The function $h$ has the following form on $(-z_1, -z_2) \cup (z_2,z_1)$:
\begin{equation}\label{a1_5}
    h(x) = - \frac{1}{4} x^2 + \Re \int_{z_1}^x \xi_{1+}(s) ds,
\end{equation}
where $\xi_1$ is a solution of the Pastur equation.
The local eigenvalue correlations in the bulk of the spectrum 
in the large $n$ limit are described by the sine kernel. 
The bulk of the spectrum is the open interval $(-z_1,z_1)$
for $a<1$, and the union of the two open intervals, $(-z_1,-z_2)$
and $(z_2,z_1)$, for $a\ge 1$ ($z_2=0$ for $a=1$).

We have the following result:

\begin{theo}
For every
$x_0$ in the bulk of the spectrum  we have that
\begin{align*}
    \lim_{n\to\infty}
    &\frac{1}{n\rho(x_0)}
    \hat{K}_n \left(x_0 + \frac{u}{n\rho(x_0)}, x_0 + \frac{v}{n \rho(x_0)}\right)
    = \frac{\sin  \pi(u-v)}{\pi(u-v)}.
    \end{align*}
\end{theo}

At the edge of the spectrum the local eigenvalue correlations are described
in the large $n$ limit by the Airy kernel:

\begin{theo}
For every $u, v \in \R$ we have
\begin{align*} 
\lim_{n \to \infty}
    &\frac{1}{(\rho_1 n)^{2/3}} \hat{K}_n\left(z_1 + \frac{u}{(\rho_1 n)^{2/3}}, z_1 + \frac{v}{(\rho_1 n)^{2/3}}\right)
    = \frac{\Ai(u)\Ai'(v)-\Ai'(u)\Ai(v)}{u-v}.
    \end{align*}
A similar limit holds near the edge point $-z_1$ and also near the edge points $\pm z_2$
if $a>1$.
\end{theo}

As is usual in a critical case, there is a family of limiting kernels
that arise when $a$ changes with $n$ and $a \to 1$ as $n \to \infty$
in a critical way. These kernels are constructed out of Pearcey integrals
and therefore they are called Pearcey kernels.
The Pearcey kernels were first described by Br\'ezin and Hikami \cite{BH4,BH5}.
A detailed proof of the following result was recently given by
Tracy and Widom \cite{TW3}.

\begin{figure}[t]
\centerline{\includegraphics[height=5cm,width=5cm]{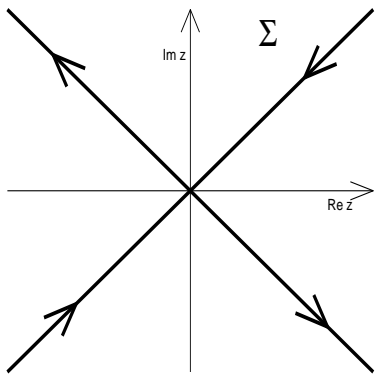}}
\caption{The contour $\Sigma$ that appears in the definition of $q(y)$.}
\label{fig:contourSigma}
\end{figure}

\begin{theo} \label{maintheo}
We have for every fixed $b \in \mathbb R$,
\begin{equation} \label{limitkernel}
    \lim_{n \to \infty} \frac{1}{n^{3/4}} K_n \left( \frac{x}{n^{3/4}}, \frac{y}{n^{3/4}};  1 + \frac{b}{2\sqrt{n}} \right)
    = K^{cusp}(x,y;b)
\end{equation}
where $K^{cusp}$ is the Pearcey kernel
\begin{equation} \label{Pearceykernel}
    K^{cusp}(x,y;b) =
    \frac{p(x) q''(y) - p'(x) q'(y) + p''(x) q(y) - bp(x) q(y)}{x-y}
       \end{equation}
with
\begin{equation} \label{Pearceyintegrals}
    p(x) = \frac{1}{2\pi} \int_{-\infty}^{\infty} e^{-\frac{1}{4}s^4 - \frac{b}{2} s^2 + isx} ds
    \qquad \mbox{ and } \qquad
   q(y) = \frac{1}{2\pi} \int_{\Sigma} e^{\frac{1}{4} t^4 + \frac{b}{2} t^2 + ity} dt.
   \end{equation}
The contour $\Sigma$ consists of the four rays $\arg y = \pm \pi/4, \pm 3\pi/4$,
with the orientation shown in Fig.{\rm\ref{fig:contourSigma}}.
\end{theo}

The functions (\ref{Pearceyintegrals}) are called Pearcey integrals \cite{Pea}.
They are solutions of the third order differential equations $p'''(x) = xp(x) + bp'(x)$
and $q'''(y) = - yq(y) + bq'(y)$, respectively.

Theorem \ref{maintheo} implies that local eigenvalue statistics of eigenvalues near $0$
are expressed in terms of the Pearcey kernel. For example we have the following
corollary of Theorem \ref{maintheo}.
\begin{cor}
The probability that a matrix of the ensemble {\rm (\ref{es_1})}, {\rm (\ref{es_1a})},
with $a = 1 + bn^{-1/2}/2$ has no eigenvalues in the interval
$[c n^{-3/4}, d n^{-3/4}]$ converges, as $n \to \infty$,
to the Fredholm determinant of the integral operator with kernel
$K^{cusp}(x,y;b)$ acting on $L^2(c,d)$.
\end{cor}
Similar expressions hold for the probability to have one, two, three, \ldots, eigenvalues
in an $O(n^{-3/4})$ neighborhood of $x=0$.

Tracy and Widom \cite{TW3} and Adler and van Moerbeke \cite{AvM3} gave
differential equations for the gap probabilities associated
with the Pearcey kernel and with the more general Pearcey process
which arises from considering the non-intersecting Brownian motion
model at several times near the critical time. See also \cite{OR}
where the Pearcey process appears in a combinatorial model on
random partitions.

Br\'ezin and Hikami and also Tracy and Widom used a double integral representation
for the kernel in order to establish Theorem \ref{maintheo}. We will describe the 
approach of \cite{BK4}, based on 
the Deift/Zhou steepest descent method for the Riemann-Hilbert problem for multiple Hermite polynomials.
This method is less direct than the steepest descent method for integrals. However,
an approach based on the Riemann-Hilbert problem may be applicable to more general situations,
where an integral representation is not available. This is the case, for example,
for the general (non-Gaussian) unitary random matrix ensemble with external source,
(\ref{es_1}),
with a general potential $V$.

The proof of the theorems above is based on the construction of a parametrix of the RHP,
and we will describe this construction for the cases $a>1$, $a<1$, and $a=1$.

\section{Construction of a parametrix in the case $a>1$}

Consider the Riemann surface given by equation (\ref{PasturEquation})
for $a>1$,
see the left surface on Fig.\ref{Riemann_surface}.
There are three roots to this equation, which behave at infinity as
\begin{equation}\label{rs_2}
\begin{aligned}
\xi_1(z)&=z -\frac{1}{z}+ O\left(\frac{1}{z^3}\right), \qquad
\xi_{2,3}(z)&=\pm a +\frac{1}{2z} +O\left(\frac{1}{z^2}\right). 
\end{aligned}
\end{equation}
We need the integrals of the $\xi$-functions,
\begin{equation} \label{wkb5a}
    \la_k(z) = \int^z \xi_k(s) ds, \qquad k =1,2,3,
\end{equation}
which we take so that $\la_1$ and $\la_2$
are analytic on $\mathbb C \setminus (-\infty, z_1]$
and $\la_3$ is analytic on $\mathbb C \setminus (-\infty, -z_2]$.
Then, as $z\to \infty$,
\begin{equation}\label{wkb6}
\begin{aligned}
\la_1(z)&=\frac{z^2}{2} - \ln z +l_1 + O\left(\frac{1}{z^2}\right),\qquad
\la_{2,3}(z)&=\pm az +\frac{1}{2}\ln z+l_{2,3}+ O\left(\frac{1}{z}\right), 
\end{aligned}
\end{equation}
where $l_1$, $l_2$, $l_3$ are some constants, which we choose as follows.
We choose $l_1$ and $l_2$ such that
\begin{equation}\label{wkb6a}
\la_1(z_1) = \la_2(z_1) = 0, 
\end{equation}
and then $l_3$ such that
\begin{equation}\label{wkb6b}
\la_3(-z_2) = \la_{1+}(-z_2) = \la_{1-}(-z_2) - \pi i. 
\end{equation}

{\it First Transformation of the RH Problem.}
Using the functions $\lambda_j$ and the constants $l_j$, $j=1,2,3$, we define
\begin{equation} \label{defT}
\begin{aligned}
   T(z) &= \diag\left(e^{-nl_1}, e^{-nl_2}, e^{-nl_3}\right)Y(z) 
\diag\left(e^{n(\la_1(z)-\frac{1}{2}z^2)}, e^{n(\la_2(z) - az)}, e^{n(\la_3(z)+az)} \right).
\end{aligned}
\end{equation}
Then  $T_+(x) = T_-(x) j_T(x)$, $x \in \mathbb R$,
where for $x\in[z_2,z_1]$,
\begin{equation} \label{defT_1}
\begin{aligned}
j_T &=\begin{pmatrix}
e^{n(\la_1-\la_2)_+} & 1 & e^{n(\la_{3} - \la_{1-})} \\
0 & e^{n(\la_1-\la_2)_-} & 0 \\
0 & 0 & 1
\end{pmatrix}
\end{aligned}
\end{equation}
and for $x\in[-z_1,-z_2]$,
\begin{equation} \label{defT_2}
\begin{aligned}
j_T&=\begin{pmatrix}
e^{n(\la_1-\la_3)_+} & e^{n(\la_{2+}-\la_{1-})} & 1 \\
0 & 1 & 0 \\
0 & 0 & e^{n(\la_1-\la_3)_-}
\end{pmatrix}.
\end{aligned}
\end{equation}

\begin{figure}[ht]
\scalebox{0.8}{\includegraphics{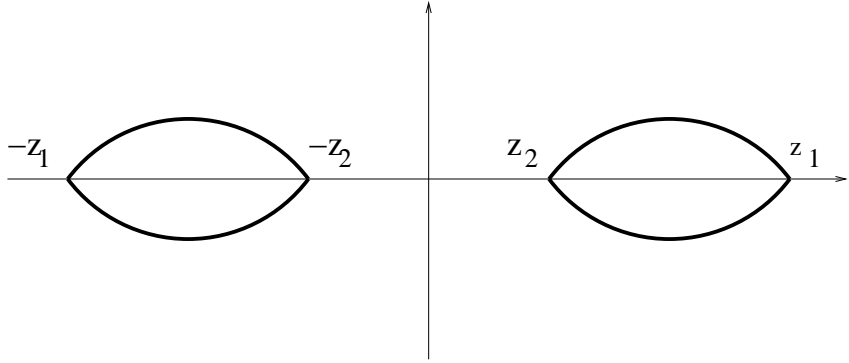}}
\caption{The lenses for $a>1$.}
\label{rs_a2}
\end{figure}

{\it Second transformation of the RH Problem: opening of lenses.}
The lens structure is shown on Fig.\ref{rs_a2}.
Set in the right lens,
\begin{equation} \label{defT_3}
S(z)=\left\{
\begin{aligned}
{}&T(z)\begin{pmatrix}
1 & 0  & 0 \\
-e^{n(\la_1(z)-\la_2(z))} & 1 & -e^{n(\la_3(z)-\la_2(z))} \\
0 & 0 & 1
\end{pmatrix} \\
{}&\;\text{\rm in the upper lens region},\\
{}&T(z)
\begin{pmatrix}
1 & 0 & 0 \\
e^{n(\la_1(z)-\la_2(z))} & 1 & -e^{n(\la_3(z)-\la_2(z))} \\
0 & 0 & 1
\end{pmatrix}\\
{}&\;\text{\rm in the lower lens region},
\end{aligned}\right.
\end{equation}
and, respectively, in the left lens,
\begin{equation} \label{defT_4}
S(z)=\left\{
\begin{aligned}
{}&T(z)\begin{pmatrix}
1 & 0  & 0 \\
0 & 1 & 0 \\
-e^{n(\la_1(z)-\la_3(z))} & -e^{n(\la_2(z)-\la_3(z))} & 1
\end{pmatrix} \\
{}&\;\text{\rm in the upper lens region},\\
{}&T(z)
\begin{pmatrix}
1 & 0 & 0 \\
0 & 1 & 0 \\
e^{n(\la_1(z)-\la_3(z)))} & -e^{n(\la_2(z)-\la_3(z))} & 1
\end{pmatrix}\\
{}&\;\text{\rm in the lower lens region}.
\end{aligned}\right.
\end{equation}
Then 
\begin{equation} \label{defT_5}
\begin{aligned}
S_+(x)&=S_-(x)j_S(x);\qquad
j_S(x)&=
\begin{pmatrix}
0 & 1 & 0 \\
-1 & 0 & 0 \\
0 & 0 & 1
\end{pmatrix},\quad x\in[z_2,z_1],
\end{aligned}
\end{equation}
and
\begin{equation} \label{defT_6}
\begin{aligned}
S_+(x)&=S_-(x)j_S(x);\qquad
j_S(x)&=
\begin{pmatrix}
0 & 0 & 1 \\
0 & 1 & 0 \\
-1 & 0 & 0
\end{pmatrix},\quad x\in[-z_1,-z_2].
\end{aligned}
\end{equation}
In addition, $S$ has jumps on the boundary of the lenses,
which are exponentially small away of the points $\pm z_{1,2}$.
The RH problem for $S$ is approximated by the model RH problem.

{\it Model RH problem}
\begin{itemize}
\item $M$ is analytic on $\C\setminus ([-z_1,-z_2]\cup[z_2,z_1])$,
\item
\begin{equation} \label{defT_7}
\begin{aligned}
M_+(x)=M_-(x)j_S(x),\quad
 x\in (-z_1,-z_2)\cup (z_2,z_1),
\end{aligned}
\end{equation}
\item as $z\to\infty$,
\begin{equation} \label{defT_8}
M(z)=I+O\left(\frac{1}{z}\right),
\end{equation}
\end{itemize}
where the jump matrix is
\begin{equation} \label{defT_9}
j_S(x)=
\left\{
\begin{aligned}
{}&\begin{pmatrix}
0 & 1  & 0 \\
-1 & 0 & 0 \\
0 & 0 & 1
\end{pmatrix},\quad  x\in (z_2,z_1)\\
{}&\begin{pmatrix}
0 & 0 & 1 \\
0 & 1 & 0 \\
-1 & 0 & 0
\end{pmatrix},\quad x\in(-z_1,-z_2).
\end{aligned}\right.
\end{equation}
Solution to the model RH problem has the form:

\vskip 3mm

\begin{equation} \label{defT_10}
M(z)=A(z)B(z)C(z),
\end{equation}
where
\begin{equation} \label{defT_11}
A(z)=\textrm{diag}\left( 1,-\frac{i}{\sqrt 2},-\frac{i}{\sqrt 2}\right),\qquad
B(z)=
\begin{pmatrix}
\xi_1^2(z)-a^2 & \xi_2^2(z)-a^2 & \xi_3^2(z)-a^2 \\
\xi_1(z)+a & \xi_2(z)+a & \xi_3(z)+a \\
\xi_1(z)-a & \xi_2(z)-a & \xi_3(z)-a 
\end{pmatrix}
\end{equation}
and
\begin{equation} \label{defT_12}
C(z)=\textrm{diag}\left( \frac{1}{\sqrt{Q(\xi_1(z))}},
\frac{1}{\sqrt{Q(\xi_2(z))}},\frac{1}{\sqrt{Q(\xi_3(z))}}
\right)
\end{equation}
where
\begin{equation} \label{defT_13}
Q(z)=z^4-(1+2a^2)z^2+(a^2-1)a^2.
\end{equation}

{\it Parametrix at  edge points}. We consider  small disks $D(\pm z_j,r)$ with radius $r > 0$
and centered at the edge points,
and look for a local parametrix $P$ defined on the union of
the four disks such that
\begin{itemize}
\item $P$ is analytic on $D(\pm z_j, r) \setminus(\R\cup\Ga)$,
\item
\begin{equation}\label{lp1}
P_+(z)= P_-(z)j_S(z),\quad z\in (\R\cup\Ga) \cap D(\pm z_j, r),
\end{equation}
\item as $n\to\infty$,
\begin{equation}\label{lp1b}
P(z)=\left(I+O\left(\frac{1}{n}\right) \right) M(z)
\quad \text{\rm uniformly for $z \in \partial D(\pm z_j,r)$}.
\end{equation}
\end{itemize}

\begin{figure}[ht]
\scalebox{0.6}{\includegraphics{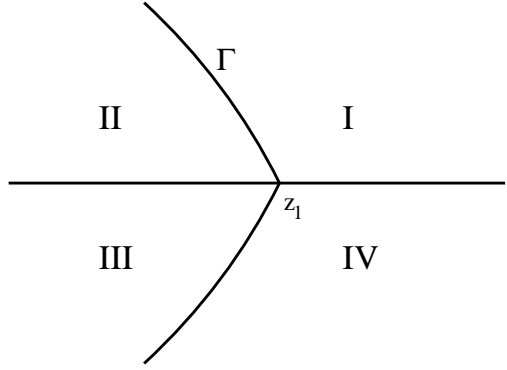}}
\caption{Partition of a neighborhood of the edge point.}
\label{ep_1}
\end{figure}

We consider here the edge point $z_1$ in detail.
We note that as $z \to z_1$,
\begin{equation} \label{rhoconst4}
\begin{aligned}
    \la_1(z) &= q(z-z_1) + \frac{2\rho_1}{3} (z-z_1)^{3/2} + O(z-z_1)^{2} \\
    \la_2(z) &= q(z-z_1) - \frac{2\rho_1}{3} (z-z_1)^{3/2} + O(z-z_1)^{2}
\end{aligned}
\end{equation}
so that
\begin{equation} \label{rhoconst5}
    \la_1(z) - \la_2(z) = \frac{4 \rho_1}{3} (z-z_1)^{3/2} + O(z-z_1)^{5/2}
\end{equation}
as $z \to z_1$. Then it follows that
\begin{equation} \label{beta}
\beta(z) = \left[\frac{3}{4}(\la_1(z)-\la_2(z))\right]^{2/3}
\end{equation}
is analytic at $z_1$, real-valued on the real axis near $z_1$
and $\beta'(z_1) = \rho_1^{2/3} > 0$. So $\beta$ is a conformal map from
$D(z_1, r)$ to a convex neighborhood of the origin, if $r$
is sufficiently small (which we assume to be the case).
We take $\Gamma$ near $z_1$ such that
\[ \beta(\Gamma \cap D(z_1,r)) \subset
\{z \mid \arg(z) = \pm 2\pi/3 \}. \]
Then $\Gamma$ and $\R$ divide the disk $D(z_1,r)$ into
four regions numbered I, II, III, and IV, such
that $0 < \arg \beta(z) < 2\pi/3$, $2\pi/3 < \arg \beta(z) < \pi$,
$-\pi < \arg \beta(z) < -2\pi/3$, and $-2\pi/3 < \arg \beta(z) < 0$ for $z$
in regions I, II, III, and IV, respectively, see Fig.\ref{ep_1}.

Recall that the jumps $j_S$ near $z_1$ are given as
\begin{equation} \label{lp2}
\begin{aligned}
j_S&=
\begin{pmatrix} 0 & 1 & 0 \\
-1 & 0 & 0 \\ 0 & 0 & 1 \end{pmatrix}
\qquad \text{\rm on } [z_1-r, z_1) \\
j_S&=
\begin{pmatrix} 1 & 0 & 0 \\
e^{n(\la_1-\la_2)} & 1 & e^{n(\la_3-\la_2)} \\
0 & 0 & 1 \end{pmatrix}
\; \text{\rm on the upper boundary of the lens in $D(z_1,r)$ } \\
j_S&=
\begin{pmatrix} 1 & 0 & 0 \\
e^{n(\la_1-\la_2)} & 1 & -e^{n(\la_3-\la_2)} \\
0 & 0 & 1 \end{pmatrix}
\; \text{\rm on the lower boundary of the lens in $D(z_1,r)$} \\
j_S&=
\begin{pmatrix} 1 & e^{n(\la_2-\la_1)} & e^{n(\la_3-\la_1)} \\
0 & 1 & 0 \\ 0 & 0 & 1 \end{pmatrix}
\qquad \text{\rm on } (z_1, z_1+r].
\end{aligned}
\end{equation}

We write
\begin{equation} \label{tildeP}
\tilde{P} =
\left\{ \begin{array}{ll}
P
\begin{pmatrix}
1 & 0 & 0 \\ 0 & 1 & -e^{n(\la_3-\la_2)} \\
0 & 0 & 1 \end{pmatrix}
& \text{\rm in regions I and IV} \\
P & \text{\rm in regions II and III.}
\end{array} \right.
\end{equation}
Then the jumps for $\tilde{P}$ are
$\tilde{P}_+ = \tilde{P}_- j_{\tilde{P}}$
where
\begin{equation} \label{lp5}
\begin{aligned}
j_{\tilde{P}}&=
\begin{pmatrix} 0 & 1 & 0 \\
-1 & 0 & 0 \\ 0 & 0 & 1 \end{pmatrix}
\qquad \text{\rm on } [z_1-r, z_1) \\
j_{\tilde{P}}&=
\begin{pmatrix} 1 & 0 & 0 \\
e^{n(\la_1-\la_2)} & 1 & 0 \\
0 & 0 & 1 \end{pmatrix}
\; \text{\rm on the upper side of the lens in } D(z_1,r) \\
j_{\tilde{P}}&=
\begin{pmatrix} 1 & 0 & 0 \\
e^{n(\la_1-\la_2)} & 1 & 0 \\
0 & 0 & 1 \end{pmatrix}
\; \text{\rm on the lower side of the lens in } D(z_1,r) \\
j_{\tilde{P}}&=
\begin{pmatrix} 1 & e^{n(\la_2-\la_1)} & 0 \\
0 & 1 & 0 \\ 0 & 0 & 1 \end{pmatrix}
\qquad \text{\rm on } (z_1, z_1+r].
\end{aligned}
\end{equation}
We also need the matching condition
\begin{equation} \label{match2}
\tilde{P}(z) = \left(I+O\left(\frac{1}{n}\right) \right) M(z)
\quad \text{\rm uniformly for $z \in \partial D(z_1,r)$}.
\end{equation}
The RH problem for $\tilde{P}$ is essentially a
$2\times 2$ problem, since the jumps (\ref{lp5}) are non-trivial only in
the upper $2\times 2$ block. A solution can be constructed in
a standard way out of Airy functions. The Airy function $\Ai(z)$
solves the equation $y'' = zy$ and for any $\varepsilon >0$, in the
sector $\pi + \varepsilon \leq \arg z \leq \pi - \varepsilon$, it has
the asymptotics as $z \to \infty$,
\begin{equation}\label{ep3}
\Ai(z)=\frac{1}{2\sqrt\pi z^{1/4}}e^{-\frac{2}{3}z^{3/2}}
\left(1+O(z^{-3/2})\right).
\end{equation}
The functions $\Ai(\om z)$, $\Ai(\om^2 z)$, where
$\om=e^{\frac{2\pi i}{3}}$, also solve the equation $y''=zy$, and we
have the linear relation,
\begin{equation}\label{ep4}
\Ai(z)+\om\Ai(\om z)+\om^2\Ai(\om^2 z)=0.
\end{equation}
Write
\begin{equation}\label{ep6}
y_0(z)=\Ai(z), \quad y_1(z)=\om\Ai(\om z),
\quad y_2(z)=\om^2\Ai(\om^2 z),
\end{equation}
and we use these functions to define
\begin{equation}\label{ep5}
\Phi(z)=
\left\{
\begin{aligned}
{}&\begin{pmatrix}
y_0(z) & -y_2(z) & 0 \\
y_0'(z) & -y_2'(z) & 0 \\
0 & 0 & 1
\end{pmatrix},\quad \mbox{for $0 < \arg z < 2\pi/3$},\\
{}&\begin{pmatrix}
-y_1(z) & - y_2(z) & 0 \\
-y_1'(z) & -y_2'(z) & 0 \\
0 & 0 & 1
\end{pmatrix}, \quad \mbox{for $2\pi/3 < \arg z < \pi$}, \\
{}&\begin{pmatrix}
-y_2(z) & y_1(z) & 0 \\
-y_2'(z) & y_1'(z) & 0 \\
0 & 0 & 1
\end{pmatrix}, \quad \mbox{for $-\pi < \arg z < -2\pi/3$}, \\
{}&\begin{pmatrix}
y_0(z) & y_1(z) & 0 \\
y_0'(z) & y_1'(z) & 0 \\
0 & 0 & 1
\end{pmatrix},\quad \mbox{for $-2\pi/3 < \arg z < 0$}.
\end{aligned}\right.
\end{equation}
Then
\begin{equation} \label{ep7}
\tilde{P}(z) = E_n(z) \Phi(n^{2/3} \beta(z))
\diag\left(e^{\frac{1}{2} n (\la_1(z)-\la_2(z))},
e^{-\frac{1}{2}n(\la_1(z) - \la_2(z))}, 1\right)
\end{equation}
where $E_n$ is an analytic prefactor that takes care of the matching
condition (\ref{match2}). Explicitly, $E_n$ is given by
\begin{equation} \label{ep8}
E_n = \sqrt{\pi} M
\begin{pmatrix} 1 & -1 & 0 \\ -i & - i & 0 \\ 0 & 0 & 1
\end{pmatrix}
\begin{pmatrix} n^{1/6} \beta^{1/4} & 0 & 0  \\ 0 & n^{-1/6}
  \beta^{-1/4} & 0 \\ 0 & 0 & 1
\end{pmatrix}.
\end{equation}
A similar construction works for a parametrix $P$
around the other edge points.

{\it Third transformation.}
In the third and final transformation we put
\begin{equation} \label{tt1}
\begin{aligned}
R(z) & = S(z) M(z)^{-1}
    \quad \text{\rm for $z$ outside the disks $D(\pm z_j, r)$, $j=1,2$} \\
R(z) & = S(z) P(z)^{-1}
    \quad \text{\rm for $z$ inside the disks.}
\end{aligned}
\end{equation}

\begin{figure}
\scalebox{0.8}{\includegraphics{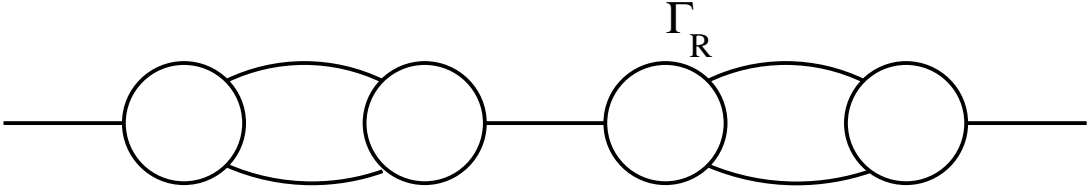}} \caption{The contour
$\Gamma_R$ for $R$.} \label{fig5}
\end{figure}

Then $R$ is analytic on $\C \setminus \Ga_R$, where $\Ga_R$ consists of
the four circles $\partial D(\pm z_j, r)$, $j=1,2$, the parts of $\Gamma$
outside the four disks, and the real intervals $(-\infty, -z_1-r)$,
$(-z_2+r, z_2-r)$, $(z_1+r,\infty)$, see Fig.\ref{fig5}.
There are jump relations
\begin{equation} \label{tt2}
    R_+ = R_- j_R
\end{equation}
where
\begin{equation} \label{tt3}
\begin{aligned}
j_R &= M P^{-1}
    \quad \text{\rm on the circles, oriented counterclockwise} \\
j_R &= M j_S M^{-1}
    \quad \text{\rm on the remaining parts of $\Gamma_R$.}
\end{aligned}
\end{equation}
We have that $j_R = I +O(1/n)$ uniformly on
the circles, and 
$j_R = I + O(e^{-cn})$ for some $c > 0$ as $n \to \infty$, uniformly on the
remaining parts of $\Gamma_R$. So we can conclude
\begin{equation} \label{tt4}
    j_R(z) = I + O\left(\frac{1}{n}\right)
    \quad \text{\rm as $n \to \infty$, uniformly on $\Gamma_R$.}
\end{equation}
As $z\to \infty$, we have
\begin{equation} \label{tt5}
    R(z) = I + O(1/z).
\end{equation}

From (\ref{tt2}), (\ref{tt4}), (\ref{tt5}) and the fact that we can deform
the contours in any desired direction, it follows that
\begin{equation} \label{tt6}
    R(z) = I + O\left(\frac{1}{n(|z|+1)}\right) \quad \mbox{as } n \to \infty.
\end{equation}
uniformly for $z \in \C \setminus \Ga_R$, see \cite{Dei,DKMVZ1,DKMVZ2,Kui}.

\section{Construction of a parametrix in the case $a<1$}

\subsection{$\la$-functions}
Consider the Riemann surface given by equation (\ref{PasturEquation})
for $a<1$,
see the right surface on Fig.\ref{Riemann_surface}.
There are three roots to this equation, which behave at infinity as in
(\ref{rs_2}).
We need the integrals of the $\xi$-functions, which we define as
\begin{equation} \label{deflambda}
\begin{aligned}
    \lambda_1(z) &= \int_{z_1}^z \xi_1(s) ds, \\
    \lambda_2(z) &= \int_{z_1}^z \xi_2(s) ds, \\
    \lambda_3(z) &= \int_{-z_{1+}}^z \xi_3(s) ds + \lambda_{1-}(-z_1),
\end{aligned}
\end{equation}
The path of integration for $\lambda_3$ lies in $\mathbb C \setminus
((-\infty,0] \cup [-iz_2,iz_2])$, and it starts at the point $-z_1$
on the upper side of the cut.
All three $\lambda$-functions are defined on their respective sheets of
the Riemann surface with an additional cut along the negative real
axis. Thus $\lambda_1, \lambda_2, \lambda_3$ are defined and analytic
on $\mathbb C \setminus (-\infty, z_1]$, $\mathbb C \setminus
((-\infty,z_1] \cup [-iz_2,iz_2])$, and $\mathbb C \setminus
((-\infty,0] \cup [-iz_2,iz_2])$, respectively.
Their behavior at infinity is
\begin{equation} \label{lambdaatinfinity}
\begin{aligned}
    \lambda_1(z) &= \frac{1}{2}z^2 - \log z + \ell_1 + O(1/z) \\
    \lambda_2(z) &= az + \frac{1}{2} \log z + \ell_2 + O(1/z) \\
    \lambda_3(z) &= -az + \frac{1}{2} \log z + \ell_3 + O(1/z)
\end{aligned}
\end{equation}
for certain constants $\ell_j$, $j=1,2,3$.
The $\lambda_j$'s satisfy the following jump relations
\begin{equation} \label{jumpslambda}
\begin{array}{ll}
    \lambda_{1 \mp}  = \lambda_{2\pm} &\textrm{ on } (0,z_1), \\
    \lambda_{1-}  = \lambda_{3+}  & \textrm{ on } (-z_1,0), \\
    \lambda_{1+}  = \lambda_{3-} -\pi i & \textrm{ on } (-z_1,0), \\
    \lambda_{2\mp}  = \lambda_{3\pm}  & \textrm{ on } (0, iz_2), \\
    \lambda_{2\mp}  = \lambda_{3\pm} - \pi i & \textrm{ on } (-iz_2,0), \\
    \lambda_{1+} = \lambda_{1-} - 2\pi i & \textrm{ on } (-\infty, -z_1), \\
    \lambda_{2+} = \lambda_{2-} + \pi i & \textrm{ on } (-\infty, 0), \\
    \lambda_{3+} = \lambda_{3-} + \pi i & \textrm{ on } (-\infty, -z_1),
\end{array}
\end{equation}
where the segment $(-iz_2, iz_2)$ is oriented upwards. 

\subsection{First transformation $Y \mapsto U$}
We define for $z \in \mathbb C \setminus (\mathbb R \cup [-iz_2, iz_2])$,
\begin{equation} \label{defU}
    U(z) = \diag(e^{-n\ell_1}, e^{-n \ell_2}, e^{-n\ell_3})
    Y(z) \diag(e^{n (\lambda_1(z) - \frac{1}{2} z^2)}, e^{n(\lambda_2(z) - az)},
        e^{n(\lambda_3(z) + az)}).
\end{equation}
This coincides with the first transformation for $a>1$.
Then $U$ solves the following RH problem.

\begin{itemize}
\item $U : \mathbb C \setminus (\mathbb R \cup [-iz_2, iz_2]) \to \mathbb C^{3 \times 3}$ is analytic.
\item $U$ satisfies the jumps
\begin{equation} \label{jumpU0}
    U_+ = U_-
    \begin{pmatrix}
        e^{n(\lambda_{1+}-\lambda_{1-})} & e^{n(\lambda_{2+}-\lambda_{1-})}
        & e^{n(\lambda_{3+}-\lambda_{1-})} \\
        0 & e^{n(\lambda_{2+}-\lambda_{2-})} & 0 \\
        0 & 0 & e^{n(\lambda_{3+}-\lambda_{3-})}
        \end{pmatrix}
        \qquad \textrm{ on } \mathbb R,
\end{equation}
and
\begin{equation}
   \label{jumpU5}
    U_+ = U_-
    \begin{pmatrix}
        1 & 0 & 0 \\
        0 & e^{n(\lambda_{2+}-\lambda_{2-})} & 0 \\
        0 & 0 & e^{n(\lambda_{3+}-\lambda_{3-})}
    \end{pmatrix}
    \qquad \textrm{on } [-iz_2,iz_2].
\end{equation}
\item $U(z) = I + O(1/z)$ as $z \to \infty$.
\end{itemize}

\subsection{Second transformation $U \mapsto T$: global opening of a lens
on $[-iz_2,iz_2]$}
The second transformation is the opening of a lens on the interval $[-iz_2,iz_2]$.
We consider a contour $\Sigma$, which goes first from $(-iz_2)$ to $iz_2$ around the point
$z_1$, and then from $iz_2$ to $(-iz_2)$ around the point
$-z_1$, see Fig.16, and such that for $z\in\Sigma$,
\begin{figure}[h] \label{figure3}
  \scalebox{0.8}{\includegraphics[width=12cm]{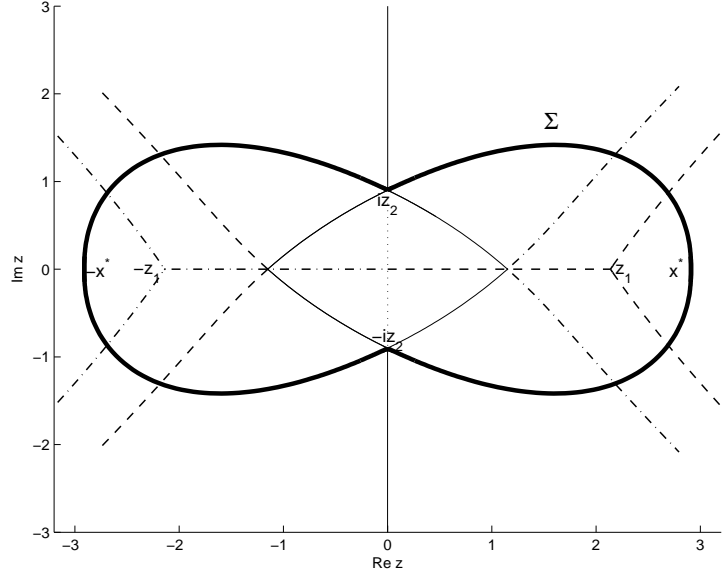}}
        \caption{Contour $\Sigma$ which is such that $\Re \lambda_2 < \Re \lambda_3$
        on the part of $\Sigma$ in the left half-plane and $\Re \lambda_2 > \Re \lambda_3$
        on the part of $\Sigma$ in the right half-plane.}
       \end{figure}
\begin{equation} \label{ut_1}
\pm\left( \Re \lambda_2(z) - \Re \lambda_3(z)\right)>0,\quad \pm\Re z>0.
\end{equation}
Observe that inside the curvilinear quadrilateral marked by a solid line on Fig.16, 
$\pm\left( \Re \lambda_2(z) - \Re \lambda_3(z)\right)<0$, hence the contour $\Sigma$ has
to stay outside of this quadrilateral. We set $T = U$ outside $\Sigma$, and inside $\Sigma$ we set
\begin{equation} \label{defT_a}
\begin{aligned}
    T& = U \begin{pmatrix}
    1 & 0 & 0 \\
    0 & 1 & 0 \\
    0 & - e^{n(\lambda_2-\lambda_3)} & 1 \end{pmatrix}
    \qquad \textrm{for } \Re z < 0 \textrm{ inside } \Sigma, \\
    T&= U \begin{pmatrix}
    1 & 0 & 0 \\
    0 & 1 & - e^{n(\lambda_3-\lambda_2)} \\
    0 & 0 & 1 \end{pmatrix}
    \qquad \textrm{for } \Re z > 0 \textrm{ inside } \Sigma.
\end{aligned}
\end{equation}

\subsection{Third transformation $T \mapsto S$: opening of a lens
on $[-z_1,z_1]$} We open a lens on $[-z_1,z_1]$, inside of
$\Sigma$, see Fig.17, and we define $S = T$ outside of the lens and
\begin{equation} \label{defS}
\begin{aligned}
    S &= T
    \begin{pmatrix} 1 & 0 & 0 \\ 0 & 1 & 0 \\
    - e^{n(\lambda_1-\lambda_3)} & 0 & 1 \end{pmatrix}
    \qquad \textrm{in upper part of the lens in left half-plane,} \\
    S & = T
    \begin{pmatrix} 1 & 0 & 0 \\ 0 & 1 & 0 \\
    e^{n(\lambda_1-\lambda_3)} & 0 & 1 \end{pmatrix}
    \qquad \textrm{in lower part of the lens in left half-plane,} \\
    S & = T
    \begin{pmatrix} 1 & 0 & 0 \\
    -e^{n(\lambda_1-\lambda_2)} & 1 & 0 \\ 0 & 0 & 1 \end{pmatrix}
    \qquad
    \textrm{in upper part of the lens in right half-plane,} \\
    S & = T
    \begin{pmatrix} 1 & 0 & 0 \\
    e^{n(\lambda_1-\lambda_2)} & 1 & 0 \\ 0 & 0 & 1 \end{pmatrix}
    \qquad
    \textrm{in lower part of the lens in right half-plane.}
\end{aligned}
\end{equation}

  \begin{figure}[h] \label{figure5}
  \scalebox{0.8}{\includegraphics[width=16cm,height=8cm]{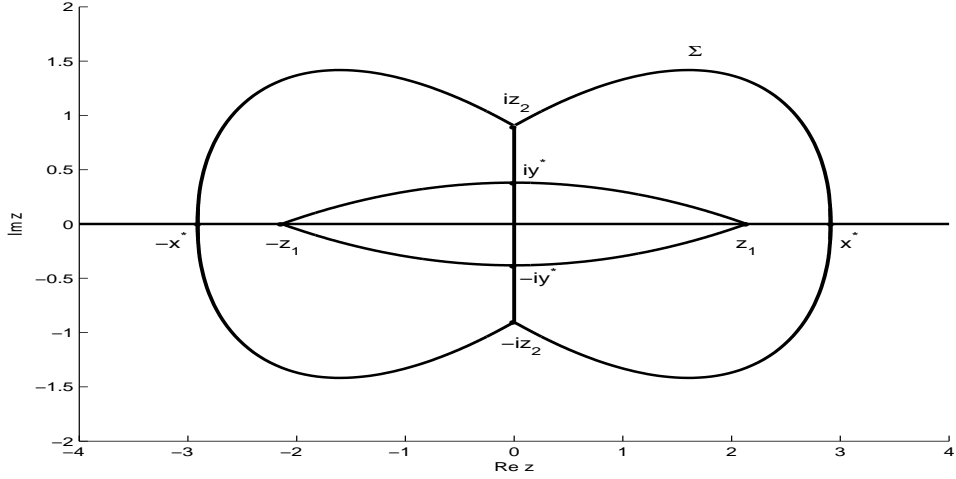}}
        \caption{Opening of a lens around $[-z_1,z_1]$. }
    \end{figure} 

Then $S$ satisfies the following RH problem:

\begin{itemize}
\item $S$ is analytic outside the real line, the vertical segment $[-iz_2,iz_2]$,
the curve $\Sigma$, and the upper and lower lips of the lens around $[-z_1,z_1]$.
\item $S$ satisfies the following jumps on the real line
\begin{align} \label{jumpS1}
    S_+ & = S_-
    \begin{pmatrix}
        1 & e^{n(\lambda_{2+}- \lambda_{1-})} & e^{n(\lambda_{3+}-\lambda_{1-})} \\
        0 & 1 & 0 \\
        0 & 0 & 1
        \end{pmatrix}  \qquad \textrm{on } (-\infty,-x^*] \\
    \label{jumpS2}
        S_+ & = S_-
    \begin{pmatrix}
        1 & 0 & e^{n(\lambda_{3+}-\lambda_{1-})} \\
        0 & 1 & 0 \\
        0 & 0 & 1
        \end{pmatrix} \qquad \textrm{on } (-x^*, -z_1] \\
    \label{jumpS3}
        S_+ & = S_-
        \begin{pmatrix} 0 & 0 & 1\\
        0 & 1 & 0 \\ -1 & 0 & 0 \end{pmatrix}
        \qquad \textrm{on } (-z_1,0) \\
    \label{jumpS4}
        S_+ & = S_-
    \begin{pmatrix} 0 & 1 & 0 \\
        -1 & 0 & 0 \\ 0 & 0 & 1 \end{pmatrix}
    \qquad \textrm{on } (0, z_1)  \\
      \label{jumpS5}
        S_+ & = S_-
      \begin{pmatrix}
        1 & e^{n(\lambda_2-\lambda_1)} & 0 \\
        0 & 1 & 0 \\
        0 & 0 & 1
        \end{pmatrix}
        \qquad \textrm{on } [z_1, x^*) \\
    \label{jumpS6}
        S_+ & = S_-
      \begin{pmatrix}
        1 & e^{n(\lambda_2-\lambda_1)} & e^{n(\lambda_3-\lambda_1)} \\
        0 & 1 & 0 \\
        0 & 0 & 1
        \end{pmatrix}
        \qquad \textrm{on } [x^*, \infty).
\end{align}
$S$ has the following jumps on the segment $[-iz_2,iz_2]$,
\begin{align} \label{jumpS7}
    S_+ & = S_-
    \begin{pmatrix}
        1 & 0 & 0 \\
        0 & 0 & 1 \\
        0 & -1 & e^{n(\lambda_{3+}-\lambda_{3-})}
        \end{pmatrix}  \qquad \textrm{on } (-iz_2, -iy^*) \\
        \label{jumpS8}
    S_+ & = S_-
    \begin{pmatrix}
        1 & 0 & 0 \\
        0 & 0 & 1 \\
        e^{n(\lambda_1-\lambda_{3-})} & -1 & e^{n(\lambda_{3+}-\lambda_{3-})}
        \end{pmatrix}  \qquad \textrm{on } (-iy^*,0) \\
        \label{jumpS9}
    S_+ & = S_-
    \begin{pmatrix}
        1 & 0 & 0 \\
        0 & 0 & 1 \\
        -e^{n(\lambda_1-\lambda_{3-})} & -1 & e^{n(\lambda_{3+}-\lambda_{3-})}
        \end{pmatrix}  \qquad \textrm{on } (0,iy^*) \\
    \label{jumpS10}
    S_+ & = S_-
    \begin{pmatrix}
        1 & 0 & 0 \\
        0 & 0 & 1 \\
        0 & -1 & e^{n(\lambda_{3+}-\lambda_{3-})}
        \end{pmatrix}  \qquad \textrm{on } (iy^*, iz_2).
\end{align}
The jumps on $\Sigma$ are
\begin{align}
    \label{jumpS11}
    S_+ & =  S_-
    \begin{pmatrix}
    1 & 0 & 0 \\
    0 & 1 & 0 \\
    0 & e^{n(\lambda_2-\lambda_3)} & 1
    \end{pmatrix}
    \qquad \textrm{on }  \{z \in \Sigma \mid \Re z < 0 \} \\
    \label{jumpS12}
    S_+ & =  S_-
    \begin{pmatrix}
    1 & 0 & 0 \\
    0 & 1 & e^{n(\lambda_3-\lambda_2)} \\
    0 & 0 & 1
    \end{pmatrix}
    \qquad \textrm{on } \{z \in \Sigma \mid \Re z > 0 \}.
\end{align}
Finally, on the upper and lower lips of the lens, we find jumps
\begin{align}
    \label{jumpS13}
    S_+ & = S_-
    \begin{pmatrix}
    1 & 0 & 0 \\ 0 & 1 & 0 \\ e^{n(\lambda_1-\lambda_3)} & 0 & 1
    \end{pmatrix}
    \qquad \textrm{on the lips of the lens in the left half-plane} \\
    \label{jumpS14}
    S_+ & =  S_-
    \begin{pmatrix}
    1 & 0 & 0 \\ e^{n(\lambda_1-\lambda_2)} & 1 & 0 \\ 0 & 0 & 1
    \end{pmatrix}
    \qquad \textrm{on the lips of the lens in the right half-plane.}
\end{align}
\item $S(z) = I + O(1/z)$ as $z \to \infty$.
\end{itemize}
As $n \to \infty$, the jump matrices have limits. Most of the limits are
the identity matrix, except for the jumps on $(-z_1,z_1)$, see (\ref{jumpS3})
and (\ref{jumpS4}), and on $(-iz_2, iz_2)$, see (\ref{jumpS7})--(\ref{jumpS10}).
The limiting model RH problem can be solved explicitly. The solution is similar 
to the case $a>1$, and it is given by formulas (\ref{defT_10})-(\ref{defT_13}), with cuts
of the  function $\sqrt{P(z)}$ on the intervals $[-z_1,z_1]$ and $[-iz_2,iz_2]$.

\subsection{Local parametrix at the branch points for $a<1$.} 
Near the branch points the model solution $M$ will not be a good approximation to $S$.
We need a local analysis near each of the branch points.
In a small circle around each of the branch points, the parametrix $P$
should have the same jumps as $S$, and on the boundary of the
circle $P$ should match with $M$ in the sense that
\begin{equation} \label{matching}
    P(z) = M(z) \left(I + O(1/n)\right)
\end{equation}
uniformly for $z$ on the boundary of the circle.

The construction of $P$ near the real branch points $\pm z_1$ makes use
of Airy functions and it is the same as the one given above
for the case $a > 1$.
The parametrix near the imaginary branch points $\pm i z_2$
is also constructed with Airy functions. We give the construction
near $i z_2$. There are three contours, parts of $\Sigma$, meeting
at $iz_2$: left, right and vertical, see Fig.17.
We want an analytic $P$ in a neigborhood of $i z_2$
with jumps
\begin{equation} \label{jumpsP}
\begin{aligned}
    P_+ & = P_-
    \begin{pmatrix}
    1 & 0 & 0 \\ 0 & 1 & 0 \\
    0 & e^{n(\lambda_2-\lambda_3)} & 1
    \end{pmatrix}
    \qquad \textrm{on  left contour} \\
    P_+ & = P_-
    \begin{pmatrix}
    1 & 0 & 0 \\ 0 & 1 & e^{n(\lambda_3-\lambda_2)} \\
    0 & 0 & 1 \end{pmatrix}
    \qquad \textrm{on right contour} \\
    P_+ & = P_-
    \begin{pmatrix}
    1 & 0 & 0 \\ 0 & 0 & 1 \\
    0 & -1 & e^{n(\lambda_{3+}-\lambda_{3-})}
    \end{pmatrix}
    \qquad \textrm{on vertical part.}
\end{aligned}
\end{equation}
In addition we need the matching condition (\ref{matching}).
Except for the matching condition (\ref{matching}),
the problem is a $2\times 2$ problem. We define 
\begin{equation} \label{deffz}
    f(z) = \left[\frac{3}{4} (\lambda_2-\lambda_3)(z) \right]^{2/3}
\end{equation}
such that
\[ \arg f(z) = \pi/3, \qquad \textrm{ for } z = iy, \, y > z_2. \]
Then $s = f(z)$ is a conformal map, which maps $[0,iz_2]$ into the
ray $\arg s = - \frac{2\pi}{3}$, and which maps
the parts of $\Sigma$ near $iz_2$ in the right and left half-planes
into the rays $\arg s =0$ and $\arg s = \frac{2\pi}{3}$,
respectively.
The local parametrix
has the form,
\begin{equation} \label{ansatzPz}
    P(z) = E(z) \Phi\left(n^{2/3} f(z)\right)
    \begin{pmatrix} 1 & 0 & 0 \\ 0 & e^{\frac{1}{2} n (\lambda_2-\lambda_3)} & 0 \\
    0 & 0 & e^{-\frac{1}{2}n (\lambda_2-\lambda_3)}
    \end{pmatrix}
\end{equation}
where $E$ is analytic. The model matrix-valued function $\Phi$ is defined
as
\begin{equation} \label{defPhi}
\begin{aligned}
    \Phi&= \begin{pmatrix}
    1 & 0 & 0 \\ 0 & y_0 & - y_2 \\ 0 & y_0' & -y_2'
    \end{pmatrix}
    \qquad \textrm{for } 0 < \arg s < 2\pi/3, \\
    \Phi&= \begin{pmatrix}
    1 & 0 & 0 \\ 0 & y_0 & y_1 \\ 0 & y_0' & y_1'
    \end{pmatrix}
    \qquad \textrm{for } -2\pi/3 < \arg s < 0, \\
    \Phi&= \begin{pmatrix}
    1 & 0 & 0 \\ 0 & - y_1 & - y_2 \\ 0 & -y_1' & -y_2'
    \end{pmatrix}
    \qquad \textrm{for } 2\pi/3 < \arg s < 4\pi/3, 
\end{aligned}
\end{equation}
where $y_0(s) = \Ai(s)$, $y_1(s) = \omega \Ai(\omega s)$,
$y_2(s) = \omega^2 \Ai(\omega^2 s)$ with $\omega = 2\pi/3$
and $\Ai$ the standard Airy function. In order to achieve
the matching (\ref{matching}) we define the prefactor $E$ as
\begin{equation} \label{defE}
    E =  M L^{-1}
\end{equation}
with
\begin{equation} \label{defL}
    L = \frac{1}{2 \sqrt{\pi}} \begin{pmatrix} 1 & 0 & 0 \\
        0 & n^{-1/6} f^{-1/4} & 0 \\
        0 & 0 & n^{1/6} f^{1/4} \end{pmatrix}
         \begin{pmatrix} 1 & 0   & 0 \\
        0 & 1 & i \\ 0 & -1 & i \end{pmatrix}
\end{equation}
where $f^{1/4}$ has a branch cut along the vertical segment
$[0, iz_2]$ and it is real and positive where $f$ is
real and positive.
The matching condition (\ref{matching}) now follows from the
asymptotics of the Airy function and its derivative.
 
A similar construction gives the parametrix in the
neighborhood of $-iz_2$.

\subsection{Fourth transformation $S \mapsto R$}
Having constructed $N$ and $P$, we define the final transformation by
\begin{equation}
\begin{aligned}
R(z) & = S(z) M(z)^{-1} \qquad \textrm{away from the branch points,} \\
R(z) & = S(z) P(z)^{-1} \qquad \textrm{near the branch points}.
\end{aligned}
\end{equation}
Since jumps of $S$ and $N$ coincide on the interval $(-z_1, z_1)$
and the jumps of $S$ and $P$ coincide inside the disks around the
branch points, we obtain that $R$ is analytic outside a system of
contours as shown in Fig.18.

  \begin{figure}[h] \label{figure7}
  \centerline{
  \includegraphics[width=16cm,height=8cm]{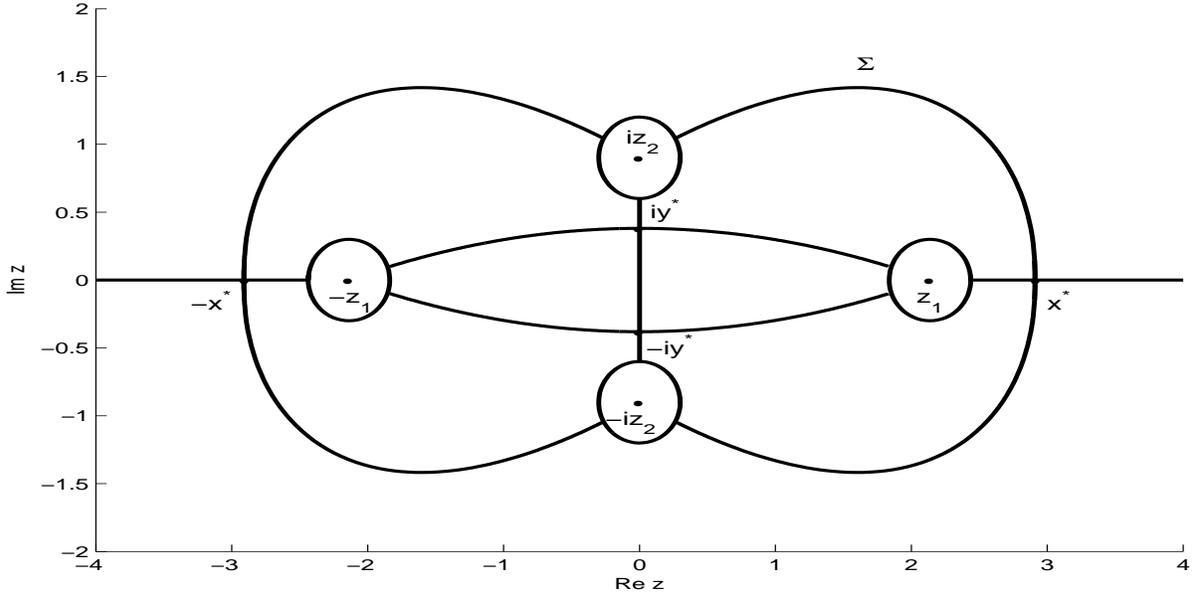}}
        \caption{$R$ has jumps on this system of contours.}
    \end{figure}

On the circles around the branch points there is a jump
\begin{equation} \label{jumpsR1}
    R_+ = R_- (I + O(1/n)),
 \end{equation}
which follows from the matching condition (\ref{matching}).
On the remaining contours, the jump is
\begin{equation} \label{jumpsR2}
    R_+ = R_- (I + O(e^{-cn}))
\end{equation}
for some $c > 0$.
Since we also have the asymptotic condition $R(z) = I + O(1/z)$
as $z \to \infty$, we may conclude as  that
\begin{equation} \label{Rasymp}
    R(z) = I + O\left(\frac{1}{n(|z|+1)} \right) \qquad \mbox{ as } n \to \infty,
\end{equation}
uniformly for $z \in \mathbb C$.

\section{Double scaling limit at $a=1$}

This section is based on the paper \cite{BK4}.

\subsection{Modified Pastur equation}
The analysis in  the cases $a > 1$ and $0 < a < 1$
was based on the Pastur equation (\ref{PasturEquation}), and it would be natural
 to use (\ref{PasturEquation}) also in the case $a=1$.
Indeed, that is what we tried to do, and we found that it works for $a\equiv 1$, but
in the double scaling regime $a= 1 + \frac{b}{2\sqrt{n}}$
with $b \neq 0$, it led to problems that we were unable to resolve
in a satisfactory way.
A crucial feature of our present approach is a {\it modification} of
the equation (\ref{PasturEquation}) when $a$ is close to $1$,
but different from $1$. At $x=0$ we wish to have a {\it double
branch point} for all values of $a$ so that the structure
of the Riemann surface is as in the middle figure of Figure \ref{Riemann_surface}
for all $a$.

For $c > 0$, we consider the Riemann surface for the equation
\begin{equation} \label{RSeq}
    z = \frac{w^3}{w^2 -c^2}
\end{equation}
where $w$ is a new auxiliary variable.
The Riemann surface has branch points at $z^* = \frac{3\sqrt{3}}{2} c$,
$-z^*$ and a double branch point at $0$.
There are three inverse functions $w_k$, $k=1,2,3$, that
behave as $z \to \infty$ as
\begin{equation} \label{wasym}
    \begin{aligned}
    w_1(z) &= z - \frac{c^2}{z} + O\left(\frac{1}{z^3}\right) \\
    w_2(z) &= c + \frac{c^2}{2z} + O\left(\frac{1}{z^2}\right) \\
    w_3(z) &= -c + \frac{c^2}{2z} + O\left(\frac{1}{z^2}\right)
    \end{aligned} \end{equation}
and which are defined and analytic on $\mathbb C \setminus [-z^*,z^*]$,
$\mathbb C \setminus [0, z^*]$ and $\mathbb C \setminus [-z^*,0]$,
respectively.

Then we define the modified $\xi$-functions
\begin{equation} \label{defxik}
    \xi_k = w_k + \frac{p}{w_k}, \qquad \mbox{for } k=1,2,3,
\end{equation}
which we also consider on their respective Riemann sheets.
In what follows we take
\begin{equation} \label{defc}
    c = \frac{a + \sqrt{a^2+8}}{4} \quad
        \mbox{ and } \quad p = c^2-1.
\end{equation}
Note that $a = 1$ corresponds to $c=1$ and $p=0$. In that case
the functions coincide with the solutions of the equation
(\ref{PasturEquation}) that we used in our earlier works.
From (\ref{RSeq}), (\ref{defxik}), and (\ref{defc}) we obtain
the modified Pastur equation
\begin{equation} \label{modPE}
    \xi^3 - z \xi^2  + (1-a^2) \xi + a^2 z
        + \frac{(c^2-1)^3}{c^2 z} = 0,
\end{equation}
where $c$ is given by (\ref{defc}). This equation has three solutions,
with the following behavior at infinity:
\begin{equation}\label{sch6}
\begin{aligned}
\xi_1(z)&=z-\frac{1}{z}+O\left(\frac{1}{z^3}\right),\\
\xi_{2,3}(z)&=\pm a+\frac{1}{2z}+O\left(\frac{1}{z^2}\right),\\
\end{aligned}
\end{equation}
and the cuts as in the middle figure of Fig.\ref{Riemann_surface}.
At zero the functions $\xi_k$ have the asymptotics,
\begin{equation} \label{xikf2g2} \xi_k(z) =
    \left\{ \begin{array}{ll} \di
     - \omega^{2k} z^{1/3}  f_2(z) -
    \omega^k z^{-1/3} g_2(z) + \frac{z}{3}
        & \mbox{ for } \Im z > 0, \\[10pt] \di
     - \omega^k z^{1/3}  f_2(z) -
    \omega^{2k} z^{-1/3} g_2(z) + \frac{z}{3}
        & \mbox{ for } \Im z < 0,
    \end{array} \right. \end{equation}
where the functions $f_2(z),g_2(z)$ are analytic at the origin and
real for real $z$, with
\begin{equation} \label{f2g2at0}
    f_2(0) = c^{2/3} + \frac{1}{3}c^{-4/3}(c^2-1),
    \qquad
    g_2(0) = c^{-2/3}(c^2-1).
\end{equation}
We define then the functions $\la_k$ as
\begin{equation} \label{deflambdak}
    \lambda_k(z) = \int_{0+}^z \xi_k(s) ds
\end{equation}
where the path of integration starts at $0$ on the upper
side of the cut and is fully contained (except for the initial point)
in $\mathbb C \setminus (-\infty,z^*]$, and we define the first transformation
of the RHP by the same formula (\ref{defT}) as for the case $a>1$.
For what follows, observe that the $\la$-functions have the following asymptotics
at the origin:
\begin{equation} \label{lambdakf3g3}
    \lambda_k(z) =
    \left\{ \begin{array}{ll} \di
     - \frac{3}{4} \omega^{2k} z^{4/3}  f_3(z)
     - \frac{1}{2} \omega^k z^{2/3} g_3(z) + \frac{z^2}{6}
        & \mbox{ for } \Im z > 0, \\[10pt] \di
    \lambda_{k-}(0) - \frac{3}{4} \omega^k z^{4/3}  f_3(z)
    - \frac{1}{2}  \omega^{2k} z^{2/3} g_3(z) + \frac{z^2}{6}
        & \mbox{ for } \Im z < 0,
    \end{array} \right. \end{equation}
where the function $f_3$ and $g_3$ are analytic at the origin and
\begin{equation} \label{f3g3at0}
f_3(0) = f_2(0) = c^{2/3} + \frac{1}{3}c^{-4/3}(c^2-1),
\quad
g_3(0) = 3 g_2(0) = 3 c^{-2/3}(c^2-1),
\end{equation}

  \begin{figure}[h] \label{fig_15}
  \centerline{
  \scalebox{0.8}{\includegraphics{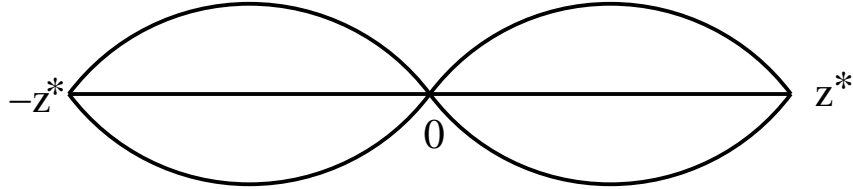}}}
        \caption{Lens structure for the double scaling limit.}
    \end{figure}

The second transformation, the opening of
lenses is given by formulas (\ref{defT_3}), (\ref{defT_4}). 
The lens structure is shown on Fig.19. The model solution
is defined as 
\begin{equation} \label{defM}
    M(z) = \begin{pmatrix}
    M_1(w_1(z)) & M_1(w_2(z)) & M_1(w_3(z)) \\
    M_2(w_1(z)) & M_2(w_2(z)) & M_2(w_3(z)) \\
    M_3(w_1(z)) & M_3(w_2(z)) & M_3(w_3(z))
    \end{pmatrix} \end{equation}
where $M_1$, $M_2$, $M_3$ are the three scalar valued functions
\begin{equation} \label{defM123}
    M_1(w) = \frac{w^2-c^2}{w \sqrt{w^2-3c^2}}, \quad
    M_{2}(w) = \frac{-i}{\sqrt{2}} \frac{w +c}{w \sqrt{w^2-3c^2}}, \quad
    M_{3}(w) = \frac{-i}{\sqrt{2}} \frac{w -c}{w \sqrt{w^2-3c^2}}.
    \end{equation}
The construction of a parametrix $P$ at the edge points $\pm z^*$
can be done with Airy functions in the same way as for $a>1$.

\subsection{Parametrix at the origin}

\begin{figure}[h]
\includegraphics[height=6cm,width=6cm]{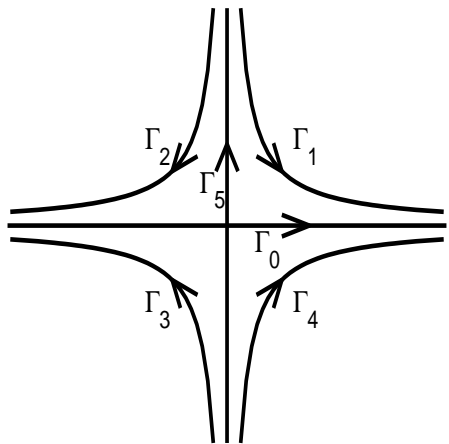}
\caption{The contours $\Gamma_j$, $j=0,1,\ldots,5$, equivalent to those in
(\ref{pi2}), that are used in the definition of the Pearcey integrals $p_j$.}
\label{fig:contoursGamma}
\end{figure}

The main issue is the construction of a parametrix at the origin,
and this is where the Pearcey integrals come in.
The Pearcey differential equation
$  p'''(\zeta) = \zeta p(\zeta) + b p'(\zeta)$
admits solutions of the form
\begin{equation} \label{pint}
    p_j(\zeta) = \int_{\Gamma_j} e^{-\frac{1}{4}s^4 - \frac{b}{2} s^2 +is\zeta} ds
\end{equation}
for $j=0,1,2,3,4,5$, where
\begin{equation}\label{pi2}
\begin{array}{ll}
\Gamma_0=(-\infty,\infty),\qquad &
\Gamma_1=(i\infty,0]\cup[0,\infty),\\
\Gamma_2=(i\infty,0]\cup[0,-\infty),\qquad &
\Gamma_3=(-i\infty,0]\cup[0,-\infty),\\
\Gamma_4=(-i\infty,0]\cup[0,\infty),\qquad &
\Gamma_5=(-i\infty,i\infty)
\end{array}
\end{equation}
or any other contours that are homotopic to them as for example given
in Fig.\ref{fig:contoursGamma}. The formulas (\ref{pi2}) also determine the orientation of
the contours $\Gamma_j$.

Define $\Phi = \Phi(\zeta;b)$ in six sectors by
\begin{align} \label{defPhi1}
    \Phi & =
    \begin{pmatrix} - p_2 & p_1 & p_5 \\
    -p_2' & p_1' & p_5' \\
    -p_2'' & p_1'' & p_5'' \end{pmatrix}
    \qquad \textrm{ for } 0 < \arg \zeta < \pi/4 \\
    \Phi & = \label{defPhi2}
    \begin{pmatrix} p_0 & p_1 & p_4 \\
    p_0' & p_1' & p_4' \\
    p_0'' & p_1'' & p_4'' \end{pmatrix}
    \qquad \textrm{ for } \pi/4 < \arg \zeta < 3\pi/4 \\
    \Phi & = \label{defPhi3}
    \begin{pmatrix} - p_3 & -p_5 & p_4 \\
    -p_3' & -p_5' & p_4' \\
    -p_3'' & -p_5'' & p_4'' \end{pmatrix}
    \qquad \textrm{ for } 3\pi/4 < \arg \zeta < \pi \\
    \Phi & = \label{defPhi4}
    \begin{pmatrix}  p_4 & -p_5 & p_3 \\
    p_4' & -p_5' & p_3' \\
    p_4'' & -p_5'' & p_3'' \end{pmatrix}
    \qquad \textrm{ for } -\pi < \arg \zeta < -3\pi/4 \\
    \Phi & = \label{defPhi5}
    \begin{pmatrix} p_0 & p_2 & p_3 \\
    p_0' & p_2' & p_3' \\
    p_0'' & p_2'' & p_3'' \end{pmatrix}
    \qquad \textrm{ for } -3\pi/4 < \arg \zeta < -\pi/4 \\
    \Phi & = \label{defPhi6}
    \begin{pmatrix} p_1 & p_2 & p_5 \\
    p_1' & p_2' & p_5' \\
    p_1'' & p_2'' & p_5'' \end{pmatrix}
    \qquad \textrm{ for } -\pi/4 < \arg \zeta < 0
    \end{align}
We define the local parametrix $Q$ in the form
\begin{equation} \label{constrdefQ}
    Q(z) = E(z) \Phi(n^{3/4} \zeta(z); n^{1/2} b(z)) e^{n \Lambda(z)} e^{-n z^2/6}, \qquad
    \Lambda = \diag(\lambda_1, \lambda_2, \lambda_3),
\end{equation}
where $E$ is an analytic prefactor, and 
\begin{equation} \label{defzeta}
    \zeta(z) = \zeta(z;a) = z \left[ f_3(z;a) \right]^{3/4}
\end{equation}
and
\begin{equation} \label{defb}
    b(z) = b(z;a) = \frac{g_3(z;a)}{f_3(z;a)^{1/2}}.
\end{equation}
The functions $f_3$, $g_3$ appear in (\ref{lambdakf3g3}), in the asymptotics
of the $\la$-functions.
In (\ref{defzeta}) and (\ref{defb}) the branch of
the fractional powers is chosen which is real and positive
for real values of $z$ near $0$. The prefactor $E(z)$ is defined as
\begin{equation} \label{defE_a}
    E(z) = -\sqrt{\frac{3}{2\pi}} i e^{-n b(z)^2/8}
    M(z) K(\zeta(z))^{-1}
    \begin{pmatrix} n^{1/4}&0&0\\0&1&0\\0&0&n^{-1/4}\end{pmatrix},
    \end{equation}
where
\begin{equation} \label{defK} K(\zeta) = \left\{
    \begin{array}{ll}
    \begin{pmatrix} \zeta^{-1/3}&0&0\\0&1&0\\0&0&\zeta^{1/3} \end{pmatrix}
    \begin{pmatrix} -\omega & \omega^2& 1\\
    -1 & 1 & 1 \\ -\omega^2 & \omega & 1\end{pmatrix} &
    \qquad \mbox{ for } \Im \zeta > 0, \\[20pt]
    \begin{pmatrix} \zeta^{-1/3}&0&0\\0&1&0\\0&0&\zeta^{1/3} \end{pmatrix}
    \begin{pmatrix} \omega^2 & \omega& 1\\
    1 & 1 & 1 \\ \omega & \omega^2 & 1\end{pmatrix} &
    \qquad \mbox{ for } \Im \zeta < 0.
    \end{array} \right.
\end{equation}

\subsection{Final transformation}
\begin{figure}[h]
\scalebox{0.7}{\includegraphics{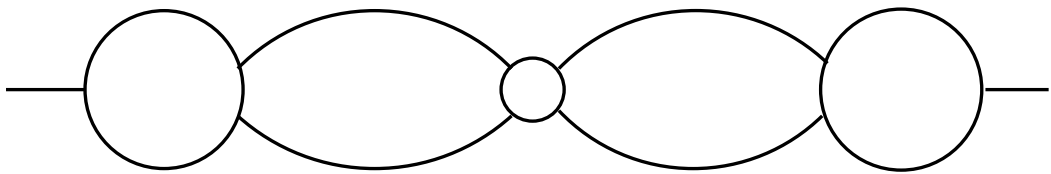}}
\caption{The contour $\Sigma_R$. The matrix-valued function $R$ is analytic
on $\mathbb C \setminus \Sigma_R$. The disk around $0$ has radius $n^{-1/4}$
and is shrinking as $n \to \infty$. The disks are oriented counterclockwise and the
remaining parts of $\Sigma_R$ are oriented from left to right.}
\label{fig7}
\end{figure}

We fix $b \in \mathbb R$ and let $a = 1 + \frac{b}{2\sqrt{n}}$ and we
define
\begin{equation} \label{defR}
    R(z) =
    \left\{ \begin{array}{ll}
        S(z) M(z)^{-1}, & \textrm{ for $z \in \mathbb C \setminus \Sigma_S$ outside the disks
        $D(0,n^{-1/4})$ and $D(\pm \frac{3\sqrt{3}}{2},r)$}, \\[10pt]
        S(z) P(z)^{-1}, & \textrm{ for } z \in D(\pm \frac{3\sqrt{3}}{2},r) \setminus \Sigma_S , \\[10pt]
        S(z) Q(z)^{-1}, & \textrm{ for } z \in D(0, n^{-1/4}) \setminus \Sigma_S.
        \end{array} \right.
        \end{equation}
Then $R(z)$ is analytic inside the disks and also across the real interval
between the disks. Thus, $R(z)$ is analytic outside the  contour $\Sigma_R$
shown in Fig.\ref{fig7}.
On the contour $\Sigma_R$ the function $R(z)$ has jumps, so that 
$R_+(z) = R_-(z)j_R(z)$, where
\begin{align} \label{jR1}
    j_R(z) &= I + O(n^{-1}) \qquad \mbox{uniformly for } \left|z \mp \frac{3\sqrt{3}}{2}\right| = r,  \\
    \label{jR2}
   j_R(z)  &= I + O(n^{-1/6}) \quad \mbox{ uniformly for } |z| = n^{-1/4},
   \end{align}
and there exists $c > 0$ so that
\begin{align}
   \label{jR3}
    j_R(z) & = I + O\left(\frac{e^{-cn^{2/3}}}{1+|z|^2}\right) \quad \mbox{ uniformly for $z$ on
    the remaining parts of $\Sigma_R$.}
    \end{align}
Also, as $z\to\infty$, we have $R(z)=I+O(1/z)$. This implies that
\begin{equation} \label{asympR1}
    R(z) = I + O\left(\frac{n^{-1/6}}{1+|z|}\right)
\end{equation}
uniformly for $z \in \mathbb C \setminus \Sigma_R$.

\section{Concluding remarks}

The Riemann-Hilbert approach is a new powerful method for random matrix models
and orthogonal polynomials. In this paper we reviewed the approach and some of its applications.
Let us mention some recent developments. The RH approach to orthogonal polynomials with complex 
exponential weights is considered in the recent work of Bertola and
Mo \cite{BM}. The RHP for discrete orthogonal polynomials and its applications is developed
in the monograph \cite {BKMM} of Baik, Kriecherbauer, McLaughlin and Miller.
Applications of random matrix models to the exact solution of the
six-vertex model with domain wall boundary conditions are considered in 
the works of Zinn-Justin \cite{ZJ3} and Colomo and Pronko \cite {CP}.
The RH approach to the six-vertex model with domain wall boundary conditions 
is developed in the work  of Bleher and Fokin \cite{BF}. The RHP for a
two matrix model is considered in the works of Bertola, Eynard, Harnad \cite{BEH2} and
Kuijlaars and McLaughlin \cite{KuM2}. The universality results for the scaling limit of correlation functions 
for orthogonal and symplectic ensembles of random matrices are obtained in the works of
Stojanovic \cite{Sto}, Deift and Gioev \cite {DG1},\cite {DG2}, Costin, Deift and  Gioev \cite {CDG}, 
Deift, Gioev, Kriecherbauer, and Vanlessen \cite{DGKV}.

\end{document}